\documentclass[astrosymb,trackchanges,twocolumn,twocolappendix,times,svgnames]{aastex701}
\usepackage[utf8]{inputenc}
\usepackage[T1]{fontenc}
\usepackage{amsmath,amssymb}
\usepackage{hyperref}
\usepackage{booktabs}
\usepackage{xcolor}
\usepackage{graphicx}
\usepackage{natbib}
\shorttitle{Drivers of ionizing escape in the Sunburst Arc}
\shortauthors{T. E. Rivera-Thorsen et al.}
\hypersetup{
 pdfauthor={T. Emil Rivera-Thorsen},
 pdftitle={The Sunburst Arc with JWST.},
 pdfkeywords={},
 pdfsubject={},
 pdfcreator={Emacs 30.1 (Org mode 9.8-pre)},
 pdflang={English}}
 
\begin{document}
\title{
  The Sunburst Arc with JWST. IV. The importance of interaction,
  turbulence, and feedback for Lyman-continuum escape} % \maketitle

\received{November 5, 2025}
\submitjournal{\apj}

\author[orcid=0000-0002-9204-3256, gname="T. Emil", sname="Rivera-Thorsen"]{
 T. Emil Rivera-Thorsen}
\affiliation{The Oskar Klein Centre, Department of Astronomy, Stockholm University, AlbaNova, 10691 Stockholm, Sweden}
\email[show]{trive@astro.su.se}
\correspondingauthor{T. Emil Rivera-Thorsen}

\author[orcid=0000-0003-1815-0114]{Brian Welch}
\affiliation{International Space Science Institute, Hallerstrasse 6, 3012 Bern, Switzerland}
\email{brian.welch@issibern.ch}

\author[orcid=0000-0001-6251-4988]{Taylor Hutchison}
\affiliation{Center for Research and Exploration in Space Science and Technology, NASA/GSFC, Greenbelt, MD 20771}
\email{taylor.hutchison@nasa.gov}

\author[orcid=0000-0001-8587-218X]{Matthew J. Hayes}
\affiliation{The Oskar Klein Centre, Department of Astronomy, Stockholm University, AlbaNova, 10691 Stockholm, Sweden}
\email{matt@astro.su.se}

\author[orcid=0000-0002-7627-6551]{Jane R. Rigby}
\affiliation{Astrophysics Science Division, Code 660, NASA Goddard Space Flight Center, 8800 Greenbelt Rd., Greenbelt, MD 20771, USA}
\email{jane.rigby@nasa.gov}

\author[orcid=0000-0001-6505-0293]{Keunho Kim}
\affiliation{IPAC, California Institute of Technology, 1200 E. California Blvd., Pasadena CA, 91125, USA}
\email{keunho11@ipac.caltech.edu}

\author[orcid=0000-0003-1343-197X]{Suhyeon Choe}
\affiliation{The Oskar Klein Centre, Department of Astronomy, Stockholm University, AlbaNova, 10691 Stockholm, Sweden}
\email{suhyeon.choe@astro.su.se}

\author[orcid=0000-0001-5097-6755]{Michael Florian}
\affiliation{Steward Observatory, University of Arizona, 933 North Cherry Avenue, Tucson, AZ 85721, USA}
\email{florianm@arizona.edu}

\author[orcid=0000-0003-1074-4807]{Matthew B. Bayliss}
\affiliation{Department of Physics, University of Cincinnati, Cincinnati, OH 45221, USA}
\email{matthew.bayliss@uc.edu}

\author[orcid=0000-0002-3475-7648]{Gourav Khullar}
\affiliation{Department of Astronomy, University of Washington, Physics-Astronomy Building, Box 351580, Seattle, WA 98195-1700, USA}
\affiliation{eScience Institute, University of Washington, Physics-Astronomy Building, Box 351580, Seattle, WA 98195-1700, USA}
\email{gkhullar@uw.edu}

\author[orcid=0000-0002-7559-0864]{Keren Sharon}
\affiliation{Department of Astronomy, University of Michigan, 1085 S. University Ave, Ann Arbor, MI 48109, USA}
\email{kerens@umich.edu}

\author[orcid=0000-0003-2200-5606]{Håkon Dahle}
\affiliation{Institute of Theoretical Astrophysics, University of Oslo, P.O. Box 1029, Blindern, NO-0315 Oslo, Norway}
\email{hdahle@astro.uio.no}

\author[orcid=0000-0002-0302-2577]{John Chisholm}
\affiliation{Department of Astronomy, University of Texas at Austin, 2515 Speedway, Austin, Texas 78712, USA}
\email{chisholm@austin.utexas.edu}

\author[orcid=0000-0003-3302-0369]{Erik Solhaug}
\affiliation{Department of Astronomy and Astrophysics, University of Chicago, 5640 South Ellis Avenue, Chicago, IL 60637, USA}
\affiliation{Kavli Institute for Cosmological Physics, University of Chicago, Chicago, IL 60637, USA}
\email{eriksolhaug@uchicago.edu}

\author[orcid=0000-0002-2862-307X,sname="Owens",gname="Riley"]{M. Riley Owens}
\affiliation{Department of Astronomy, University of California, Berkeley, Berkeley, CA 94720, USA}
\email{m.riley.owens@gmail.com}

\author[orcid=0000-0003-1370-5010]{Michael D. Gladders}
\affiliation{Department of Astronomy and Astrophysics, University of Chicago, 5640 South Ellis Avenue, Chicago, IL 60637, USA}
\affiliation{Kavli Institute for Cosmological Physics, University of Chicago, Chicago, IL 60637, USA}
\email{gladders@astro.uchicago.edu}

\begin{abstract}
  % Mapping the escape paths of Lyman-continuum (LyC) photons from starforming galaxies is a problem sensitive to physical properties on scales from the CGM and down to individual \ion{H}{2} region. Studying the smallest scales of the LyC escape paths is difficult; at redshifts low enough to allow the required spatial resolution,  mapping of LyC is made difficult or impossible by Galactic \ion{H}{1} absorption and the technical limitations of existing facilities.
  At present,  the best opportunity for detailed Lyman Continuum escape studies is in gravitationally lensed galaxies at $z \gtrsim 2$. Only one such galaxy currently exists in the literature with sufficient magnification and spatial resolution: The Sunburst Arc at redshift $z=2.37$.
Here, we present rest-frame optical JWST NIRSpec integral field observations of the Sunburst Arc that  cover a large fraction of the source plane. From this dataset, we generate precise maps of ISM kinematics, dust geometry, ionization, and chemical enrichment. We extract a stacked spectrum of five gravitationally lensed images of the Lyman-Continuum leaking star cluster, as well as a $\mu^{-1}$ weighted, integrated spectrum of most of the galaxy, enabling a direct comparison to other LyC leakers in the literature.
We find that the galaxy rotates but also shows strong, possibly dominant,  signatures of turbulence,  which are indicative of recent or ongoing major interaction. The cluster that leaks ionizing photons shows little difference from the rest of the galaxy in terms of kinematics or dust coverage, but dramatically elevated ionization, indicating that photoionization is the predominant mechanism that creates paths for  LyC escape. We propose that tidal stripping of \ion{H}{1} gas due to an interaction may have removed a large portion of the neutral ISM around the LyC emitting cluster,  making it easier for the cluster to completely ionize the rest.
% In this work, we present maps of kinematic and ionization properties of the Sunburst Arc and its Lyman-continuum leaking cluster, physically resolved down to scales of \(\sim 10\) pc.
% We present JWST/NIRSpec Integral Field Spectroscopy of three pointings along the arc, selected to cover the largest possible fraction of the arc while at the same time providing the best possible detail in areas of particular interest. We simultaneously fit of a selection of strong, nebular emission lines, and produce spaxel-wise maps of emission, kinematics, and derived ISM properties often used as diagnostics in the literature. We also derive global values of same properties, corrected for differential magnification, enabling direct comparison to unresolved galaxies from the literature.
% We find that the Sunburst Arc source galaxy overall is a very strong emission line galaxy with \(O_{32}\) and \(\Delta\)~[\ion{S}{2}] corroborating previously stated hypotheses that LyC is escaping through a very highly ionized channel aimed in our direction. Kinematically, we find the galaxy to be a highly perturbed rotating galaxy, with peculiar regions of broader lines, possibly originating from additional, Nitrogen-enriched, outflowing components.
\end{abstract}


\section{Introduction}
\label{sec:introduction}

A few hundred million years after the Big Bang, ionizing photons, termed Lyman-Continuum (LyC) radiation, from the ﬁrst stars began to ionize the then-neutral IGM \citep[e.g.,][]{hashimoto2018nature,haardt2012,fauchergiguere2020} in what is called the Epoch of Reionization (EoR). Passageways must thus have been present through the neutral ISM of these galaxies, which allowed LyC to leak out into the surrounding IGM, but it is not yet clear how these passageways were carved. In order to have reionized the universe, an estimated fraction of $\sim$10–20\% of ionizing photons must have escaped their galaxies \citep{robertson2015,naidu2020a}, with some studies suggesting it might be more \citep{becker2021,davies2021}.
The ionizing escape fraction cannot be directly measured in the EoR, because the IGM is opaque to ionizing photons at redshifts $z \ga 4$ \citep{inoue2014}. All studies of escaping LyC must use lower-redshift galaxies to understand the underlying mechanisms.  Such Lyman-Continuum Emitting galaxies (LCE) are rare in the local Universe, but become more common at higher redshifts, as both cosmic star formation activity and ionizing background grow larger. Since the first detection of a Lyman-Continum Emitting galaxy \citep[LCE;][]{bergvall2006}, a small number of LCEs have been detected in the local Universe \cite[see e.g.][]{bergvall2006, leitet2011, puschnig2017, ostlin2021, komarova2024, leitherer2016}; and at slightly higher redshifts, \(0.1 \lesssim z \lesssim 0.5\) a considerable number of Green Pea \citep{cardamone2009} and similar galaxies have found to be LCEs \cite[e.g.][]{borthakur2014, jaskot2019, wang2021, malkan2021b, flury2022a}. At redshifts $z \ga 0.5$, about \(\sim 65 \) LCEs have been identified, depending on delineation criteria for candidates \citep{vanzella2012, mostardi2015, debarros2016, shapley2016, vanzella2016, bian2017, chisholm2018a, fletcher2018, steidel2018, vanzella2018, riverathorsen2019, riverathorsen2022, saha2020, marqueschaves2021, marqueschaves2022, ji2020, flury2022a, pahl2021, saxena2022, kerutt2024}. 

Since it is not possible to directly measure galaxy LyC emission during the EoR, much work has been put into the identification and characterization of secondary observables which correlate with LyC escape, and the dual, intertwined questions of a) how they statistically scale with \(f_{\text{esc}}^{\text{LyC}}\), and b) what they reveal about the mechanisms that carve out the paths of escape for the ionizing radiation. Very simplified, these mechanisms fall into the families of radiative versus mechanical feedback, which give rise to different escape scenarios.  In a spherically symmetric, dust-free toy model of a young star cluster enveloped in  an isotropic, neutral ISM, photoionization over time leads to a falling neutral fraction of the ISM in a growing Strömgren sphere which may eventually grow larger than the boundary of the ISM. This scenario is often called the ``ionization bounded'' scenario, where there is still an intact neutral envelope present, and the ``density bounded scenario'' when the Strömgren sphere has grown larger than the gas envelope \citep[see e.g.][for an overview]{zackrisson2013}. In such a scenario, \(f_{\text{esc}}^{\text{LyC}}\) depends solely on the column density of the residual neutral gas fraction in the ISM.

In contrast, radiative pressure and mechanical feedback such as stellar winds or momentum deposited from Supernovae, can result in large-scale ISM outflows, in which Rayleigh-Taylor instabilities will result in a breakup of this expanding bubble into a clumpy, highly anisotropic, expanding medium, in which some lines of sight will be transparent to ionizing photons, while others are optically thick. This is often referred to as the ``perforated ionization bounded scenario'' or the ``picket fence scenario'' \citep[e.g.][]{heckman2011,zackrisson2013, james2014,jaskot2014}; in this limit,  \(f_{\text{esc}}^{\text{LyC}}\) is regulated solely by the global or line-of-sight (LOS) covering fraction of the neutral gas. Real-life scenarios are of course much more complex, but the escape fraction can still be considered as regulated by the covering fraction of neutral gas clumps and the residual neutral column in the inter-clump medium. 

A number of observables either correlate with, or directly encode, these two properties. One of the most direct measurements of the line-of-sight covering fraction is the equivalent width of \ion{H}{1} absorption lines \citep{heckman2011,jones2013,gazagnes2018,chisholm2018a}, although for practical reasons, neutral metal lines are often used as proxies for these \citep[e.g.][]{riverathorsen2017,alexandroff2015,jones2013,borthakur2014,heckman2011}. However, absorption lines are very observationally expensive, especially at high redshifts and thus often not practical for EoR studies.
While neutral absorption lines directly trace the covering fraction, the properties of the Lyman-\(\alpha\) line directly traces the neutral column density in the inter-clump medium, although the strongly resonant nature of the line allows it to traverse open channels without any direct open lines of sight \citep[e.g.][]{gronke2016letter,dijkstra2016,dijkstrarev,verhamme2015,verhamme2008,verhamme2006}. A number of line properties have been found to correlate with LyC escape, such as e.g. velocity offset of the red peak or separation of the dual peaks \citep{verhamme2008,verhamme2015,jaskot2024a,jaskot2024b}, or the steepness of the red peak \citep{kakiichi2021}. The presence of a narrow line component at systemic velocity has been predicted \citep{behrens2014,verhamme2015,riverathorsen2017} and confirmed \citep{riverathorsen2019} to indicate the presence of a line of sight of extremely low \(N_{\text{H I}} \lesssim 10^{13} \text{cm}^{-2}\). However, due to the predominantly neutral IGM at the EoR, Ly\(\alpha\) emission cannot be depended on to trace ionizing escape at that time. 

Other observables correlate either with high ionizing production rate, high escape fractions, or with physical conditions that bring about these conditions. In the latter category is the finding that LyC escape seems to correlate with broad emission line components, which in turn indicate the presence of large scale bulk outflows which can rupture and thin out the neutral ISM and create escape channels \citep[see e.g.][and references therein]{mainali2022}. However, in a study of low-redshift extreme emission line galaxies (EELGs), \cite{jaskot2019} found that outflows tended to be weaker at higher ionization, which they interpret as the onset of catastrophic cooling of neutral gas clumps which subsequently are photoionized by the still-proximate ionizing sources; marking the transition to a regime dominated by the density bounded scenario. LyC escape has also been found to correlate with star formation surface density \citep[e.g.][and references therein]{jaskot2024a}.
% Of course, galaxies do not consist of single star clusters surrounded by an isotropic ISM; and any real-life scenario consists of contributions from both types of scenario in a much more complex geometry. Still, these scenarios form a good basic framework for building an understanding of LyC escape.
% Of course, real escape scenarios are much more complex than these toy examples, but together, they do span most more complex scenarios, in which the When considering more realistic escape scenarios, one can think of the 

Besides properties internal to the source galaxy, the impact of the environment is a topic of active debate and research. Major mergers and IGM accretion events are predicted to spark strong star formation episodes, which in turn would increase the ionizing photon production, and feedback from this star formation could lead to increased escape fractions \citep[e.g.][]{kostyuk2024}. Major interactions could also lead to strong turbulence, which in simulations have shown to be able to create ionizing escape paths through the ISM \citep[e.g.,][]{kakiichi2019}. 

Observations have been ambiguous as to the importance of mergers. \cite{zhu2025} find from visual morphological classification of a sample of 23 \(z \sim 3\) LCEs in GOODS-South that 20 of them are mergers, a dramatic overrepresentation relative to the general merger fraction at \(z \sim 3\) \citep{duan2025,puskas2025}. In their simulated high-redshift IllustrisTNG galaxies, \cite{kostyuk2024} find that not only star formation induced by mergers, but also tidal stripping of the surrounding \ion{H}{1} envelope plays a crucial role in facilitating LyC escape. Interestingly, this effect was observed directly by \cite{lereste2024}, who used 21 cm interferometry to directly map \ion{H}{1} in the local-Universe, interacting LCE Haro 11. These authors found that tidal stripping by the companion galaxy had offset large portions of the \ion{H}{1} envelope away from the star forming knots in that galaxy, helping to clear out the ionizing escape paths. 
% \textcolor{ForestGreen}{Continue here with something about more realistic escape scenarios and their physical and observable fingerprints, list key papers in the literature, and illustrate why we need spatially resolved observations to properly understand this.}
% In addition, the relation between obsereved line-of-sight (LoS) escape fraction, and the global escape fraction relevant for the cosmology of the EoR is non-trivial and highly dependent on the relative importance of various mechanisms.

Studying LCEs on the small physical scales required to map the ionizing escape paths is complicated by the fact that at redshifts low enough to allow the required spatial resolution, LyC emission is either completely absorbed by Galactic \ion{H}{1} or, where the redshift is large enough to allow a window of escaping wavelengths, the localization of the LyC source is limited by the \(2\farcs5\) aperture of the Cosmic Origins Spectrograph; the only instrument which can effectively detect LyC at these redshifts. Samples at low redshifts redshifts are also limited by the small Cosmic volume that makes up the local Universe, as well as the low prevalence of Lyman Continuum Emitters (LCEs) in the present-day Universe. 

At higher redshifts, where LyC has been shifted into the detectable wavelength ranges of HST imaging instruments, spatial detail is lost to distance.
However, magnification by gravitational lensing can in fortuitous cases image galaxies at cosmological distances at a level of detail that is  usually only seen in the local Universe. At the same time, at these distances, the ionizing radiation has been redshifted into ranges where it is comfortably observable in the WFC3 UVIS channel of HST, at physical de-magnified resolutions down to a few tens of parsecs or occasionally even better. As a result, such gravitationally lensed LCE galaxies currently present the best available opportunity for this kind of  high-detail mapping of ionizing escape paths. 

However, such lensed LCEs are rare. While searches are ongoing, at the time of writing, only one lensed LCE of sufficient brightness for detailed spatially and spectrally resolved studies is found in the literature; the  \object{Sunburst Arc} \citep{dahle2016, riverathorsen2019} at redshift \(z = 2.37\). The Sunburst Arc was discovered during optical follow-up of Sunyaev-Zeldovich galaxy cluster candidates in the ESA Planck Survey \citep{plancksz2014} by \citet{dahle2016}, who found it to be the brightest known gravitationally lensed galaxy with an integrated magnitude of \(m_R = 17.82\). Lensing by a redshift $z = 0.44$ foreground cluster generates a total of 12 total or partial images \citep{dahle2016,sharon2022,pignataro2021,diego2022,solhaug2025}. \citet{sunburst2017} found that the galaxy displayed a unique, triple-peaked Ly\(\alpha\) emission profile, which is predicted to arise from Ly\(\alpha\) radiative transfer in an expanding shell of H\textsc{i} perforated by a narrow channel devoid of H\textsc{i} along our line of sight \citep{behrens2014,verhamme2015}. \citet{sunburst2017} estimated the neutral column density along this line of sight to be extremely low, \(\log N_{\text{H\textsc{i}}} [\text{cm}^{-2}] \lesssim 13\), while one optical depth for LyC corresponds to \(\log N_{\text{H\textsc{i}}} \approx 17.2\); orders of magnitude higher. 
%The galaxy has been observed to have a comparatively low molecular gas fraction \citep{solimano2021}. 
% \textcolor{blue}{\emph{Continue a little bit more about the SBA and what has been found out about it relevant to the LyC escape angle \citep{mainali2022,kim2023,owens2024}}}
The galaxy is dominated by a very young \citep[3--5 Myr][]{chisholm2019,vanzella2020,pascale2023,mestric2023,riverathorsen2024} and massive \citep[$M_{\star} \approx M_{\text{dyn}} \lesssim 10^{7} M_{\odot}$][]{vanzella2020,riverathorsen2024} Lyman Continuum emitting \citep[][]{riverathorsen2019} star cluster, from hereon referred to as ``the LCE''. This cluster is dense and massive and contains a large population of Wolf-Rayet stars and Very Massive Stars \citep{pascale2023,mestric2023,riverathorsen2024}, and a strong enrichment in both doubly \citep{pascale2023} and singly \citep{welch2025} ionized Nitrogen, as has also been observed in a number of high-redshift galaxies with JWST \citep[e.g.][and references therein]{ji2025arxiv}. The LCE cluster has a very steep rest-frame UV slope of $\beta = -2.8$ \citep{kim2023}, and \cite{mainali2022} found strong broad emission line components in ground-based, rest-frame optical spectroscopy, which those authors interpreted as the presence of strong bulk outflows driven by stellar winds from the young stellar population of the cluster. Based on this, \citeauthor{mainali2022} concluded that mechanical feedback was the main driver of ionizing escape in the Sunburst Arc, presumably by driving the expansion and rupturing the surrounding neutral ISM.  The present NIRSpec IFU observations provide an opportunity to spatially map the kinematic properties which have previously only been attainable in integrated slit spectra through atmospheric distortion, among other things enabling us to test this scenario more thoroughly.

This paper assumes a standard \(\Lambda\)-CDM cosmology with \(H_0 = 70\) km s\(^{-1}\) Mpc\(^{-1}\), $\Omega_{M} = 0.3\text{, and }\Omega_{\Lambda} = 0.7$. Unless otherwise noted, all images are oriented with N  up and E left. We use the abbreviations [\ion{O}{1}] for [\ion{O}{1}] \(\lambda\) 6300, [\ion{O}{3}] for [\ion{O}{3}] \(\lambda\) 5008, [\ion{O}{2}] for [\ion{O}{2}] \(\lambda \lambda\) 3727+29, [\ion{N}{2}] for [\ion{N}{2}] \(\lambda\) 6584, [\ion{Ne}{3}] for [\ion{Ne}{3}] \(\lambda\) 3869, [\ion{S}{2}] for [\ion{S}{2}] \(\lambda \lambda\) 6717+31.

\section{Observations}
\label{sec:observations}

The JWST \citep{gardner2023} data analyzed in this paper have also been included in the previous papers in this series; \cite{riverathorsen2024} \defcitealias{riverathorsen2024}{Paper~I} \citepalias{riverathorsen2024}, \cite{choe2025}\defcitealias{choe2025}{Paper~II}\citepalias{choe2025}, and \cite{welch2025}\defcitealias{welch2025}{Paper~III}\citepalias{welch2025}. The observations were obtained with the NIRSpec instrument \citep{boeker2023} in the IFS mode with the G140H and G235H gratings, and and with the NIRCam instrument \citep{rieke2023}.  These observations were  obtained over the period from April 4, 2023, to April 10, 2023, as part of program GO-2555 (PI Rivera-Thorsen). 

\begin{figure*}[ht!]
\centering
\includegraphics[width=.995\linewidth]{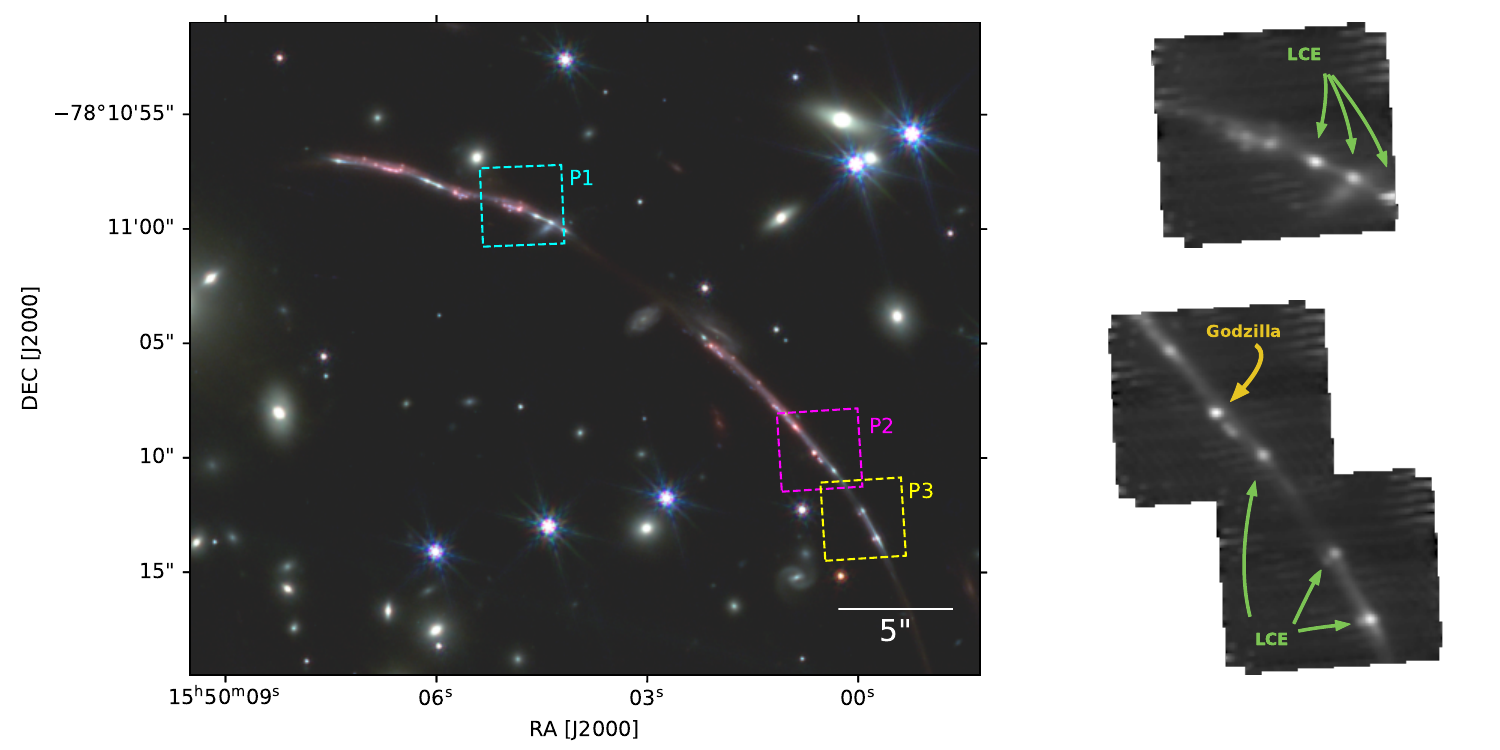}
\caption{\label{fig:nircamrgb}Left: NIRCam color composite of the N and NW segments of the Sunburst Arc and member galaxies of the lensing foreground cluster. The approximate footprints of NIRSpec IFU pointings are overlaid in cyan (P1), magenta (P2), and yellow (P3). The RGB channels show the rest-frame near-IR in the NIRCam filter F444W, rest-frame green in F200W, and rest-frame blue/optical in F115W, respectively. The image is oriented with N up, E to the left. Right: Median NIRSpec IFU cube of P1 (upper) and P2+P3 (lower). Six lensed images of the LCE cluster plus the object "Godzilla" \citep{diego2022,choe2025} are marked. All three panels are shown using square root -stretched color scales, to strike a balance between bright and faint features.}
\end{figure*}

The full field was observed with NIRCam in the filters F115W, F150W, F200W, F277W, F356W, and F444W. Four pointings were observed with NIRSpec; three on-target pointings and one off-target for background subtraction. Each of the NIRSpec pointings were observed in both the F100L/G140H and F170L/G235H settings, together covering the entire rest-frame Optical range at \(2900 \AA \la \lambda_0 \la 9700 \AA\).

\autoref{fig:nircamrgb} shows an RGB composite of the NIRCam F444W, F200W, and F115W imaging observations of the N and NW arc segments. Approximate footprints of the NIRSpec IFU pointings are overlaid in cyan, magenta, and yellow.

The NIRSpec pointings were designed such that P1 covers the largest possible part of the galaxy in one pointing; P2 is centered around the peculiar clump in its center, denoted ``Tr'' in \citet{vanzella2020}, or ``Godzilla'' by \citet{diego2022}; and P3 is placed to cover two strongly magnified images of the LCE cluster, to optimize the amount of information we can get about this cluster and its closest surroundings. 
Since the lensing shear of P1 is approximately perpendicular to the lensing shear in P2 and P3 (according to the lensing model of \citealt{sharon2022}), this choice of pointings optimizes the amount of spatial information that can be recovered. 

\begin{figure}[htbp]
\centering
\includegraphics[width=0.8\columnwidth]{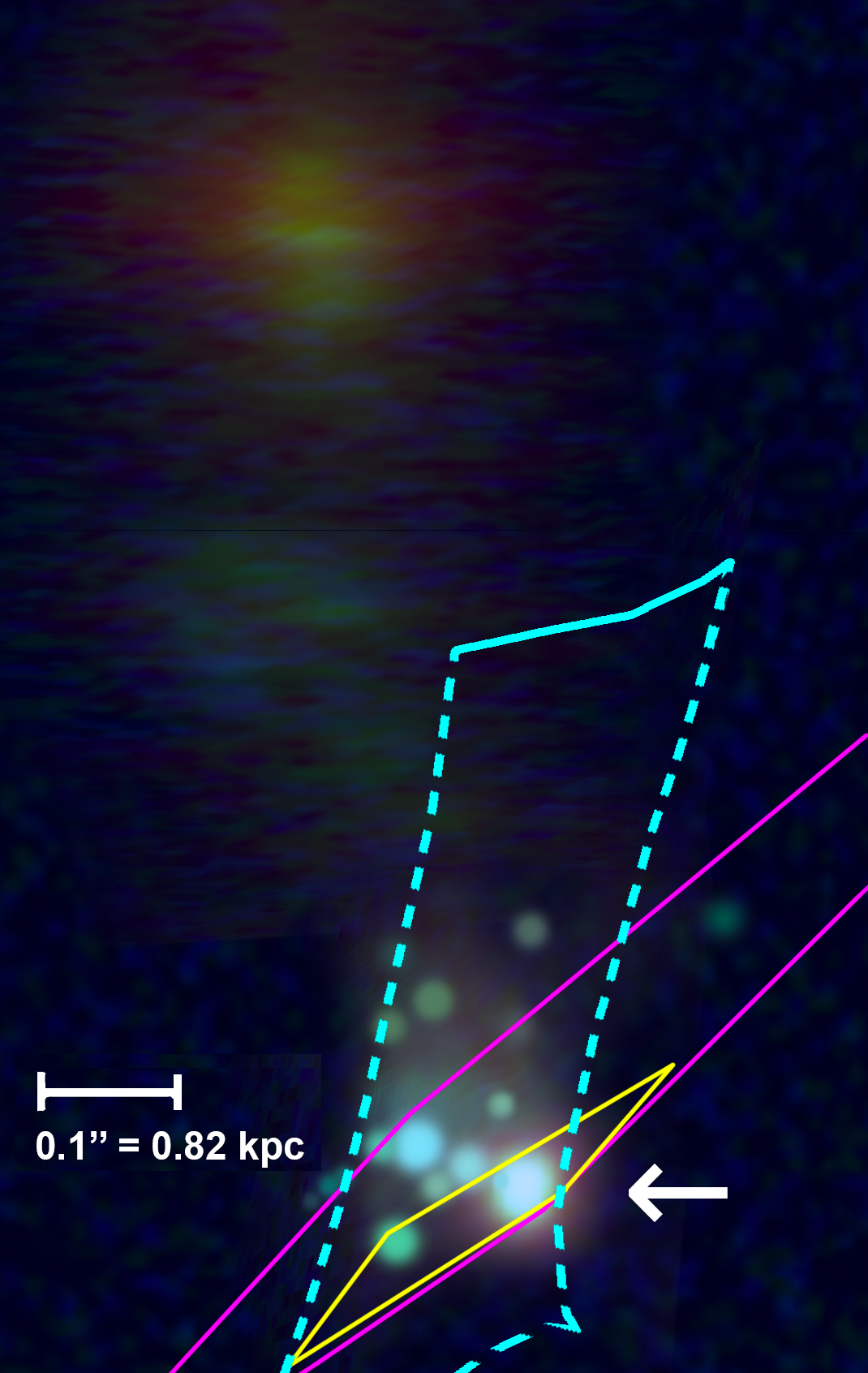}
\caption{\label{fig:artist}An artist's impression approximate view of the de-lensed source galaxy, with the approximate de-lensed footprints of the three IFU pointings overlaid along with an approximate physical scale bar. While the upper, diffuse areas are ray traced from image plane to source plane, the approximate position, color and brightness of the observed clumps/clusters seen in the bottom parts are added by hand. The arrow points to the LCE cluster. Adapted from \citet{sharon2022}.}
\end{figure}

Fig.\ \ref{fig:artist} shows an artistic approximation of the de-lensed Sunburst Galaxy based on HST imaging, identical to the left panel of Fig.\ 11 in \citet{sharon2022}, except here we have overlaid the approximate de-lensed IFU footprints; the direction in which each distorted square is shortest is the direction of its maximal spatial resolution. The figure consists of an image of the full, de-lensed arc as seen in the W and SE arc segments in \citet[see][Fig.\ 2]{sharon2022}, onto which are superimposed by hand the more precise positions, colors, and brightnesses of the better resolved clumps found in the N and NW arc segments. The redder part in the N is a likely interacting companion galaxy, while the blue-green blurry parts in the center are likely to be a tidal bridge or similar feature.

NIRSpec Integral Field Unit (IFU) data were reduced following the methods presented in \citet{rigby23_overview}. We used the JWST data reduction pipeline version 1.11.4 \citep{bushouse23_pipeine1p11p4}, with the calibration reference files from context \texttt{pmap\_1105}. As discussed in \citet{rigby23_overview}, the electronics of the NIRSpec detectors undergo small-scale fluctuations in temperature, which create  pattern noise. We correct this detector pattern noise using the NSClean software \citep{rauscher23_nsclean}.

The raw NIRSpec data contains a large number of outlier pixels due to cosmic rays. The standard pipeline includes an outlier rejection step; however, this step does not adequately remove outliers from our final data products. We therefore use the baryon-sweep code \citep{baryonsweep} to remove remaining outliers and artifacts. \cite{hutchison23_sigmaclip} found  that a combination of the standard pipeline outlier rejection and the baryon-sweep method produces the cleanest final data products.

To take full advantage of the magnifying effects of gravitational lensing, we produce two complementary data cubes, one with the standard \(0 \farcs 1\) spaxel 
% \(N_{22}\) 
resolution, and another with a finer \(0 \farcs 05\) spaxels. We use the \texttt{Spec3} pipeline's \texttt{cube\_build} to produce these cubes, by setting the \texttt{scalexy} parameter. 

Single-spaxel spectra of point-like sources from either of JWST's integral field units (MIRI MRS and the NIRspec IFU) are susceptible to a ``wiggle'' artifact showing as faint, very broad waves in the continuum , which arises in each case because the detector does not critically sample the point spread function \citep{law23_3ddrizzle}. These wiggles are much broader than emission lines and easily removed in continuum subtraction, but they do give rise to these Moiré-like patterns in ratio maps. For the NIRSpec IFU, drizzling this undersampled data to even smaller output spaxels exaggerates the artifact (when using a standard 4-point dither as was done in this program). As noted by \citet{law23_3ddrizzle}, using a larger extraction aperture around point sources mitigates this artifact, even in the \(0\farcs05\) NIRSpec data cubes. We find that the benefit from improved spatial resolution for extended sources outweighs the additional artifacts seen in point-like sources within this dataset. Hence, we have opted to analyze the 0.\arcsec05 data cubes.

\section{Methods}
\label{sec:orgaa62de9}
\subsection{Correction of Uncertainties}
\label{sec:errcorr}
Fitting of emission lines in a spectrum requires a good characterization of the uncertainty spectrum.  Underestimated uncertainties will lead to disproportionately strong weighting of the wings of strong emission lines, which in turn will lead to over-estimation of line widths, under-estimation of peak intensity, and poor recovery of the integrated line flux.  We find that the data products produced by the pipeline have uncertainties under-estimated by a factor of  \(\sim 4\) for some spatial regions.  Here we describe this measurement and how we rescale the estimated uncertainties. 
% \emph{\textcolor{red}{TODO:} I seem to remember some communications from STScI about this?}.

We estimated the statistical uncertainty in every wavelength bin in the Level 3 datacubes as follows. We first manually constructed masks covering regions of empty sky. These masks were placed to minimize line emission from the source galaxy, but were also kept well clear of edges of each pointing, since these regions are noisier and not representative of the noise properties of the spaxels covering the target. The masks are shown in \autoref{fig:noisemasks}. 

For each wavelength bin, we then took the standard deviation of the data within this mask, after first sigma-clipping the data by \(5\sigma\) to avoid artifacts like hot pixels and cosmic rays from skewing the distribution. We smoothed the resulting error spectrum by a running median kernel to remove any remaining individual bins of inflated error.

For each spaxel, we then assumed a new uncertainty spectrum, as the greater of the sky-region--derived uncertainty and the pipeline-provided spectrum.  This  ensured that individual wavelength bins and spaxels with higher uncertainty still were treated correctly, while the floor of the uncertainty is set by the measured standard deviation in sourceless pixels.

\subsection{Stacked spectrum of the LyC leaking knot  \label{sec:lcestack}}
\label{sec:org7ed81fc}
Of the twelve gravitationally lensed images of the LCE cluster, five are fully captured by the NIRSpec IFU pointings.  Since these are all physically the same object, we combined their spectra by extracting each in a \(3 \times 3\) pixels aperture, normalized each spectrum by its median value to standardize their differently magnified fluxes, before stacking them by taking a flux-weighted average. All quantities derived from this spectrum are thus given in relative \(f_{\nu}\) or \(f_{\lambda}\) units. 

\begin{figure*}
\centering
\includegraphics[width=.9\linewidth]{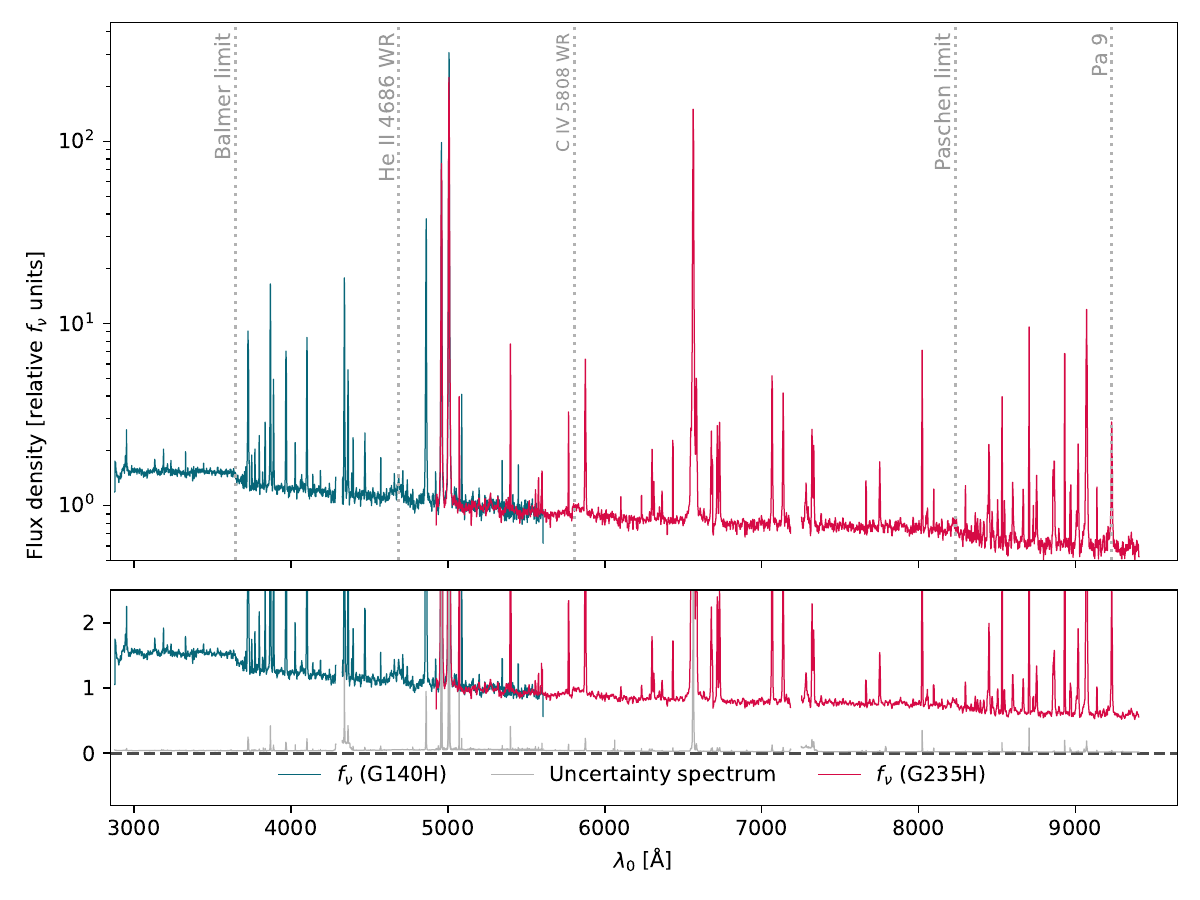}
\caption{\label{fig:leakerspec}Restframe blue/optical spectrum of the LCE cluster in the Sunburst Arc. The upper panel shows the flux on a log \(y\)-axis to display features on a wide range of scales, while the lower panel shows continuum and uncertainty spectrum on a linear scale. As is visible in \autoref{fig:lineid1}, the two features slightly offset from each other at \(\lambda_0 \sim 5100\) Å are due to outliers.}
\end{figure*}

\autoref{fig:leakerspec} shows the full stacked, rest-frame blue and optical spectrum of the LCE cluster. The spectrum contains a large number of emission lines; we have identified 59 nebular lines, of which \(\sim\) 30 are Hydrogen recombination lines (including about 20 Balmer lines and 10 Paschen lines).  The spectra also feature a set of broad, blended, stellar Wolf-Rayet line complexes that were discussed in \citet{riverathorsen2024}.  

In \autoref{app:lines}, we give some more details about the unstacked and stacked LCE spectra. In \autoref{fig:unstacked}, we show the individual median-normalized, but otherwise unaltered, extraced spectra of each of the five gravitationally lensed images, along with the stacked spectrum itself for comparison. The similarity between the spectra is remarkable, and lends strong support to our claim that the images are indeed of the same object. Next, \autoref{fig:lineid1} and \autoref{fig:lineid2} show visual identification of the emission lines tabulated in \autoref{tab:linefluxes}. 
% identify all these features.
To enable conversion to physical units, we report the measured and
demagnifed flux of H$\beta$ for the image of the LCE that appears in image 4 of the lensed galaxy: the measured flux is $1.01 \pm 0.03 \times 10^{-17} \text{erg s}^{-1} \text{cm}^{-1}$, and the demagnfied flux is \(6.6 \pm 0.2 \times 10^{-19} \text{erg s}^{-1} \text{cm}^{-1}\), assuming the calculated
magnfication of \(\mu= 15.3\) from K. Sharon et al. (2022). 

\subsection{\texttt{CubeFitter.jl}: A new software package for fitting line fluxes and kinematics in spectral cubes}
\label{sec:org8326217}

For spectral line modeling and generation of maps of line flux and kinematics, we used the Julia package \texttt{CubeFitter.jl} \citep{riverathorsen2025cubefitter}\footnote{\url{https://github.com/thriveth/CubeFitter.jl}},
%\footnote{\url{https://github.com/thriveth/CubeFitter.jl}}
a software package that quickly and flexibly models emission lines in spectroscopic datacubes.
CubeFitter is written in the \texttt{Julia} programming language \citep{bezanson2017julia}, to strike a balance between computation speed and simple, flexible development.
Julia has proven for our tasks to be \(\sim \times 100\) faster than Python, while taking up a number of lines of code and a level of coding complexity similar to Python code of equal functionality. \texttt{CubeFitter} relies on a number of existing Julia packages for its core functionality, most importantly the general curve fitting package \texttt{GModelFit.jl}\footnote{\url{https://gcalderone.github.io/GModelFit.jl}} for model building infrastructure and fitting functionality, \texttt{Measurements.jl}\footnote{\url{https://juliaphysics.github.io/Measurements.jl}} for propagation of uncertainties, and optionally \texttt{VoronoiBinning.jl}\footnote{\url{https://github.com/Michael-Reefe/VoronoiBinning.jl}}, a Julia implementation of the original \texttt{Vorbin} package \citep{capellari2003}.

The \texttt{CubeFitter.jl} package loads a FITS datacube
%\footnote{Currently, only NIRSpec IFU cubes are implemented, but support for MIRI and MUSE, as well as a generic spectral cube type, are in preparation.}
into a data structure with relevant, partially user-provided, metadata such as instrument resolving power, data units, and more. The program ships with a list of emission lines, mainly taken from Drew Chojnowski's online line list\footnote{\url{http://astronomy.nmsu.edu/drewski/tableofemissionlines.html}}, in turn compiled mainly from data from NIST \citep{nist_asd}. A few additional lines are taken from other places, see the package documentation.  Alternatively,  the user can pass their own line list to the code. The package incorporates the tables of spectral resolution for NIRSpec\footnote{available for download at \url{https://jwst-docs.stsci.edu/jwst-near-infrared-spectrograph/nirspec-instrumentation/nirspec-dispersers-and-filters\#gsc.tab=0}. These are pre-launch estimates; although in-orbit measurements have recently been made \citep{shajib2025arxiv}, they have not yet yielded data products suitable for implementation in code like this.}.

An optional function allows the user to subtract the continuum under the emission lines. The function, \texttt{cont\_subt}, estimates the continuum by masking out a wavelength range around a set of the strongest emission lines, sigma clipping the remaining data bins to 3 \(\sigma\), and taking a running median within a broad kernel (51 pixels by default, but settable through an option argument to the function). This running median is then interpolated into the masked pixels and subtracted from the original spectrum, and the procedure is repeated for better masking and sigma clipping. Caution should be given to pick a median window width suitable for the data; specifically, it will subtract all features with a width \(\lesssim \) this window.

\texttt{CubeFitter} can either extract individual spaxels, or co-add arbitrary rectangular cut-outs or image segments into one-dimensional spectra. When co-adding spectra, the uncertainties are expected to be random and independent in neighboring spaxels and wavelength bins, such that they can be propagated by addition in quadrature in the standard way.

Emission lines in an extracted spectrum are fitted simultaneously with Gaussian profiles, assuming that all lines follow the same kinematic structure with a shared redshift and velocity width, leaving line flux as the free parameter for each line. By default, the code attempts to fit all lines in the line list.  Optionally, a custom subset of these can be passed, which enables fitting of multiple subsets or individual lines to different kinematic parameters. The code can also fit any line or set of lines with locked kinematic properties, or fit the kinematic properties from a subset of lines and use that fit as a kinematic template for the rest.  These features allow for very flexible modeling strategies.

Data regions with undetected lines contribute no signal to the fit, increase computation time, and unnecessarily increase uncertainties of the resulting fit parameters. To avoid this, the code performs a quick, numerical estimate of the signal-to-noise ratio of each line in each spaxel, and discards any lines below a (settable) S/N threshold, inserting a \texttt{NaN} value in the resulting flux maps for the given line.
The code has rudimentary support for modeling two kinematic components in each line, but at the time of writing this, it is not yet possible to combine this with the use of a custom selection of lines to compute the kinematics.

\subsection{Emission line maps}
\label{sec:org4cb7c0f}

\subsubsection{Line fitting strategy   \label{sec:fitstrat}}
\label{sec:org461451a}
For each spaxel and filter/grating setting, we first subtracted the continuum as described in the previous section. We ascertained by eye that no strong, broad (\(\text{FWHM}\gtrsim 2000 \text{ km s}^{-1}\)) nebular lines were present, such that any broad emission lines present would be of a stellar nature. We chose a kernel width such that broad stellar components, if present, were modeled and subtracted along with the continuum and only narrow (\(\text{FWHM} \lesssim 2000 \text{ km s}^{-1}\)) nebular emission lines remained. These lines can in each spaxel be expected to originate in the same collections of gas clouds and thus to a reasonable approximation to share kinematic properties.

We modeled all emission lines as single Gaussian profiles, with a shared redshift and velocity width. We constructed this model as follows:
We picked a number of the strongest lines on which to base the kinematics. In the case of F100LP/G140H, we used [\ion{O}{2}]$\lambda\lambda 3727,29$, H$\beta$, and [\ion{O}{3}]$\lambda \lambda 4960,5008$; and for F170LP/G235H we used [\ion{O}{3}]$\lambda \lambda 4960,5008$, H$\alpha$, and [\ion{N}{2}]$\lambda \lambda 6548,84$ (blended features must be fitted together to get correct fluxes). These were then used to build one single model, which was fit to the observed data. We have not attempted to account for the possible presence of multiple gas components in each line, and we have not attempted to account for possible kinematic differences between various ionization zones. A more detailed and in-depth kinematic modeling of the gas is deferred to a follow-up paper.

Before fitting the spectrum of each spaxel, we removed data chunks with negligible line flux, to avoid empty chuncks of data degrading the signal-to-noise of the resulting fits. To this end, we numerically estimated the integrated flux and standard error within a \(\pm 500\) km s\(^{-1}\) window around each line. If this yielded a S/N < 0.5, the line was excluded from modeling in this spaxel.

After simultaneously fitting strong and blended lines together in one model, we cycled through a number of weaker lines and modeled each individually, locking each to the kinematic properties found from the strong lines and allowing only the line flux to vary.

\subsubsection{Masking \label{sec:masking}}
\label{sec:org1427a4a}
Because of the strongly elongated shape of the arc, a large number of spaxels in each datacube do not contain any emission from the source galaxy. These spaxels may however still spuriously give rise to fluxes above the S/N limit set during line fitting. We have thus constructed masks to remove pixels based on the following criteria:

\begin{enumerate}
\item Since the kinematics in most pixels is dominated by the [\ion{O}{3}] 4960,5008 Å doublet, we produced a stacked [\ion{O}{3}] S/N map by assuming a fixed line ratio of \(R=2.919\) \citep{vanhoof2018} and, once this fixed line ratio was corrected for, taking the mean of the four line maps (each doublet member in each of the two settings). Uncertainties were propagated in the standard way. We then created a S/N map, and excluded all spaxels with \(S/N < 3\) for this stacked [\ion{O}{3}] map.
\item Because there was still a significant number of spurious spaxels present in the resulting mask, we produced a polygon mask by hand, using the S/N mask as a guide. This mask is shown in \autoref{fig:manualmask}.
\end{enumerate}

These masks have been applied to all maps shown subsequently, as well as all statistical treatment of source spaxels, unless otherwise stated. For many lines and line-derived property maps, additional spaxels outside these masks also contain missing values due to lower S/N ratio in one or more of the involved lines, but spaxels within the masks have been removed everywhere.

\subsubsection{Combined maps  \label{sec:combimap}}
\label{sec:org34f462e}
The F100LP/G140H and F170LP/G235H observations resulted in separate datacubes which we did not attempt to combine directly, but produced separate line and kinematic maps of each cube. This resulted in duplicates of the maps of kinematic properties, as well as flux maps of the [\ion{O}{3}] \(\lambda \lambda\) 4960,5008 Å lines which are included in both grating/filter settings. We combined these maps by a simple average of all spaxels present in both frames of the relevant quantity, with standard errors propagated by summation in quadrature. Whenever a pixel had a missing value in one of the maps, this pixel in the combined map would be set equal to the one existing value.

\subsection{Dust attenuation maps\label{sec:dustion}}

We have accounted for dust attenuation in the galaxy using a standard starburst attenuation law \citep{calzetti2000}. We have computed \(E(B-V)\) using the Balmer decrement of H\(\alpha\)/H\(\beta\) following standard methods as outlined in e.g. \citet{dominguez2013},
and using intrinsic line ratios from \citet{osterbrock} assuming \(T_{e}=10^{4} K \) and \(n_{e} = 10^{2}\ cm^{-3}\). In the very high density region surrounding the ``Godzilla'' region \citep[see][]{choe2025}, the emission from partially ionized regions \citep{dai2024arxiv} or broad H\(\alpha\) emission from the power-law winds of the extremely lensed Godzilla object may lead to an elevated H\(\alpha\)/H\(\beta\) line ratio and thus an overestimation of \(E(B-V)\); a more thorough treatment of dust attenuation in this region was done by \citet{choe2025}, who found a strongly elevated \(E(B-V) \approx 0.9\) from H\(\alpha\) in this clump, while a selection of other Balmer and Paschen lines unaffected by this effect yielded a value of only half that. We therefore advise caution when interpreting the values found from H\(\alpha\) near the Godzilla region. However, the rest of the areas covered by the IFU are more typical for warm starburst ISM, and thus a H\(\alpha\) based determination of \(E(B-V)\) is appropriate for this region.

From these dust reddening maps, we then produced maps of dereddened line flux of all fitted emission line maps. However, the NIRSpec IFU pixel size undersamples the PSF, even in the pseudospaxels resulting from drizzling the 4-point dithering pattern used for the present observations. This may lead to a fringing-like stripe pattern when dividing line maps (see e.g. figures in Sect. \ref{sec:ionabund}. 
% because the dithering pattern does not sample the PSF completely, some of these line ratio maps leave a striping pattern in the data.
% , such as e.g. the \(O_{32}\) and \(\Delta\)[\ion{S}{2}] maps in \autoref{fig:dustion}. 
We have therefore used the dereddened maps only when it was necessary, and we caution the reader that this residual stripe pattern is not physical.

\subsection{Determining ionization and abundances \label{sec:ionabund}}
\label{sec:orgd844ea5}
In addition to dust and kinematic properties, we have computed spatially resolved maps of commonly-used diagnostics of  ionization conditions and chemical abundance.  These are the line ratios  \(\text{R3} = \log([\text{O \sc{iii}}]\,\lambda\,5008/\text{H}\beta), \text{O1} = \log([\text{O \sc{i}}]/H\alpha, \text{N2} = \log([\text{N \sc{ii}}]/H\alpha), \text{ and } \text{S2} = \log([\text{S \sc{ii}}])\), as well as the ionization diagnostic \(O_{32} = \log([\text{O \textsc{iii}}]/[\text{O \textsc{ii}}])\), and the diagnostic line ratio \(S2 = \log([\text{S \textsc{ii}}]/\text{H}\alpha)\). 
From S2, we also computed a spatially resolved map of the sulfur deficiency \(\Delta[ \text{S \textsc{ii}}]\) described in Sect. \ref{sec:ionprop} below. O3 and \(\Delta[\text{S \sc{ii}}]\) are shown in figures in Sect.\ref{sec:dustion}; the rest of these maps are not shown, but were included in the analysis of the next section. 

\subsection{Reconstructing global ISM properties of the source galaxy}
The advantage of spatial resolution motivates this paper.  However, in order to contextualize the results with samples of unlensed galaxies, we need to compute its  integrated or global properties, as it would appear if not lensed and observed at ordinary spatial resolution.  %

\subsubsection{Masking and de-magnification map    \label{sec:globalmask}}
\begin{figure}[htbp]
\centering
\includegraphics[width=\columnwidth]{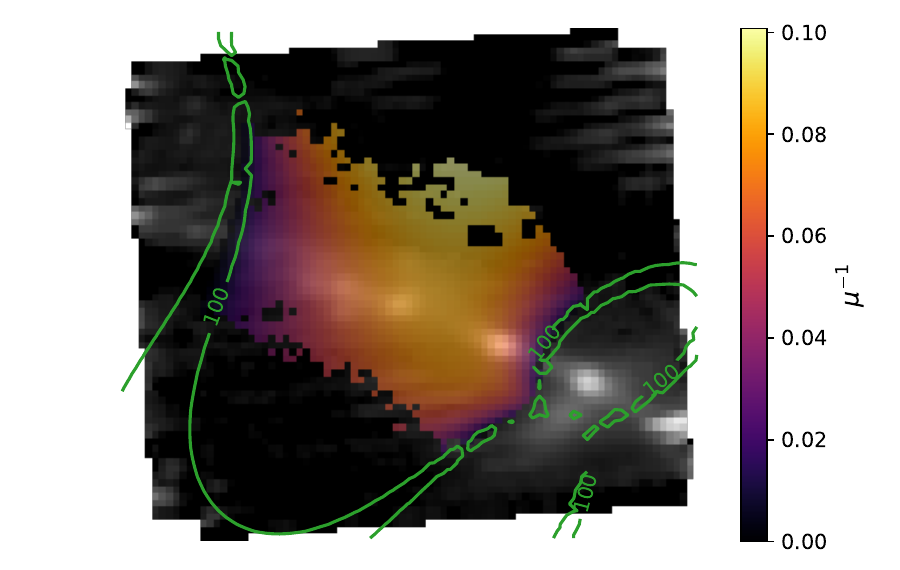}
\caption{\label{fig:kinemask}Mask used for kinematic measurements. The median map of P1 is shown in grayscale. Overlaid is the the inverse magnification map of \citet{sharon2022}, with higher values meaning that a spaxel is weighted more when deriving global properties of the galaxy. Spaxels in pure grayscale are not included in these derivations. Also shown are the  \(\mu=100\) contours from the lensing model of \citet{sharon2022}.}
\end{figure}

To emulate properties of the galaxy as they would have been observed in a slit spectrum without gravitational lensing, we extracted the global spectroscopic properties of the galaxy by the following method. First, we constructed a mask which contained only the one lensed image that covers the largest part of the source galaxy. The region, which is found in P1, is shown in \autoref{fig:kinemask}. This galaxy image is bounded in both directions along the arc by crossing critical curves, and in the off-arc direction predominantly by the [\ion{O}{3}] S/N based mask from \autoref{sec:masking}.

The location of the critical curves is derived from the lensing model of \citet{sharon2022}. To compensate for differential magnification within the mask, we used magnification maps derived from the same work. We used the Python package \texttt{Reproject} to project the magnification maps onto the WCS of the drizzled NIRSpec cube of P1. We then applied the combined S/N and critical curve mask to the magnification map as well. The inverse of the resulting masked map can then be used to compensate for differential magnification. We note that the exact delineation of the mask along the critical curve is of limited importance: Magnification is high in the vicinity of the critical curve, and since we weigh the data by the inverse magnification, a few pixels too many or too few in this region will have very limited impact on the resulting values of integrated and weighted quantities.

A more thorough treatment of a magnification-weighted, integrated spectrum and comparisons to non-resolved starburst galaxies at similar redshifts is deferred to future work. Here, we have used it to derive global values of properties derived per-spaxel in this work; including redshift, ionization parameter, [\ion{S}{2}] deficiency (see \autoref{sec:ionprop}), and more. An important caveat to these properties is that the regions included in the mask in pointing 1 are not a fully complete image of the galaxy. As is evident from \autoref{fig:artist}, the image shown in \autoref{fig:kinemask} covers most of the galaxy, but not all. All global properties are given under the assumption that the included regions are an accurate representation of the global properties of the galaxy. 

\subsubsection{Global kinematic properties}
\label{sec:org3575953}
A common means to characterize the kinematical properties of a galaxy is by comparing the \emph{shearing velocity} \(v_{\text{shear}}\) to the \emph{intrinsic velocity dispersion} \(\sigma_{0}\) \citep{glazebrook2013, herenz2016, bik2022, herenz2025}. $v_{\text{shear}}$ is defined as the difference in centroid velocity between the spaxels with the most extreme values.   \(\sigma_{0}\) is defined as the flux-weighted velocity dispersion averaged over all spaxels. Following \citet{herenz2016}, but substituting [\ion{O}{3}]\(\lambda\)5008 for H\(\alpha\), these are written as:

\begin{align}
v_{\text{shear}} &= \frac{1}{2} (v_{\text{max}} - v_{\text{min}}), \\
\sigma_{0}         &= \frac{\sum F^{\lambda5008}_{\text{spaxel}}\sigma_{\text{spaxel}}}{\sum F^{\lambda5008}_{\text{spaxel}}},
\end{align}

Following \citeauthor{herenz2016}, we have used the 5\textsuperscript{th} and 95\textsuperscript{th} percentiles of velocities to avoid outliers from skewing the results, while still sampling the extreme ends of the velocity distribution. We opted to base \(\sigma_{0}\) on [\ion{O}{3}]\(\lambda\)5008 rather than H\(\alpha\), because in our dataset, the former both intrinsically has higher S/N than the latter, and because this line at the redshift of the Sunburst Arc falls in the overlapping range between G140H and G235H and thus was observed in both, effectively doubling the exposure time. 
When computing \(\sigma_{0}\), we have not made any attempt to correct for beam smearing such as has been done in the ground based studies of e.g. \cite{glazebrook2013,herenz2016,bik2022,herenz2025}. The effective spatial resolution of JWST/NIRSpec for this lensed target is so high that the problem is negligible. Additionally, the beam smearing correction in these works was motivated by a central spike in line width in those kinematics maps caused by the atmospheric blurring, leading to an overestimation of the brightness-weighted velocity dispersion \(\sigma_0\) in these works. We do not observe such an effect, and consequently do not expect the same skewing of \(\sigma_0\).

The observed value of \(v_{\text{shear}}\) is the projection along the line of sight of the true, physical value, and needs to be corrected by a factor of \(\sin^{-1} i\), with \(i\) being the inclination angle. Finding the inclination of the galaxy requires detailed 3D modeling, something which is outside the scope of the present work. However, we can apply some statistical and geometrical considerations. If we assume a disk-shaped galaxy at an inclination \(i\), we know that the median inclination angle in a large sample of randomly oriented disks is 60\textdegree{}. If we adopt this value, this gives a correction factor to \(v_{\text{shear}}\) of \(\sin^{-1} 60\deg \approx 1.19\).

%The ratio \(v_{\text{shear}}/\sigma_{0}\) is an indicator of the global kinematic properties of resolved galaxies \citep{glazebrook2013, herenz2016, schreiber2020, bik2022, herenz2025}. 
% The observed value of \(v_{\text{shear}}\) is the projection along the line of sight of the true, physical value, and needs to be corrected by a factor of \(\sin^{-1} i\), with \(i\) being the inclination angle. Finding the inclination of the galaxy requires detailed 3D modeling, something which is outside the scope of the present work. However, we can make some statements based on statistical and geometrical considerations. If we assume a disk-shaped galaxy at an inclination \(i\), we know that the median inclination angle in a large sample of randomly oriented disks is 60\textdegree{}. If we adopt this value, this gives a correction factor to \(v_{\text{shear}}\) of \(\sin^{-1} 60\deg \approx 1.19\). This gives a corrected value of \(v_{\text{shear}}/\sigma_{0} \approx 1.60\). In \citep{schreiber2020,bik2022}, a value of \(v_{\text{shear}}/\sigma_{0} \gtrsim 1.83\) is adopted as the demarkation above which they consider a galaxy to be rotation dominated. This value is, for the Sunburst Arc galaxy, reached at angles of \(\lesssim 45\deg\) (see \autoref{sec:results}). In a random distribution, one finds about 30\% of galaxies in this inclination angle range, meaning that the galaxy, assuming no knowledge of its orientation, has about a 70\% probability of being dispersion rather than rotation dominated, and for the majority of the remaining 30\% not very strongly rotation dominated.

\subsubsection{Ionization properties and diagnostic line ratios  \label{sec:ionprop}}
\label{sec:orgc4825cb}
Using the single-galaxy mask and inverse magnification weight map, we have computed global values of diagnostic line ratios and derived properties that are often used in the literature, so that the Sunburst Arc can be directly compared to observations of unlensed galaxies at similar redshifts. These also provided the opportunity to compare the properties derived from the stacked spectrum of the LCE, to a spectrum as it would more realistically look if unlensed and unresolved. This could tell us whether the Sunburst Arc is globally similar to the LCE galaxies found in LCE surveys \citep[e.g.,][]{nakajima2020, cooke2014, fletcher2018, liu2023, wang2021, xu2023, flury2022a, flury2022b}.

We calculated each of these diagnostics by extracting a inverse magnification weighted spectrum in the single-image mask discussed in 
%taking the inverse-magnification weighted sum of each emission line within the single-image mask discussed in 
Sect. \ref{sec:globalmask}. We then used these weighted, masked and integrated line fluxes to compute derived properties. In addition to the global kinematic properties and 
% We picked a superset of 
the spatially mapped diagnostics discussed in Sect. \ref{sec:ionprop}, we computed the sulphur deficiency \(\Delta \text{[S \textsc{ii}]}\) as the difference between \(\text{S2} = \log(\text{[S \textsc{ii}]}/ \text{H}\alpha)\) and the locus described in \citet{wang2021}. 
In a similar manner, we computed global values of \(\text{N2} = \log(\text{[N \textsc{ii}]} 6584/\text{H}\alpha)\) and derived properties.

\section{Results \label{sec:results}}
\label{sec:org0173785}
We first present spatially resolved maps, then present results based on the weighed and integrated maps described in \autoref{sec:globalmask}, as well as the stacked LCE spectra described in \autoref{sec:lcestack}.
\subsection{Line flux maps}
\label{sec:orgbd52626}
\begin{figure*}[ht!]
\centering
\includegraphics[width=\textwidth]{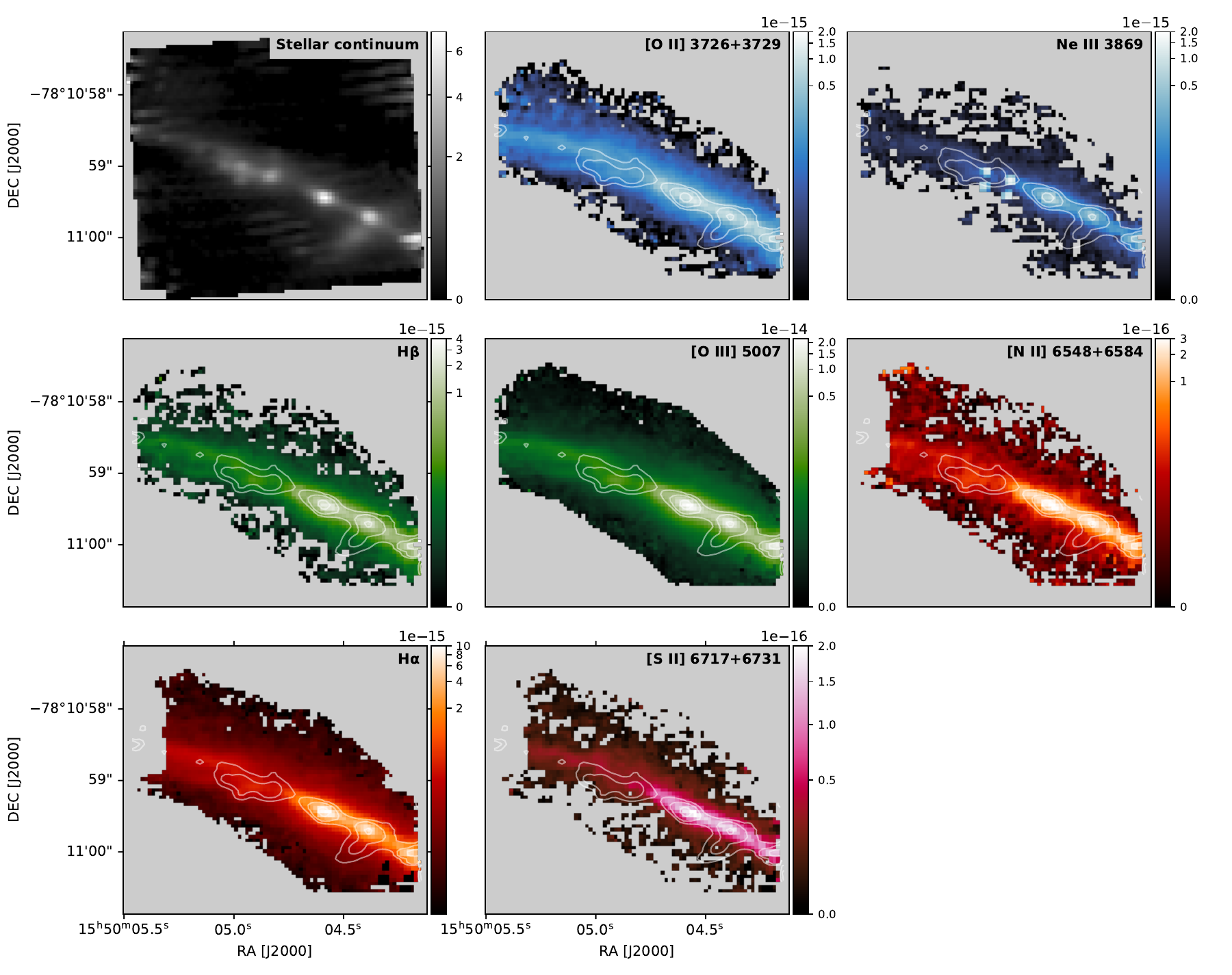}
\caption{\label{fig:linemaps1}Maps of line emission within Pointing 1, based on single-component Gaussian profile fits for each individual spaxel. Fluxes are given in ergs/(s cm\textsuperscript{2}) except stellar continuum which is given in MJy/SR. Stellar continuum is also shown as contours in the line maps to show the relative position of line emission features. The line maps are showed on a logarithmic color scale to better present extended, low surface brightness structure.  The hand drawn masks of Sect. \ref{sec:masking} are applied. The four bright spots in the center of the \ion{Ne}{3} map are an artifact.}
\end{figure*}

\begin{figure*}[ht!]
\centering
\includegraphics[width=\textwidth]{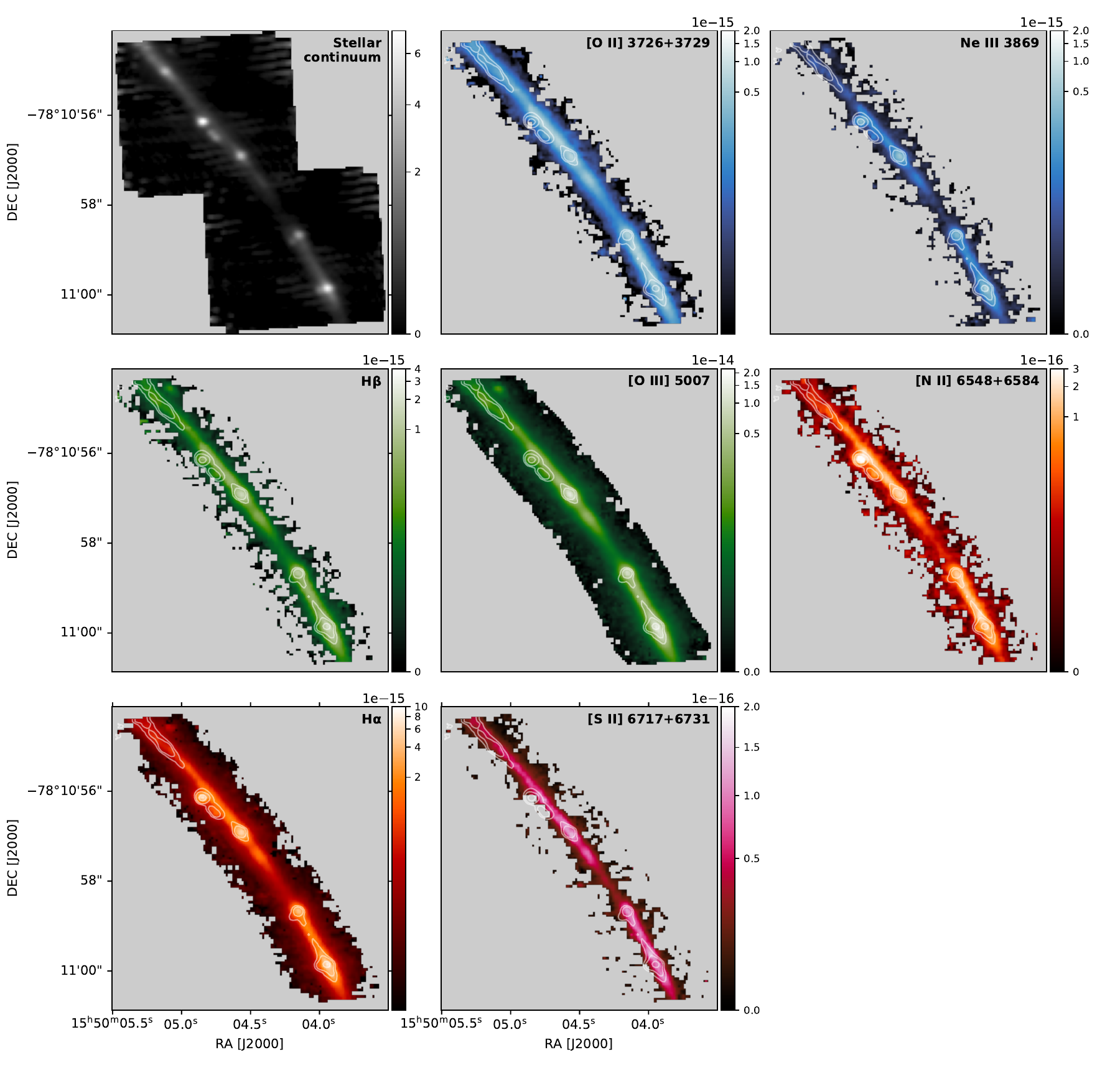}
\caption{\label{fig:linemaps23}Same as Fig.\ \ref{fig:linemaps1}, but for Pointing 2+3.}
\end{figure*}

\autoref{fig:linemaps1} and \autoref{fig:linemaps23} show the emission intensity maps for the majority of the lines used in the diagnostic maps that follow. Each line map is colored to loosely correspond to the rest-frame wavelength of the line. The masks described in \autoref{sec:masking} are applied here and in all subsequent maps, unless otherwise stated.

To optimize visual quality, we show the sum of both lines in the [\ion{N}{2}] \(\lambda \lambda\) 6548,84, although only [\ion{N}{2}] is included in further diagnostics.

\subsection{Kinematics   \label{sec:kineresults}}
\label{sec:org4d75a92}
\begin{figure*}[ht!]
\centering
\includegraphics[width=\textwidth]{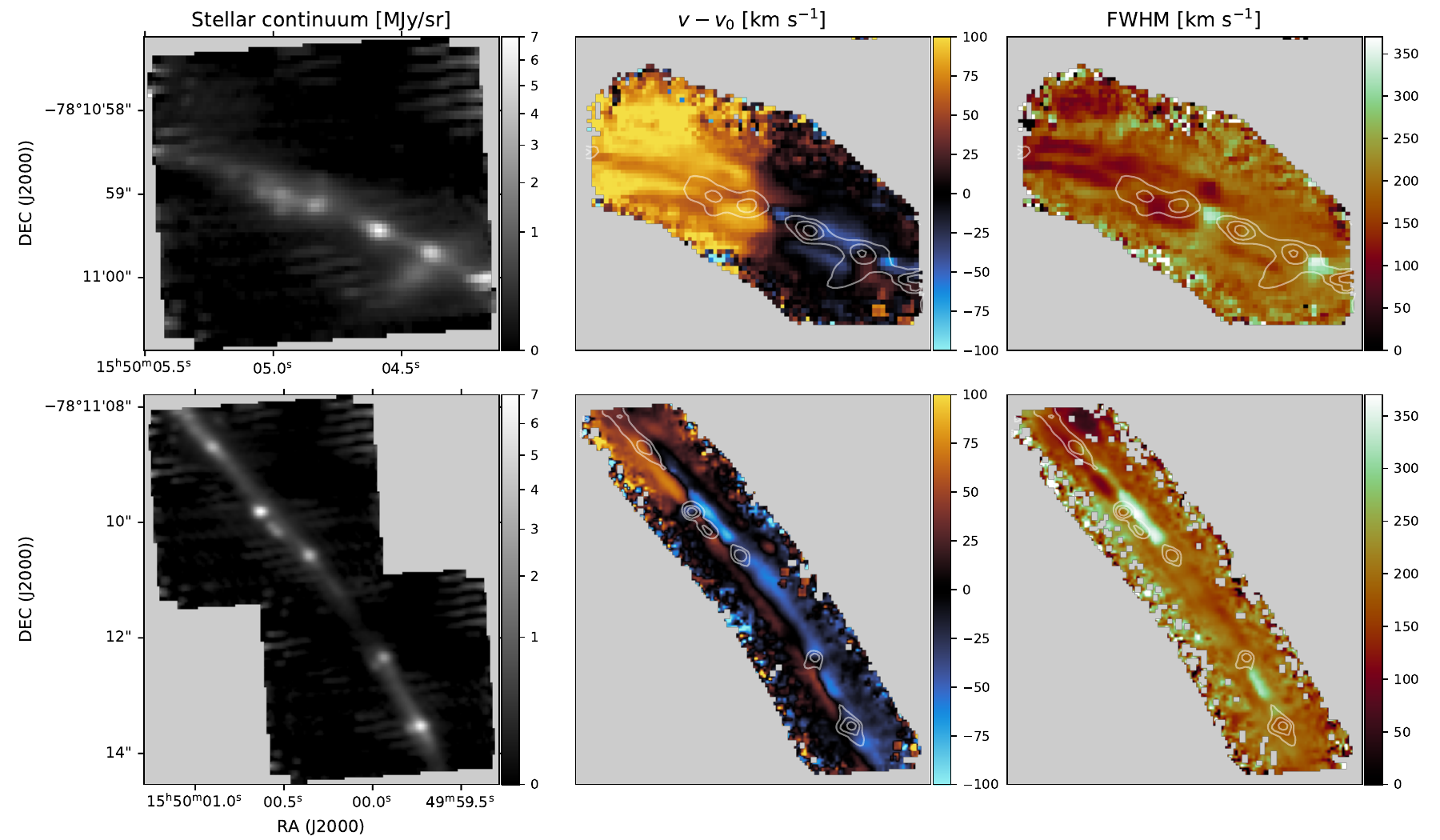}
\caption{\label{fig:kinemaps}Kinematic properties of H\(\alpha\) and [\ion{O}{3}] in P1 (upper panels) and P2+3 (lower panels). Left panels show the stellar continuum on a square root color scale for morphological comparison. \textbf{Center panels} show velocity offset from systemic velocity. Right panels show the full width at half maximum of the fitted line. On the center and right panels are overlaid contours of the stellar continuum for comparison.}
\end{figure*}

% As described in Sect. \ref{sec:fitstrat}, we modeled the emission lines in each spaxel under the assumption that they can each be fit by a single Gaussian profile, and that they emanate from the same gas phase and therefore share a common redshift and common velocity line width. Results derived from the two different filter/grating settings were then combined by a simple average as described in Sect. \ref{sec:combimap}. 
Maps of centroid velocity relative to systemic (\(v - v_0\)) and the velocity width (FWHM) resulting from the fitting described above in Sect. \ref{sec:fitstrat} are shown in \autoref{fig:kinemaps}, along with maps of stellar continuum shown in a separate column, as well as overlaid as contours on the kinematic property maps as an aid for visual orientation.

In \autoref{fig:kinemaps}, the \(v-v_0\) map of P1 (upper center) shows a clearly defined velocity gradient from E to W which, as seen in \autoref{fig:artist} roughly corresponds to the same direction in the source plane. However, this apparent simplicity might be slightly misleading. The parts of the galaxy which are redshifted relative to systemic velocity (to the E, shown in orange tones) has a somewhat more complex kinemorphology. In addition, due to their brightness, the blueshifted parts are dominant in the integrated spectra from which the systemic velocity is derived. If we focus solely on the single, almost-complete image of the galaxy (see \autoref{fig:kinemask}), the blueshifted parts make up only a small part of this. Thus, if we had compared to a simple, unweighted average of the centroid velocities, the kinematic complexities in the regions of the galaxy not containing the LCE would likely have appeared more complex than they do in \autoref{fig:kinemaps}. 
% In the map of P2+3 (lower center), the lensing shear is approximately perpendicular to that of P1, and the velocity gradient closer to being perpendicular to the direction of the arc.

The right column of \autoref{fig:kinemaps} shows the velocity-space FWHM. Of particular interest is the dramatic line broadening immediately to one side of the multiple images of the LCE but, interestingly, not actually coincident with it.
Comparing to the middle column of \autoref{fig:kinemaps}, we see that this region of high FWHM is coincident with a slight spike in outflow velocity.
%; and comparison to \autoref{fig:abund} shows that it also coincides with a spike in [\ion{N}{2}/H\(\alpha\)]. 
Visual inspection of the datacubes shows the presence in these broadened regions of a secondary, slightly blueshifted line component of width comparable to that of the single component in the surrounding regions; we do not present this line profile here but defer this to a future work (Komarova et al., in prep.). This secondary component, when fitted together with the main component to a single Gaussian profile, likely accounts for the observed line broadening.
% This offset kinematic spike might very well originate in a separate outflowing gas component which is automatically incorporated into our single-component model.
In P2+3, this region of broad emission appears particularly dramatic and elongated immediately NE of the LCE image 1.8 (in the terminology of \citealp{sharon2022}), immediately next to the object ``Godzilla''. This region is subject to extreme magnification and shearing due to a low surface brightness lensing foreground galaxy \citep{choe2025}, but despite this extreme magnification and distortion, it is still another lensed image of the same region as in P1.

\subsection{Dust and ionization}
\label{sec:orgff3dc7a}
\begin{figure*}[ht!]
\centering
\includegraphics[width=\textwidth]{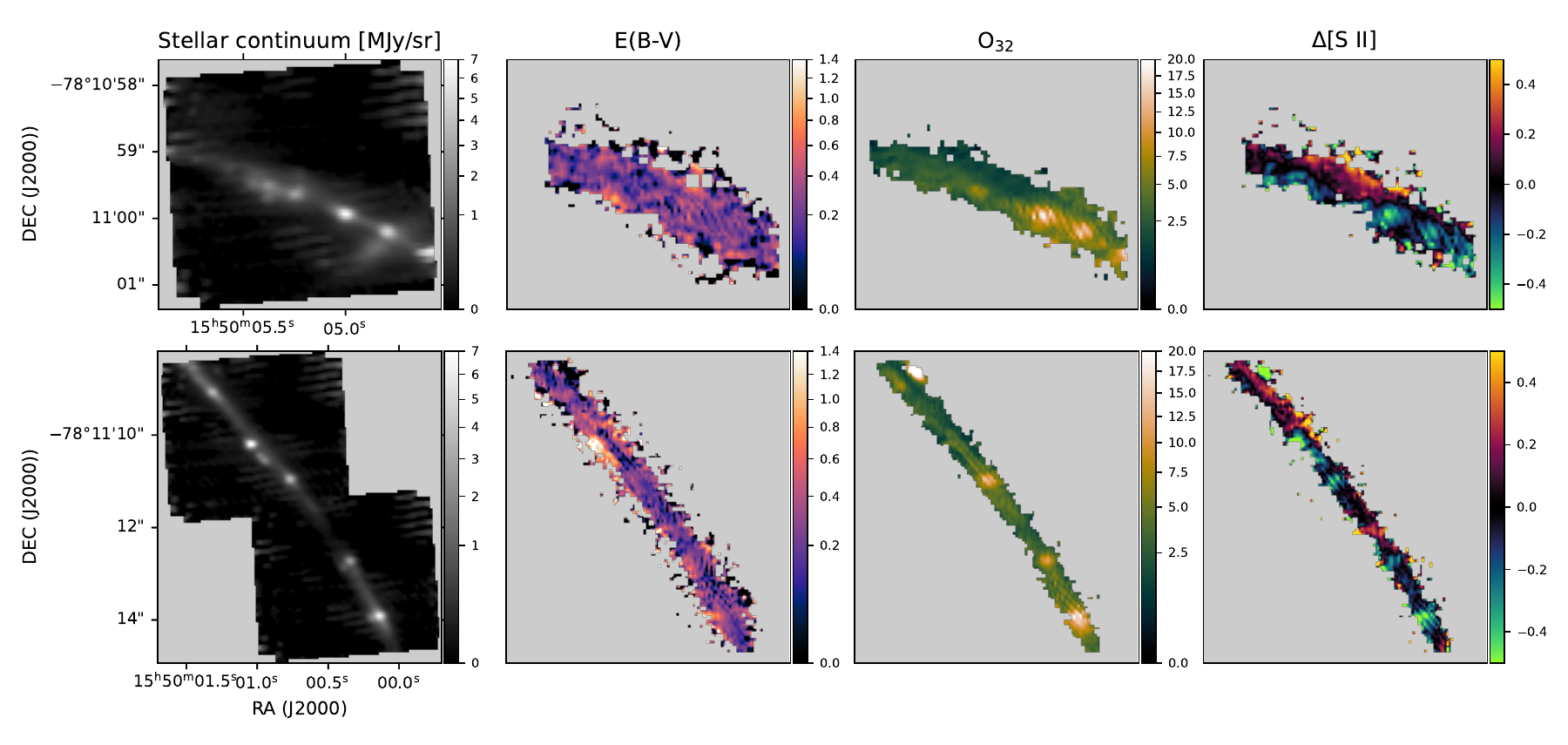}
\caption{\label{fig:dustion}Maps of dust and ionization properties for pointing 1 (upper panels) and pointing 2+3 (lower panels). As in previous figures, the leftmost panels show the stellar continuum on a square-root color scale to aid spatial orientation. The second column shows $E(B-V)$ as derived from the ratio of H$\alpha$  and H$\beta$. The third column shows $O_{32}$, and the rightmost column shows the Sulphur deficiency, $\Delta [\text{\ion{S}{2}}]$, as described in the main text. The stripe-like patterns are artifacts of the PSF being undersampled.}
\end{figure*}

In \autoref{fig:dustion}, we show spatially resolved maps of dust reddening and ionization diagnostics, again with the upper row showing the maps for P1, and the lower row showing them for P2+3; and with stellar continuum maps shown in the leftmost column.

In the center-left column, we show maps of \(E(B-V )\) derived as described in \autoref{sec:dustion} from the H$\alpha$/H$\beta$ ratio and a Calzetti attenuation law. The stripey patterns showing in these panels are an artifact of the undersampled PSF. We see that the dust attenuation is quite uniform across the arc, except in the ``Godzilla'' feature marked in \autoref{fig:nircamrgb}, and its much fainter counterimages in both P1 and the lower left of P2+3 \citep[see][for a further discussion]{choe2025}.

The third column shows maps of the ionization diagnostic \(O_{32} =\) {[}\ion{O}{3}]/[\ion{O}{2}]; the same artifacts are present here as in the maps of \(E(B-V)\). The ionization is quite elevated around the images of the LCE and in the central part of P1, we also note the feature in very N end of P2, which is associated with a very faint star cluster, yet apparently displays an extreme ionization of \(O_{32} \gtrsim 20\).

Finally, the right panels show another ionization diagnostic, \(\Delta[\text{S \textsc{ii}}]\), which has been proposed as an improved tracer of LyC escape due to the lower ionization potential of [\ion{S}{2}] compared to [\ion{O}{2}], meaning that these lines more accurately trace the presence or absence of neutral gas, where O$^{+}$ ions also are present in substantial amounts inside \ion{H}{2} regions and the photodissociation regions at their boundaries. While the [\ion{S}{2}] line may be fainter than [\ion{O}{2}], the diagnostic is based entirely on lines very close in wavelength and thus independent of dust model. For the same reason, it these stripey artifacts are suppressed in this map, although still visible. It is to our knowledge the first time it has been possible to map both log [\ion{S}{2}]/H\(\alpha\) and LyC emission in any galaxy and thus test its strength as a tracer on a local scale. We note that there is a strong similarity between the spatial distribution of high- and low-ionization regions by the two methods.

\subsection{Nitrogen spike}
\label{sec:orgeeddec1}
\begin{figure}[htbp]
\centering
\includegraphics[width=\columnwidth]{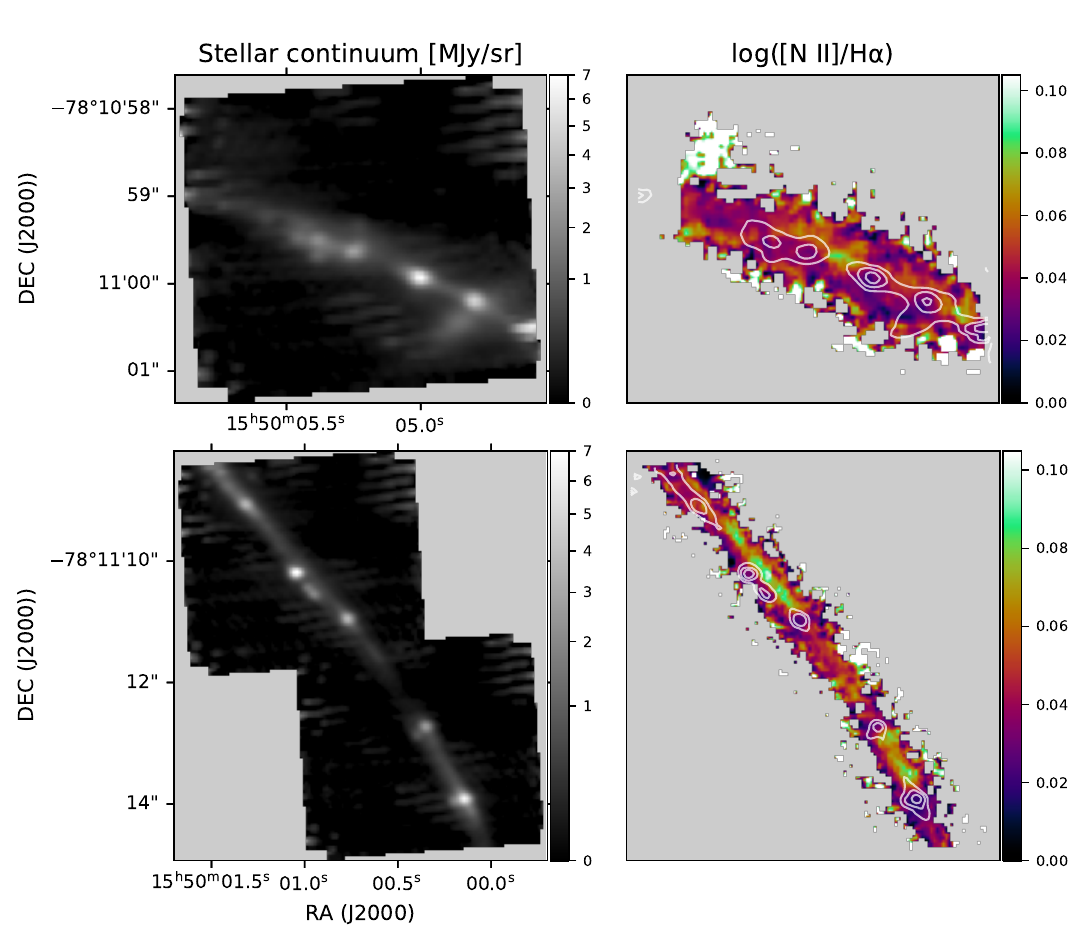}
\caption{\label{fig:abund}[\ion{N}{2}]/H$\alpha$ maps. As in previous figures, the upper row shows maps of P1, while the lower row shows maps of P2+P3.}
\end{figure}

\autoref{fig:abund} shows a maps of 
N2 = [\ion{N}{2}]/H\(\alpha\).
% the diagnostic N2. This diagnostic is powerful because of its short wavelength baseline, eliminating dust model dependence; on the other hand, it can only show the N/H abundance, not the relative abundance of N relative to other elements. 
We note the conspicuous regions of elevated [\ion{N}{2}]/H$\alpha$ at locations next to, but not coincident with, the LCE cluster. In these regions, the elevated [\ion{N}{2}] might be due to elevated N/H, or due to collisional excitation from shocks. Spatially, they closely match the regions of larger line FWHM and a slight rise in blueshift as seen in  \autoref{fig:kinemaps}; something which could support the collisional interpretation. The region is not coincident with any star clusters we have been able to identify in our NIRCam images or archival HST images; we speculate that it might be the result of an anisotropic outflow from the LCE cluster.

% The center-right column shows maps of log(N/O) derived from the empirical N2S2 method of \citep{perezmontero2017}, as described in Sect. \ref{sec:ionabund}. Interestingly, this abundance measure does not show quite the same strongly localized spikes as the N/H abundance. Finally, the rightmost column shows maps of empirical Oxygen abundance \(12 + \log(\text{O/H})\) from the S2N2 (\(y\)) method from \citep{dopita2016}. The latter two maps show a strong similarity, because they both essentially are the same quantity scaled by different functions, and with a second-order correction factor in one of them. Still, this similarity should not be taken too literally; the N/O and O/H abundances cannot physcially have identical spatial distribution while at the same time N/H shows such conspicuous variation as seen in the center-left column. Instead, it may serve as a reminder of the limitations of the empirical calibration methods.

\subsection{Global properties}
\label{sec:orgedab152}
In addition to the spatially resolved maps, we have computed values of some core diagnostics from the stacked LCE spectrum, as well as globally for the full, magnification-weighted galaxy as described in Sect. \ref{sec:globalmask}, allowing for a more apples-to-apples comparison of properties of the Sunburst Arc galaxy to other known LCE galaxies in the literature and a discussion of global vs. local properties in the galaxy and the efficiency of observational proxies for LyC escape. The properties and their measured values are tabulated in \autoref{tab:globprop}.

\begin{table*}[htbp]
\caption{\label{tab:globprop}Overview of global, inverse-magnification weighted properties of the Sunburst galaxy, compared to local properties of the LCE stack.}
\centering
\begin{tabular}{lrr}
\toprule
Quantity & Galaxy & LCE\\
\midrule
\(z_{\text{int}}\) & \(2.37014^{+ 0.00012}_{-0.00027}\) & \(2.371017\pm9\times10^{-6}\)\\
\(v_{\text{shear}}\) [km s\(^-1\)] & \(109.9\pm4.7\) & ---\\
\(\sigma_{0}\) [km s\(^-1\)] & \(82.02\pm0.14\) & ---\\
\(v_{\text{shear}}/\sigma_{0}\) & \(1.34\pm0.06\) & ---\\
O\textsubscript{32} & \(4.0 \pm 0.7\) & \(11.6 \pm 1.2\)\\
R3 (\(\log([\text{O \sc{iii}}]/ \text{H}\beta)\)) & \(0.828 \pm 0.003\) & \(0.85 \pm 0.03\)\\
O1 (\(\log([\text{O \sc i}]/\text{H}\alpha)\)) & \(-1.83 \pm 0.05\) & \(-2.12 \pm 0.05\)\\
N2 (\(\log([\text{N \sc{ii}}]/ \text{H}\alpha)\)) & \(-1.40 \pm 0.01\) & \(-0.49 \pm 0.04\)\\
S2 (\(\log([\text{S \sc ii}]/\text{H}\alpha)\)) & \(-0.940\pm0.077\) & \(-1.60\pm0.03\)\\
\(\Delta[\text{S \sc{ii}}]\) & \(0.163\pm0.218\) & \(-0.43\pm0.06\)\\
\bottomrule
\end{tabular}
\end{table*}

\section{Discussion   \label{sec:discuss}}
\label{sec:orgd3062f7}
\subsection{Kinematics    \label{sec:kinem}}
\label{sec:org5282786}
The upper center panel in \autoref{fig:kinemaps} clearly shows that the galaxy has a distinct, ordered velocity gradient. As this pointing (P1) has the most homogeneous magnification and least complex distortion pattern (see \autoref{fig:kinemask}) and thus the lensed image within this mask likely is quite similar in shape to the source gaalxy, this translates quite smoothly into a gradient in the direction from the lower right to the upper left corner of the de-lensed P1 footprint shown in cyan in \autoref{fig:artist}, the kinematic structure of a rotating disk. The picture is more complex in P2+3 (lower panels of \autoref{fig:kinemaps}), but this is expected: The shearing direction in the lower part (P3) is at a high angle to that of P1, and in the upper parts, the lensing geometry is complex and not well solved.
As discussed in Sect. \ref{sec:kineresults}, we have found a shearing velocity of about 109 km s\(^{-1}\), and a flux-weighted velocity dispersion \(\sigma_{0}=82 \text{ km s}^{-1}\). 

% The observed value of \(v_{\text{shear}}\) is the projection along the line of sight of the true, physical value, and needs to be corrected by a factor of \(\sin^{-1} i\), with \(i\) being the inclination angle. Finding the inclination of the galaxy requires detailed 3D modeling, something which is outside the scope of the present work. However, we can make some statements based on statistical and geometrical considerations. If we assume a disk-shaped galaxy at an inclination \(i\), we know that the median inclination angle in a large sample of randomly oriented disks is 60\textdegree{}. 
If we adopt the median inclination angle reached above, value, this gives a correction factor to \(v_{\text{shear}}\) of \(\sin^{-1} 60\deg \approx 1.19\), yielding a corrected value of \(v_{\text{shear}}/\sigma_{0} \approx 1.60\). In \cite{schreiber2020,bik2022}, a value of \(v_{\text{shear}}/\sigma_{0} \gtrsim 1.83\) is adopted as the demarkation above which they consider a galaxy to be rotation dominated. This value is, for the Sunburst Arc galaxy, reached at angles of \(\lesssim 45\deg\) (see \autoref{sec:results}). In a random distribution, one finds about 30\% of galaxies in this inclination angle range, meaning that the galaxy, assuming no knowledge of its orientation, has about a 70\% probability of being dispersion rather than rotation dominated, and for the majority of the remaining 30\% not very strongly rotation dominated. Additionally, the measured rotation speeds are already comparatively large; for the low inclination angles required for the galaxy to be solidly rotation dominated, the intrisinc \(v_{\text{rot}}\) would be so high as to require an unusually high mass in order to maintain its structural integrity; further lowering the probability for a rotation dominated scenario.

This paints a picture of a galaxy which is rotating, but also undergoing significant turbulence. From the FWHM maps in \autoref{fig:kinemaps}, the velocity disperson does not spatially correlate with strong star formation activity (apart from localized and directional outflows near the LCE, cf. section 4.2 and 4.4), suggesting that this strong dispersion, rather than stemming from star formation feedback, may be due to outside influences such as tidal disruptions by the Northern companion seen in \autoref{fig:artist}.

 %{\color{red} \textbf{TODO:} Here, I should discuss the classifications of Glazebrook2013, Herenz2016, Herenz2025, Bik2022 regarding the rotation-versus-dispersion dominated galaxies and their categorization into rotation-dominated and turbulence/interaction-dominated galaxies based on their kinematics. That the galaxy according to that classification is a perturbed rotator, pretty much on the cusp between the two types. Mention that it supports the hypothesis that the galaxy is interacting with the component from Sharon 2022, and perhaps mention the tidal stripping hypothesis of Le Reste 2024 (haro 11 paper). }

\subsubsection{The origin of a broad emission component in H\(\alpha\) and [\ion{O}{3}]}
\label{sec:orgf36ca74}
It has long been predicted that stellar winds and supernova feedback would play an important role in clearing the escape paths for Lyman-continuum, both in theoretical works \citep{trebitsch2017,rosdahl2018,kakiichi2021} and in observations \citep{heckman2011,chisholm2017,kim2020}. \citet{amorin2024} have found a strong preponderance of broad \((\sigma \approx 400 \text{km s}^{-1})\) emission lines in LyC-leaking Green Pea galaxies, although it has been difficult to establish any direct correlations between broad emission components and LyC escape fraction.

\citet{mainali2022} studied the rest-UV and -optical emission in the Sunburst Arc from the ground using the Magellan/MagE and Magellan/FIRE spectrographs, respectively. These authors did find a strong broad component in the emission features of H\(\alpha\) and [\ion{O}{3}] in the LCE spectra, while spectra from non-leaking regions showed a considerably fainter broad feature. However, the broad feature in \citeauthor{mainali2022} has FWHM \(\approx 325\)~km~s\(^{-1} \) or \(\sigma \approx 140\)~km~s\(^{-1}\) somewhere in between the broad- and medium-width features of \citep{amorin2024}. Compared to the ground-based spatially unresolved spectra of \citet{mainali2022}, the current NIRSpec IFU observations enable spectra to be extracted from spatial regions much more selectively, with lower contamination from nearby regions. 

\citet{riverathorsen2024,welch2025} extracted and stacked spectra from multiple lensed images of the LCE, allowing for unprecedented signal/noise ratio and absence of contamination from surrounding regions. Both these works modeled strong, rest-frame optical nebular emission lines using three Gaussian components, yielding FWHM values of \(\lesssim 100, \sim 200, \text{and} \sim 600\) km s\(^{-1}\); consistent with each other and broadly consistent with typical values of the narrow, medium and broad components found by \citet{amorin2024}.

In this work, we have only used one Gaussian component to model the emission lines in each spaxel, so our measured velocities are not directly comparable to those of the above works; in particular, faint but broad components will have limited impact on the model line width and centroid velocity. On the other hand, single component fits can still robustly show spatial variation in kinematic properties, and a brightening of a broad component will manifest itself as a larger measured line width. Looking at the FWHM maps in the right panels of \autoref{fig:kinemaps}, it is interesting to note that while we know from the stacked spectra that a broad component is present in the LCE cluster, it is not strong enough to lead to an elevation in the FWHM map; which is dominated by a component of \(\sim 200\)~km~s\(^{-1}\) consistent with the surroundings.

The line is however substantially wider in a small region immediately next to the LCE as seen in the two bright spikes in P1, reaching an FWHM of \(\sim 350\) km s\(^{-1}\). The same region appears twice in P2+3 in highly elongated form due to the strong lensing shear in these images. This region is also slightly blueshifted relative to the LCE (\(\sim 30 - 50 \text{ km s}^{-1}\)) and shows an equally conspicuous spike in the [\ion{N}{2}]/H\(\alpha\) ratio. It is not immediately clear what drives this apparent outflow, as it is spatially distinct from the LCE and there is no star cluster visible at its location. It is close enough to the LCE to have bled significantly into the MagE and FIRE slits used by \citeauthor{mainali2022}. We therefore conclude from our maps that this region, rather than the LCE itself, gives rise to the broad component in the LCE spectra in that paper. The ground-based spectra of \citet{mainali2022} lacked the spatial resolution to make this distinction. 
If correct, this interpretation of course raises the question of how often the observed broad components in LCEs from the literature originate from nearby regions, rather than from the isotropic stellar winds of the leaking clusters themselves, and in extension, whether a wider array of mechanisms are needed to explain the appearance and sustenance of ionizing escape channels.

\subsection{Dust geometry}
\label{sec:orgb08a9e8}
The leaking clump is known to have a bright, narrow Ly\(\alpha\) emission spike at systemic velocity  \citep{sunburst2017}. This is indicative of have a line of sight towards us that is virtually free of neutral gas. However, the dust reddening as traced by the Balmer decrement is not significantly lower along this line of sight compared to others in the vicinity. This lack of a hole in the dust, together with the relatively smooth and undramatic behavior of the kinematic parameters around the LCE, indicates the absence of strong mechanical feedback. If mechanical feedback is not at work, this implies that photoionization is the main mechanism creating the escape path for Ly\(\alpha\) and LyC photons, at least in the immediate surroundings of the LCE. As discussed above in Sect. \ref{sec:kinem}, the \(v_{\text{shear}}/\sigma_{0}\) suggests that the galaxy is currently or recently interacting with anothet galaxy.  As discussed below, an interaction could have helped strip the bulk of the surrounding H~\textsc{i} envelope away from the region surrounding the LCE.

\subsection{Ionization}
\label{sec:org792ffa1}
As discussed in \autoref{sec:introduction}, the \(O_{32}\) diagnostics is a good indicator of overall ionization in a given region, but not a good predictor of LyC escape.
An alternative, ionization-based tracer of LyC escape is the relative weakness of [\ion{S}{2}], first suggested by \cite{alexandroff2015}. Like the O\textsubscript{32} diagnostic, it is a measure of ionization; but the ionization potential of neutral sulphur is as low as 10.36 eV, lower than the ionization potential of hydrogen, such that it is effectively shielded from ionization in neutral regions. [\ion{S}{2}] lines are only emitted from  only present at the outer boundary of \ion{H}{2} regions. These  boundaries disappear as a region becomes fully ionized, exactly in the conditions that allow LyC escape. 
%, and has later been proven to work quite well \citep{wang2019,ramambason2020,wang2021}. Like the O\textsubscript{32} diagnostic, it is a measure of ionization; but the ionization potential of neutral sulphur is as low as 10.36 eV, lower than the ionization potential of hydrogen, such that it is effectively shielded from ionization in neutral regions. [\ion{S}{2}] lines are only emitted from  only present at the outer boundary of \ion{H}{2} regions. These  boundaries disappear as a region becomes fully ionized, making the disappearance of  [\ion{S}{2}] a potentially very interesting tracer of LyC escape. 
%\textbf{TODO:} I'm a little stuck about the differences between the O\textsubscript{32} and the SII methods here.  }

\begin{figure*}
\centering
\includegraphics[width=0.8\textwidth]{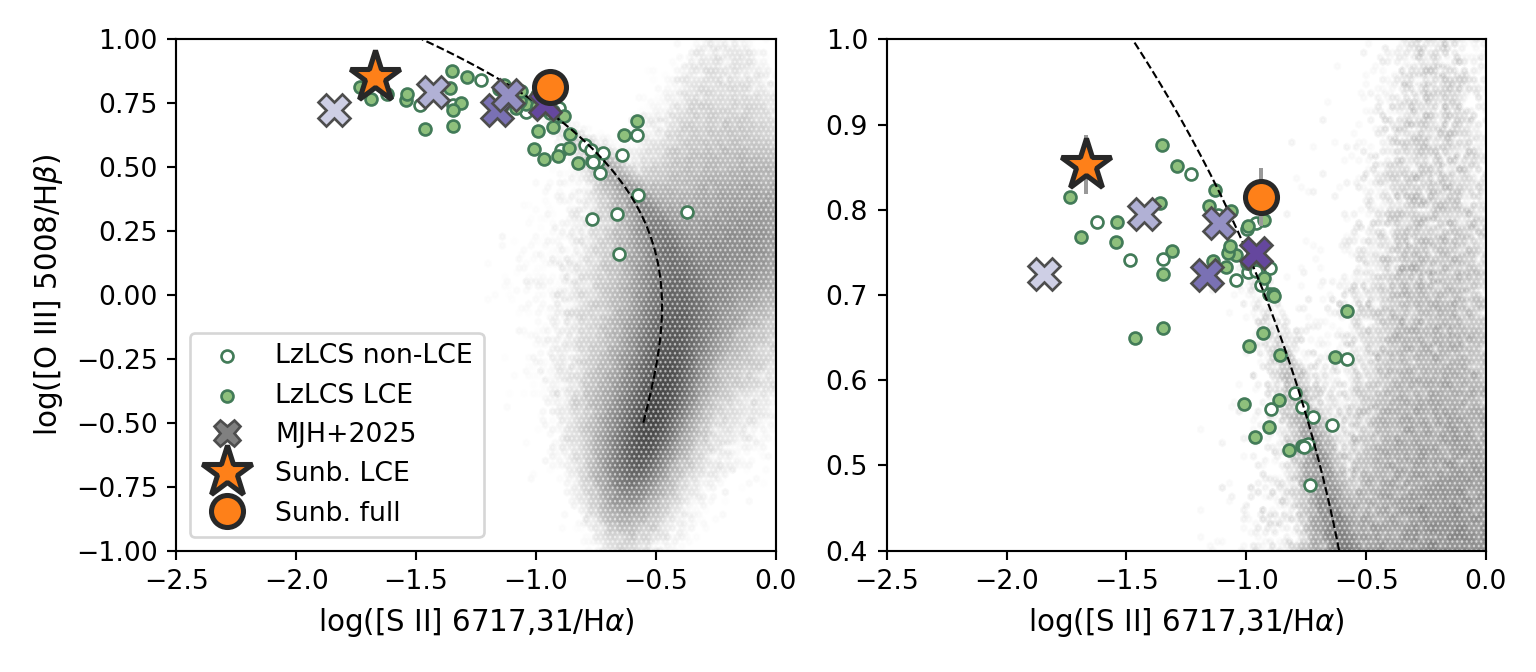}
\caption{\label{fig:siidef}Diagram of \(\log(\text{[\ion{S}{2}]/H}\alpha)\) vs. \(\log(\text{[\ion{O}{3}]/H}\beta)\) for the Sunburst LCE. The black dashed line is the SDSS/BOSS locus of \(\Delta\) [S \textsc{ii}] = 0 as seen in Fig.\ 1 of \citep{wang2021} for comparison. The heat map shows the point density for the full SDSS DR9 (note that \citeauthor{wang2021} fit their curve to a subsample of these data, hence the slight discrepancy). The green filled (open) circles show LyC leakers (non-leakers) from the LzLCS \citep{wang2021}. The purple crosses represent the high-z JWST stacks from \cite{hayes2025}, grouped by EW([\ion{O}{3}]) from lowest (darker) to higher (lighter).} 
\end{figure*}

For the Sunburst Arc, we can compare the spatially resolved maps of log [\ion{S}{2}]/H\(\alpha\) shown in the rightmost panels of \autoref{fig:dustion} to both the local values at the LCE cluster, and the global, integrated value as found in the previous section. From \autoref{fig:dustion}, it is clear to see in the rightmost panels that the [\ion{S}{2}]/H\(\alpha\) map has a strong, and highly localized, minimum at all the images of the LCE cluster. 
% As seen in \autoref{tab:globprop}, the value for the stacked spectrum is  \(\log \text{[S \scshape{ii}]/H}\alpha = -1.6 \pm 0.03\), consistent with what we see in the figure. 
From \autoref{tab:globprop}, we see that the local value at the LCE is indeed \(\log\text{[S \textsc{ii}]/H}\alpha_{\text{LCE}} = -1.60 \pm 0.03\), consistent with what is seen in the figure, while the global value is \(\log\text{[S \textsc{ii}]/H}\alpha_{\text{Int}} = -0.940 \pm 0.077\). These latter two values correspond to values of the quantity \(\Delta\){[}\ion{S}{2}] as defined in \citet{wang2021} of \(-0.43 \pm 0.06\) and \(0.163 \pm 0.218\), respectively. Comparing to the histograms and derived kernel density estimates in Fig.\ 2 in \citet{wang2021}, we see that these values put the LCE into a range in which the LzLCS contains practically no non-LCEs, while the value from the integrated spectrum places it slightly above the average from SDSS. This value falls in a tricky region in the KDEs of \citeauthor{wang2021}; it is on one hand dominated by by non-leakers, but on the other hand it is also close to the peaks of the distributions both of strong LCEs and that of all LCEs in the LzLCS. 

Thus, the LCE cluster alone has such strong \(\Delta\)[\ion{S}{2}] that it could not be mistaken for a non-leaker. In contrast, the galaxy as a whole displays a \(\Delta\)[\ion{S}{2}] level which is almost equally represenative for leakers and non-leakers. Thus, we see that while \(\Delta\) [\ion{S}{2}] seems to be a very good local tracer of ionized channels, the diagnostic can still be diluted by surrounding regions, making a leaker look like a non-leaker, as also found in \citet{wang2019,wang2021}.

The strong contrast in \(\Delta\)[\ion{S}{2}] at the LCE compared to the surroundings also agrees very well with the predictions, based on which the method was initially proposed, and supports that the method is indeed well physically motivated.

For comparison, \autoref{fig:siidef} also shows the values of \(\Delta\)~[\ion{S}{2}] from the stacked archival JWST spectra of \(4 \lesssim z \lesssim 10\) \(\simeq 1000\) star forming galaxie from \cite{hayes2025}. The stacks are grouped by EW([\ion{O}{3}]) as a proxy of ionization. While the Sunburst Arc LCE stack sits somewhere between the fourth and fifth quintile of the high-\(z\) ionization distribution, the global spectrum actually show slightly higher [\ion{S}{2}]/H\(\alpha\) ratio than typical. 
Interestingly, though, even the galaxy as a whole has slightly higher ionization than the majority of observed galaxies in the high-\(z\) Universe.

\subsection{Global ionization properties}
\label{sec:globion}
%It might be tempting to look at a galaxy like the Sunburst Arc source galaxy as a kind of ``rosetta stone'' in the translation between on one side the local properties at the production sites and through the escape channels of LyC emission, and on the other side the global, integrated properties often used as proxies for LyC escape in unresolved galaxies. To serve as such, the galaxy must be representative of the typical properties of LCE galaxies in the Universe, or at least in its own neighborhood. From the spatial maps of ionization properties in \autoref{fig:dustion}, one could expect to find a galaxy with a relatively moderate global ionization parameter, 

% One intriguing promise of having a highly detailed LCE like the Sunburst Arc is the potential to use it as a "Rosetta Stone" in the understanding of how the local properties at the LyC production site and along the path out of the galaxy translate into global properties. For that to be possible, we must also place it in the greater population of starburst galaxies and LCEs. 

We also produced the classic Baldwin, Philips, Terlevich \citep[BPT][]{baldwin1981,kewley2001,kauffmann2003,kewley2006} diagram of the Sunburst Arc based on line ratios extracted both locally in the LCE cluster stack, and globally from the \(\mu^{-1}\) weighted map described in \autoref{sec:globalmask}, left panel. For comparison, we also included the LzLCS \citep{wang2021,flury2022b}, with their LCEs shown as filled and their non-LCEs as empty circles. The shaded zone in the upper left corner of the diagram is the \cite{erb2016} Extreme Emission Line Galaxy (EELG) classification region. Interestingly, while the Sunburst Arc on the diagnostic maps in e.g. \autoref{fig:dustion} can look like an overall quite modestly ionized galaxy, once the diagnostics properties get weighted by flux as they would in a spatially integrated spectrum, it is revealed as a quite extreme galaxy, even for its redshift, which is noteworthy considering the general breakdown of the correlations between just these these properties and LyC at \(z \sim 3\).

\begin{figure*}
\centering
\includegraphics[width=0.8\textwidth]{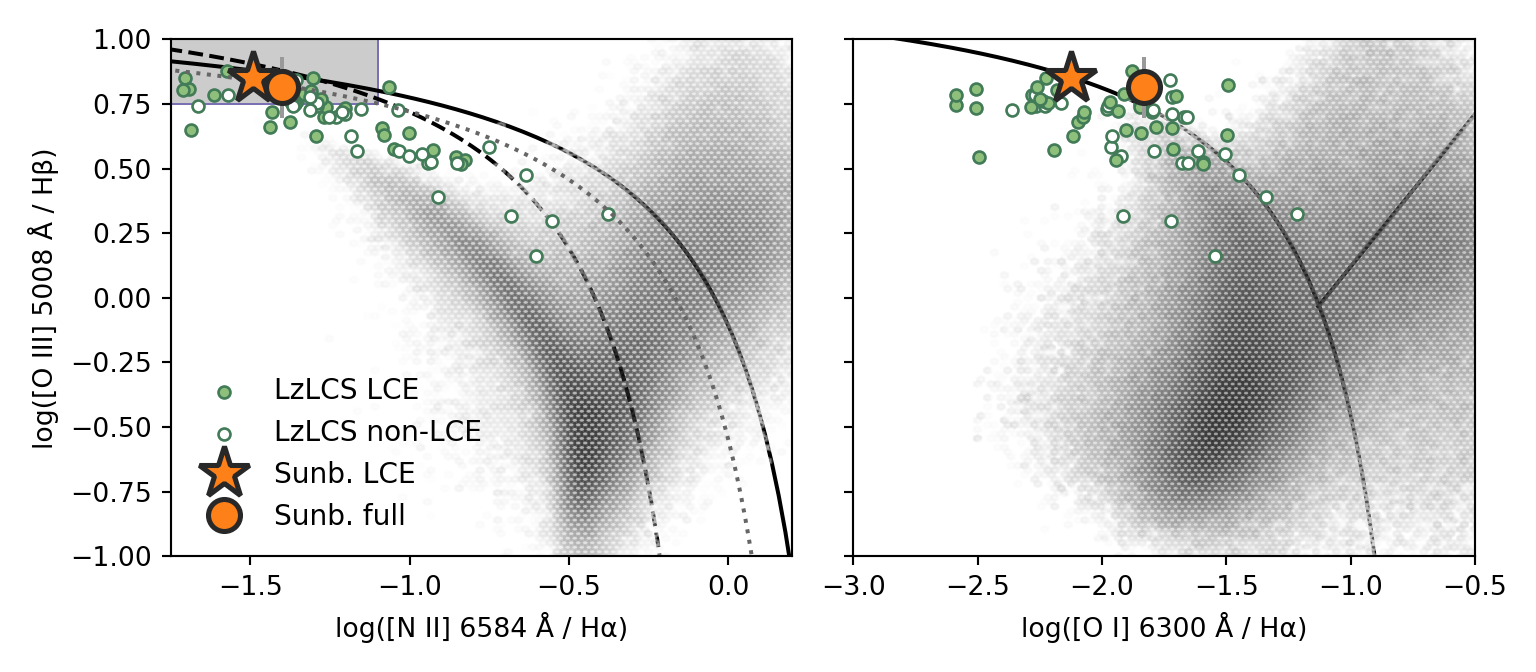}
\caption{\label{fig:bpts}Baldwin, Phillips, Terlevic \citep{baldwin1981} diagrams of the Sunburst Arc. As in Fig.\ \ref{fig:siidef}, the star denotes the value derived from the stacked LCE spectrum, while the circle denotes the value derived from the integrated, magnification-corrected spectrum; and the LzLCS galaxies are shown as open or filled green circles. The shaded region in the left panel shows the classification for Extreme Emission Line Galaxies (EELGs) from \cite{erb2016}.}
\end{figure*}

In addition to the classical BPT diagram, we have also included the [\ion{O}{1}]~$\lambda 6300$ based diagram in right panel of \autoref{fig:bpts}. This diagram shows that there is little sign of shock ionization of the ISM in the galaxy, at least not in the brightest regions in the LCE neighborhood; consistent with the a general absence of strong mechanical feedback in this region.

\subsection{LyC escape scenario}
\label{sec:org8a2129b}
\citet{sunburst2017} observed a triple-peaked Ly\(\alpha\) profile emanating from the Sunburst Arc, consisting of the classic double-peak profile observed many times in the literature, combined with a narrow peak with a peak velocity and line width almost identical to those of H\(\alpha\), telling of a component undergoing no radiative transfer effects, escaping through a channel of less then one optical depth in Ly\(\alpha\). Based on this, \cite{sunburst2017} predicted a scenario in which Ly\(\alpha\) escapes directly through a narrow perforation in an otherwise optically thick shroud of H~\textsc{i}, in a manner reminiscent of the arc's namesake meteorological phenomenon. According to our best models at the time, the total solid opening angle could be no more than than 2-5\% of the total projected area surrounding the LCE cluster, with an extremely low neutral column density of \(\log N_{\text{H\textsc{i}}} \lesssim 12.8\). If the combined opening angle were larger, practically all Ly\(\alpha\) photons were predicted to escape through the open channel without the wavelength-space redistribution leading to the observed dual peaks surrounding the central peak. Such a channel would have to be created by a highly directional effect such as a jet from a previous accretion event; and it would have to be coincidentally aligned to a high degree with our line of sight. While such a scenario is certainly possible and might indeed have been the only plausible explanation; having to invoke this level of coincidental fine-tuning is uncomfortable, and furthermore, such a puncture by jet scenario would also have cleared away the majority of dust along the line of sight, but a significant dust screen was already then observed to be present in the nebular emission in \citet{sunburst2017}, and since confirmed in models of the stellar population \citep{chisholm2019,riverathorsen2019}.

Recent modeling work by \citet{almadamonter2024} has shown that H~\textsc{i} clouds are much more ``sticky'' to Ly\(\alpha\) photons than previously believed, in the sense that Ly\(\alpha\) photons have a substantially higher probability of getting trapped in the H~\textsc{i} gas and undergoing significant spatial and frequency-space redistribution, than was believed at the time of \citet{sunburst2017,riverathorsen2019}. Consequently, a much wider range of opening angle than previously believed can give rise to the characteristic Ly\(\alpha\) profile of the Sunburst Arc; perhaps angles as large as 30-40\% of the total surrounding solid angle. In this case, there is no longer a need to invoke a purely serendipetous, á priori very unlikely alignment of a jet or similar, highly directional, mechanism with our line of sight, and with this proposition also disappears the problem of explaining how such a case of mechanical feedback would have extremely efficiently cleared a pathway of neutral gas, yet left a seemingly unaltered dust screen behind. With the new anisotropy characteristics allowed by \citet{almadamonter2024}, we once again find ourselves in territory in which photoionization combined with an asymmetric H~\textsc{i} geometry could clear away the H~\textsc{i} in a cone-like configuration, while leaving the dust screen largely unchanged, as well as leaving behind a majority of projected surrounding area still containing enough H~\textsc{i} to make it optically thick to Ly\(\alpha\), thus generating the classic double-peaked Ly\(\alpha\) profile surrounding the central peak.

%It has long been debated whether mergers or major interactions have a strong impact on LyC emission [CITATIONS]. Most discussions have focused on how mergers or major interactions could spark strong bouts of star formation in one or more of the involved systems; however, another proposition is that tidal stripping resulting from the interaction could help remove significant neutral gas reservoirs from the immediate surroundings of the brightest LyC producing star clusters. One intriguing example is the nearby starburst galaxy Haro 11, which is the closest known LyC leaker \citep{komarova2024,leitet2011,bergvall2006}. This galaxy has highly disturbed morphology and kinematics, and an interacting companion [CITATIONS]. \cite{lereste2024} obseved it in 21 cm H~\textsc{i} emission using MeerKAT, and found a large bridge of neutral gas spanning the space between the two interacting galaxies, while the regions closest to the LyC emitting clusters seem to have been stripped of most of their surrounding H~\textsc{i} envelope (see Fig.\ 1 in that paper).

We propose that such a geometry of an \ion{H}{1} envelope with a cone-like section almost devoid of neutral gas pointing out from the central LCE cluster, is best explained through a combination of tidal stripping of \ion{H}{1} by an interacting companion, combined with strong and highly efficient photoionization of the remaining gas in the regions around the line-of-sight. 
Photoionization as the most important mechanism creating the outflows is suggested by the absence of strong, broad blueshifted line components, and indeed of any major kinematic disruptions, in the kinematic maps in \autoref{fig:kinemaps}. This shows that the gas is unlikely to have been ruptured by major outflows, something which is also supported by the smooth dust cover around the LCE; 
any mechanical feedback capable of creating a channel with \(\log N_{\text{HI}} \lesssim 13\) would have to be extremely efficient and would also have removed or dramatically altered the dust cover in the same region, in contrast to what is observed.
The calm kinematic conditions, smooth dust cover, extremely low neutral column density, and extreme ionizing output of the LCE cluster, all point to photoionization as the more important mechanism creating the escape channels. 
However, as seen in \autoref{fig:dustion}, the highly ionized region surrounding the LCE is relatively small, and the rest of the galaxy much more moderately ionized. Thus, for photionization from the LCE to have ionized the ISM in our direction to such an extreme degree as observed by \cite{sunburst2017}, the ISM cover here must also intrinsically have been thin compared to in other directions.
While multiple scenarios could be conjured to explain such a strongly asymmetric neutral gas geometry, we posit that it is most likely provided by a process of tidal stripping from a major interaction similar to what is found in Haro 11, discussed in \autoref{sec:introduction}. 
Haro 11 is a local starburst galaxy and the closest known LCE \citep{bergvall2006,leitet2011,komarova2024}. It is a late-stage merger, somewhat reminiscent of the famous Antennae galaxy \citep{ostlin2015}, with faint but significant LyC emission from two different star-forming clusters \citep{komarova2024}. 
\cite{lereste2024} found from 21 cm VLA interferometry that the \ion{H}{1} gas had been displaced by the interaction, leaving only a very thin layer near the LyC escape sites, most likely playing an important part in facilitating the escape. 
% The Sunburst source galaxy appears to be in an earlier stage of interaction, 
% The Sunburst source galaxy is not entirely similar to Haro 11. While Haro 11 is in the late stage of a merger, with the main cores already very close to together in the center, the Sunburst galaxy has an interacting companion \(\sim 4\) kpc away, with a visible tidal bridge feature connecting them, likely in an early merger stage (see \autoref{fig:artist}). 
% Still, mergers and interactions are known to be able to displace gas from stars \citep[e.g.][]{bridge2010,bergvall2013},and 
Comparing \autoref{fig:artist} to Fig. 1(a) in \cite{lereste2024}, the Sunburst source galaxy has a geometry consistent with a similar escape scenario of LyC escape facilitated by tidal stripping of \ion{H}{1} gas. In the upper (N) end is the interacting companion, connected to the main galaxy with a bridge-like structure in the central part of the figure. The LCE cluster is located in the bottom (S) end of the figure, farthest away from the companion in a region that could be stripped of much of its H~\textsc{i} through such a tidal interaction. The combined effect of such tidal stripping, and the cluster being located in the outskirts of the main galaxy in a direction away from the bulk of the neutral gas, would conspire to create a preferred direction of very low column density in the neutral envelope covering the LCE cluster.
% While Haro 11 is a late stage merger with the two cores in close vicinity of each other, the Sunburst companion galaxy is clearly distinct from the main galaxy and separated from it by \(\sim 4\) kpc, indicating that it is likely in an early stage of merging. On the other hand, the Sunburst Arc shows a much higher ionizing escape fraction than Haro 11 \(f_{\text{esc,LyC}}^{\text{Haro 11}} \sim 1-2\%\), and much more efficiently cleared 
% We therefore propose that tidal stripping, while not shown to actually take place, could plausibly be occurring also in the Sunburst galaxy and could well be at least in part responsible for clearing the paths for the escaping ionizing radiation. 
Where Haro 11 is a late-stage merger, where the neutral ISM appears to have been flung out, the Sunburst galaxy appears to be in an earlier stage of interaction, where gravitational pull may be playing a more important role stripping the gas away; yet, the resulting gas arrangement is similar. 
% The high ionizing photon production rate in the LCE cluster has helped ionize the remaining neutral ISM with extreme efficiency, given the extremely low neutral column density inferred from the Ly\(\alpha\) profile. 
% The seemingly even dust cover suggests that photoionization, rather than mechanical feedback or radiative pressure, has played the main role in clearing the ionizing escape path.

\begin{figure}[htbp]
\centering
\includegraphics[width=.9\linewidth]{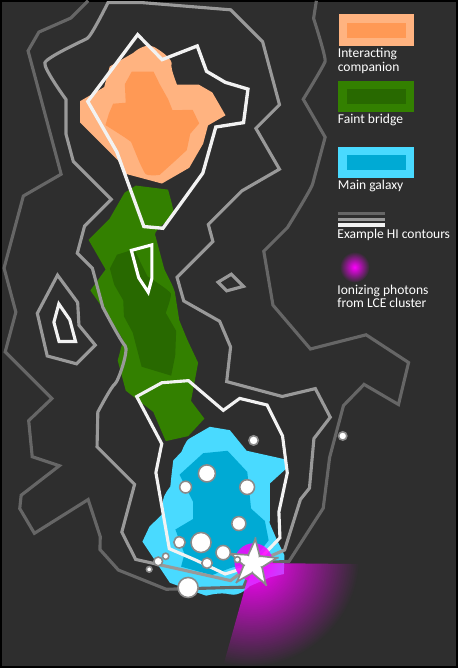}
\caption{\label{fig:expmod}Conceptual sketch of our proposed LyC escape scenario, showing the de-lensed galaxy with interacting companion and fainter bridge as seen in \autoref{fig:artist}. In this scenario, much of the H~\textsc{i} is tidally stripped from the LCE cluster area by the interaction. The strong ionizing emission from the LCE then has little trouble ionizing the remaining H~\textsc{i} cover, even down to an extremely low column density \citep{riverathorsen2017}. The blue, green, and orange patches represent the main galaxy, bridge, and companion, as seen in \autoref{fig:artist}. Copious LyC emission (magenta) fills an H~\textsc{ii} region surrounding the LCE cluster, and wells out from an opening in the H~\textsc{i} created by intense ionization of the already thin neutral gas, in a solid angle including our line of sight.}
\end{figure}

A conceptual view of this escape scenario is shown in \autoref{fig:expmod}. The figure shows in sketch form the same view as in \autoref{fig:artist}. Here, the blue, green, and orange patches represent the main source galaxy, the faint bridge, and the redder interacting companion. The main stellar clumps are drawn in the same positions as seen in that figure. Superimposed is shown in grayscale an illustrative set of \(N_{\text{HI}}\) contours as they could plausibly look in a case of recent or ongoing tidal stripping by the companion \citep[See][Fig.\ 2, for a real-world observed scenario]{lereste2024}. The LCE cluster has ionized a surrounding H~\textsc{ii} region to a degree where the surrounding neutral Hydrogen, already thin from the tidal interaction, has been completely ionized, allowing LyC photons to well out in a cone-like shape. This cone can, from the results of \citep{almadamonter2024}, be up to \(\sim 30\%\) of the projected surrounding area and still be consistent with the Ly\(\alpha\) profile reported in \citep{riverathorsen2017}.

\section{Summary and conclusion}
\label{sec:summary}

In this work, we present NIRSpec integral field spectroscopy of three pointings at the gravitationally lensed \(z = 2.37\) LyC-leaking starburst galaxy, the Sunburst Arc. For each spaxel, we have simultaneously fitted the strongest emission lines to each a single Gaussian profile, all with a shared redshift and FWHM in velocity space, yielding maps of velocity centroid, FWHM, and line fluxes at a pixel resolution of \(0\farcs05\). We constructed maps of dust reddening, ionization parameter \(O_{32}\), ionization tracer \(\Delta\)~[\ion{S}{2}], and shock- and abundance tracer [\ion{N}{2}]/H\(\alpha\).

We have extracted a stacked spectrum of 5 lensed images of the LyC emitting cluster (the LCE) found within the IFU footprints. We have extracted a magnification-corrected and  integrated spectrum of the region within the IFU footprint which covers the largest part of the unlensed galaxy, to mimic a spectrum of the galaxy as it would be observed without gravitational lensing. From these, we were able to compute diagnostic line ratios of interest for LyC escape both for the LCE stack, the full galaxy and in one case also as a 2D map. 

 %{\color{red} \textbf{TODO:} Flesh out more items for this list. All the important ones, in fact.  }

We find that:

\begin{enumerate}
%%%%%%%% MAPS MAPS MAPS MAPS MAPS
%\item Line emission maps shows strong, extended emission in many emission lines, particularly in [\ion{O}{3}]$\lambda 5008$ and 

\item Line centroid velocity maps show that the Sunburst Arc source galaxy displays a strong, though disturbed, velocity gradient, with an inclination- and lensing-uncorrected velocity span \(v_{\text{shear}} = 109.9 \pm 4.7 \text{ km s}^{-1}\), and a lensing-uncorrected intrinsic velocity dispersion \(\sigma_0 = 82.02 \pm 0.14\).

\item The kinematic maps show that the Sunburst Arc is not clearly dominated by neither rotation nor turbulence. The ratio of shearing velocity to intrinsic velocity dispersion is \(v_{\text{shear}}/\sigma_{0} = 1.34 \pm 0.06\). If the inclination is more than 45\textdegree{}, which is the case for 70\% of randomly distributed galaxies, it is dispersion-dominated. In either case, the velocity dispersion is comparatively high for a star forming galaxy, indicating substantial perturbations. We interpret this as a sign of current and/or recent interaction with the companion seen in the upper part of \autoref{fig:artist}, which may have helped spark strong star formation in the main galaxy.

\item The LCE is clearly visible in maps of ionization indicators such as \(O_{32}\) and \(\Delta[ \text{S \scshape{ii}}]\). In contrast, it is barely distinguishable in the maps of dust and kinematics. We interpret this as an indication that mechanical feedback is not strong in this region, leaving photoionization as the dominant mechanism clearing away the surrounding H~\textsc{i} envelope and facilitating LyC escape.

%%%%%%%% NOT MAPS NOT MAPS NOT MAPS NOT MAPS
\item In a stacked spectrum extracted from five gravitationally lensed images of the LCE cluster, we have identified more than 60 rest-frame optical and red emission lines, spanning from \ion{O}{3} \(\lambda\) 3132 to Pa\textsubscript{9} \(\lambda\) 9229, including Balmer lines up to H\textsubscript{20} and Paschen lines from Pa\textsubscript{9} to Pa\textsubscript{22}. The stacked spectrum includes a Balmer jump, treated in \citep{welch2025}.

\item The LCE stack has a sulfur deficiency of \(\Delta[\text{S \scshape{ii}}] = -0.43\). Compared to the LzLCS, this places the Sunburst Arc in a range almost exclusively populated by strong LyC emitters. In contrast, the integrated spectrum \(\Delta[\text{S \scshape{ii}}] = 0.163 \pm 0.218\), a value slightly above the average, in a range which in the LzLCS is populated by both strong and weak leakers and non-leakers \citep{wang2021}, showing that contributions from less ionized surrounding regions can dilute this signature beyond detectability even for a strong LyC leaker.

\item The strong, broad line emission in [\ion{O}{3}] and H\(\alpha\) observed by \citep{mainali2022} does most likely not originate in the LCE itself, but in a Nitrogen-enriched cloud in its vicinity, with a physical distance within a few tens of parsec. This could re-ignite the debates about the role of stellar winds and alternative momentum injection mechanisms in clearing the escape paths for LyC photons.

\item Based on the extremely low \ion{H}{1} column density along the LOS to the LCE clump \citep{sunburst2017}, the presence of an interacting companion to the N of the main galaxy\citep{sharon2022}, the strong spike in ionization around the LCE and lack of signs of strong mechanical feedback around the LCE cluster; we propose a LyC escape scenario in which tidal interaction with the companion has displaced the majority of the \ion{H}{1} envelope around the LCE, leaving the rest to be easily ionized by the strong ionizing emission from the LCE cluster itself.
\end{enumerate}

\begin{acknowledgements}
The authors thank Arjan Bik for useful comments and advice. We also thank the anonymous reviewer for constructive and insightful comments.
This work is based on observations made with the NASA/ESA/CSA James Webb Space Telescope. The data were obtained from the Mikulski Archive for Space Telescopes at the Space Telescope Science Institute, which is operated by the Association of Universities for Research in Astronomy, Inc., under NASA contract NAS 5-03127 for JWST. These observations are associated with program \#02555. The data may be obtained from MAST at \dataset[doi:10.17909/2g5s-2215]{https://doi.org/10.17909/2g5s-2215}.
TER-T is supported by the Swedish Research Council grant \#2022-04805.

%This work is based on observations made with the NASA/ESA/CSA James Webb Space Telescope. The data were obtained from the Mikulski Archive for Space Telescopes at the Space Telescope Science Institute, which is operated by the Association of Universities for Research in Astronomy, Inc., under NASA contract NAS 5-03127 for JWST. These observations are associated with program GO 2555.

\end{acknowledgements}
\facilities{JWST (NIRSpec, NIRCam)}
\software{Astropy \citep{astropy2013,astropy2018,astropy2022},
NumPy \citep{NumPy2020},
Matplotlib \citep{matplotlib},
SciPy \citep{SciPy2020},
Pandas \citep{pandas,Pandas2022}
LMfit \citep{lmfit2014,lmfit_zenodo}
%Uncertainties, 
Jupyter \citep{ipython2007,jupyter2021}
}

\bibliographystyle{aasjournalv7}
\bibliography{./AllPapers}

@article{NumPy2020,
	title = {Array programming with {NumPy}},
	author = {Charles R. Harris and K. Jarrod Millman and St{\'{e}}fan J. van der Walt and Ralf Gommers and Pauli Virtanen and David Cournapeau and Eric Wieser and Julian Taylor and Sebastian Berg and Nathaniel J. Smith and Robert Kern and Matti Picus and Stephan Hoyer and Marten H. van Kerkwijk and Matthew Brett and Allan Haldane and Jaime Fern{\'{a}}ndez del R{\'{i}}o and Mark Wiebe and Pearu Peterson and Pierre G{\'{e}}rard-Marchant and Kevin Sheppard and Tyler Reddy and Warren Weckesser and Hameer Abbasi and Christoph Gohlke and Travis E. Oliphant},
	year = {2020},
	month = sep,
	journal = {Nature},
	volume = {585},
	number = {7825},
	pages = {357--362},
	doi = {10.1038/s41586-020-2649-2},
	publisher = {Springer Science and Business Media {LLC}},
	url = {https://doi.org/10.1038/s41586-020-2649-2}
}

@software{Pandas2022,
	author = {The pandas development team},
	title = {pandas-dev/pandas: Pandas},
	month = nov,
	year = 2022,
	note = {If you use this software, please cite it as below.},
	publisher = {Zenodo},
	version = {v1.5.2},
	doi = {10.5281/zenodo.7344967},
	url = {https://doi.org/10.5281/zenodo.7344967}
}

@article{SciPy2020,
	author = {Virtanen, Pauli and Gommers, Ralf and Oliphant, Travis E. and Haberland, Matt and Reddy, Tyler and Cournapeau, David and Burovski, Evgeni and Peterson, Pearu and Weckesser, Warren and Bright, Jonathan and {van der Walt}, St{\'e}fan J. and Brett, Matthew and Wilson, Joshua and Millman, K. Jarrod and Mayorov, Nikolay and Nelson, Andrew R. J. and Jones, Eric and Kern, Robert and Larson, Eric and Carey, C J and Polat, {\.I}lhan and Feng, Yu and Moore, Eric W. and {VanderPlas}, Jake and Laxalde, Denis and Perktold, Josef and Cimrman, Robert and Henriksen, Ian and Quintero, E. A. and Harris, Charles R. and Archibald, Anne M. and Ribeiro, Ant{\^o}nio H. and Pedregosa, Fabian and {van Mulbregt}, Paul and {SciPy 1.0 Contributors}},
	title = {{{SciPy} 1.0: Fundamental Algorithms for Scientific Computing in Python}},
	journal = {Nature Methods},
	year = {2020},
	volume = {17},
	pages = {261--272},
	adsurl = {https://rdcu.be/b08Wh},
	doi = {10.1038/s41592-019-0686-2}
}

@article{alexandroff2015,
	author = {{Alexandroff}, Rachael M. and {Heckman}, Timothy M. and {Borthakur}, Sanchayeeta and {Overzier}, Roderik and {Leitherer}, Claus},
	title = "{Indirect Evidence for Escaping Ionizing Photons in Local Lyman Break Galaxy Analogs}",
	journal = {\apj},
	year = 2015,
	month = sep,
	volume = {810},
	number = {2},
	eid = {104},
	pages = {104},
	doi = {10.1088/0004-637X/810/2/104},
	archiveprefix = {arXiv},
	eprint = {1504.02446},
	primaryclass = {astro-ph.GA},
	adsurl = {https://ui.adsabs.harvard.edu/abs/2015ApJ...810..104A}
}

@ARTICLE{almadamonter2024,
	author = {{Almada Monter}, Silvia and {Gronke}, Max},
	title = "{Crossing walls and windows: the curious escape of Lyman-{\ensuremath{\alpha}} photons through ionized channels}",
	journal = {\mnras},
	year = 2024,
	month = oct,
	volume = {534},
	number = {1},
	pages = {L7-L13},
	doi = {10.1093/mnrasl/slae074},
	archivePrefix = {arXiv},
	eprint = {2404.07169},
	primaryClass = {astro-ph.GA},
	adsurl = {https://ui.adsabs.harvard.edu/abs/2024MNRAS.534L...7A}
}

@ARTICLE{amorin2024,
	author = {{Amor{\'\i}n}, R.~O. and {Rodr{\'\i}guez-Henr{\'\i}quez}, M. and {Fern{\'a}ndez}, V. and {V{\'\i}lchez}, J.~M. and {Marques-Chaves}, R. and {Schaerer}, D. and {Izotov}, Y.~I. and {Firpo}, V. and {Guseva}, N. and {Jaskot}, A.~E. and {Komarova}, L. and {Mu{\~n}oz-Vergara}, D. and {Oey}, M.~S. and {Bait}, O. and {Carr}, C. and {Chisholm}, J. and {Ferguson}, H. and {Flury}, S.~R. and {Giavalisco}, M. and {Hayes}, M.~J. and {Henry}, A. and {Ji}, Z. and {King}, W. and {Leclercq}, F. and {{\"O}stlin}, G. and {Pentericci}, L. and {Saldana-Lopez}, A. and {Thuan}, T.~X. and {Trebitsch}, M. and {Wang}, B. and {Worseck}, G. and {Xu}, X.},
	title = "{Ubiquitous broad-line emission and the relation between ionized gas outflows and Lyman continuum escape in Green Pea galaxies}",
	journal = {\aap},
	year = 2024,
	month = feb,
	volume = {682},
	eid = {L25},
	pages = {L25},
	doi = {10.1051/0004-6361/202449175},
	archivePrefix = {arXiv},
	eprint = {2401.04278},
	primaryClass = {astro-ph.GA},
	adsurl = {https://ui.adsabs.harvard.edu/abs/2024A&A...682L..25A}
}

@article{astropy2013,
	author = {{Astropy Collaboration} and {Robitaille}, T.~P. and {Tollerud}, E.~J. and {Greenfield}, P. and {Droettboom}, M. and {Bray}, E. and {Aldcroft}, T. and {Davis}, M. and {Ginsburg}, A. and {Price-Whelan}, A.~M. and {Kerzendorf}, W.~E. and {Conley}, A. and {Crighton}, N. and {Barbary}, K. and {Muna}, D. and {Ferguson}, H. and {Grollier}, F. and {Parikh}, M.~M. and {Nair}, P.~H. and {Unther}, H.~M. and {Deil}, C. and {Woillez}, J. and {Conseil}, S. and {Kramer}, R. and {Turner}, J.~E.~H. and {Singer}, L. and {Fox}, R. and {Weaver}, B.~A. and {Zabalza}, V. and {Edwards}, Z.~I. and {Azalee Bostroem}, K. and {Burke}, D.~J. and {Casey}, A.~R. and {Crawford}, S.~M. and {Dencheva}, N. and {Ely}, J. and {Jenness}, T. and {Labrie}, K. and {Lim}, P.~L. and {Pierfederici}, F. and {Pontzen}, A. and {Ptak}, A. and {Refsdal}, B. and {Servillat}, M. and {Streicher}, O.},
	title = "{Astropy: A community Python package for astronomy}",
	journal = {\aap},
	archiveprefix = "arXiv",
	eprint = {1307.6212},
	primaryclass = "astro-ph.IM",
	year = 2013,
	month = oct,
	volume = 558,
	eid = {A33},
	pages = {A33},
	doi = {10.1051/0004-6361/201322068},
	adsurl = {http://adsabs.harvard.edu/abs/2013A\%26A...558A..33A}
}

@article{astropy2018,
	author = {{Astropy Collaboration} and {Price-Whelan}, A.~M. and {Sip{\H{o}}cz}, B.~M. and {G{\"u}nther}, H.~M. and {Lim}, P.~L. and {Crawford}, S.~M. and {Conseil}, S. and {Shupe}, D.~L. and {Craig}, M.~W. and {Dencheva}, N. and {Ginsburg}, A. and {Vand erPlas}, J.~T. and {Bradley}, L.~D. and {P{\'e}rez-Su{\'a}rez}, D. and {de Val-Borro}, M. and {Aldcroft}, T.~L. and {Cruz}, K.~L. and {Robitaille}, T.~P. and {Tollerud}, E.~J. and {Ardelean}, C. and {Babej}, T. and {Bach}, Y.~P. and {Bachetti}, M. and {Bakanov}, A.~V. and {Bamford}, S.~P. and {Barentsen}, G. and {Barmby}, P. and {Baumbach}, A. and {Berry}, K.~L. and {Biscani}, F. and {Boquien}, M. and {Bostroem}, K.~A. and {Bouma}, L.~G. and {Brammer}, G.~B. and {Bray}, E.~M. and {Breytenbach}, H. and {Buddelmeijer}, H. and {Burke}, D.~J. and {Calderone}, G. and {Cano Rodr{\'\i}guez}, J.~L. and {Cara}, M. and {Cardoso}, J.~V.~M. and {Cheedella}, S. and {Copin}, Y. and {Corrales}, L. and {Crichton}, D. and {D'Avella}, D. and {Deil}, C. and {Depagne}, {\'E}. and {Dietrich}, J.~P. and {Donath}, A. and {Droettboom}, M. and {Earl}, N. and {Erben}, T. and {Fabbro}, S. and {Ferreira}, L.~A. and {Finethy}, T. and {Fox}, R.~T. and {Garrison}, L.~H. and {Gibbons}, S.~L.~J. and {Goldstein}, D.~A. and {Gommers}, R. and {Greco}, J.~P. and {Greenfield}, P. and {Groener}, A.~M. and {Grollier}, F. and {Hagen}, A. and {Hirst}, P. and {Homeier}, D. and {Horton}, A.~J. and {Hosseinzadeh}, G. and {Hu}, L. and {Hunkeler}, J.~S. and {Ivezi{\'c}}, {\v{Z}}. and {Jain}, A. and {Jenness}, T. and {Kanarek}, G. and {Kendrew}, S. and {Kern}, N.~S. and {Kerzendorf}, W.~E. and {Khvalko}, A. and {King}, J. and {Kirkby}, D. and {Kulkarni}, A.~M. and {Kumar}, A. and {Lee}, A. and {Lenz}, D. and {Littlefair}, S.~P. and {Ma}, Z. and {Macleod}, D.~M. and {Mastropietro}, M. and {McCully}, C. and {Montagnac}, S. and {Morris}, B.~M. and {Mueller}, M. and {Mumford}, S.~J. and {Muna}, D. and {Murphy}, N.~A. and {Nelson}, S. and {Nguyen}, G.~H. and {Ninan}, J.~P. and {N{\"o}the}, M. and {Ogaz}, S. and {Oh}, S. and {Parejko}, J.~K. and {Parley}, N. and {Pascual}, S. and {Patil}, R. and {Patil}, A.~A. and {Plunkett}, A.~L. and {Prochaska}, J.~X. and {Rastogi}, T. and {Reddy Janga}, V. and {Sabater}, J. and {Sakurikar}, P. and {Seifert}, M. and {Sherbert}, L.~E. and {Sherwood-Taylor}, H. and {Shih}, A.~Y. and {Sick}, J. and {Silbiger}, M.~T. and {Singanamalla}, S. and {Singer}, L.~P. and {Sladen}, P.~H. and {Sooley}, K.~A. and {Sornarajah}, S. and {Streicher}, O. and {Teuben}, P. and {Thomas}, S.~W. and {Tremblay}, G.~R. and {Turner}, J.~E.~H. and {Terr{\'o}n}, V. and {van Kerkwijk}, M.~H. and {de la Vega}, A. and {Watkins}, L.~L. and {Weaver}, B.~A. and {Whitmore}, J.~B. and {Woillez}, J. and {Zabalza}, V. and {Astropy Contributors}},
	title = "{The Astropy Project: Building an Open-science Project and Status of the v2.0 Core Package}",
	journal = {\aj},
	year = 2018,
	month = sep,
	volume = {156},
	number = {3},
	eid = {123},
	pages = {123},
	doi = {10.3847/1538-3881/aabc4f},
	archiveprefix = {arXiv},
	eprint = {1801.02634},
	primaryclass = {astro-ph.IM},
	adsurl = {https://ui.adsabs.harvard.edu/abs/2018AJ....156..123A}
}

@article{astropy2022,
	author = {{Astropy Collaboration} and {Price-Whelan}, Adrian M. and {Lim}, Pey Lian and {Earl}, Nicholas and {Starkman}, Nathaniel and {Bradley}, Larry and {Shupe}, David L. and {Patil}, Aarya A. and {Corrales}, Lia and {Brasseur}, C.~E. and {N{"o}the}, Maximilian and {Donath}, Axel and {Tollerud}, Erik and {Morris}, Brett M. and {Ginsburg}, Adam and {Vaher}, Eero and {Weaver}, Benjamin A. and {Tocknell}, James and {Jamieson}, William and {van Kerkwijk}, Marten H. and {Robitaille}, Thomas P. and {Merry}, Bruce and {Bachetti}, Matteo and {G{"u}nther}, H. Moritz and {Aldcroft}, Thomas L. and {Alvarado-Montes}, Jaime A. and {Archibald}, Anne M. and {B{'o}di}, Attila and {Bapat}, Shreyas and {Barentsen}, Geert and {Baz{'a}n}, Juanjo and {Biswas}, Manish and {Boquien}, M{'e}d{'e}ric and {Burke}, D.~J. and {Cara}, Daria and {Cara}, Mihai and {Conroy}, Kyle E. and {Conseil}, Simon and {Craig}, Matthew W. and {Cross}, Robert M. and {Cruz}, Kelle L. and {D'Eugenio}, Francesco and {Dencheva}, Nadia and {Devillepoix}, Hadrien A.~R. and {Dietrich}, J{"o}rg P. and {Eigenbrot}, Arthur Davis and {Erben}, Thomas and {Ferreira}, Leonardo and {Foreman-Mackey}, Daniel and {Fox}, Ryan and {Freij}, Nabil and {Garg}, Suyog and {Geda}, Robel and {Glattly}, Lauren and {Gondhalekar}, Yash and {Gordon}, Karl D. and {Grant}, David and {Greenfield}, Perry and {Groener}, Austen M. and {Guest}, Steve and {Gurovich}, Sebastian and {Handberg}, Rasmus and {Hart}, Akeem and {Hatfield-Dodds}, Zac and {Homeier}, Derek and {Hosseinzadeh}, Griffin and {Jenness}, Tim and {Jones}, Craig K. and {Joseph}, Prajwel and {Kalmbach}, J. Bryce and {Karamehmetoglu}, Emir and {Ka{l}uszy{'n}ski}, Miko{l}aj and {Kelley}, Michael S.~P. and {Kern}, Nicholas and {Kerzendorf}, Wolfgang E. and {Koch}, Eric W. and {Kulumani}, Shankar and {Lee}, Antony and {Ly}, Chun and {Ma}, Zhiyuan and {MacBride}, Conor and {Maljaars}, Jakob M. and {Muna}, Demitri and {Murphy}, N.~A. and {Norman}, Henrik and {O'Steen}, Richard and {Oman}, Kyle A. and {Pacifici}, Camilla and {Pascual}, Sergio and {Pascual-Granado}, J. and {Patil}, Rohit R. and {Perren}, Gabriel I. and {Pickering}, Timothy E. and {Rastogi}, Tanuj and {Roulston}, Benjamin R. and {Ryan}, Daniel F. and {Rykoff}, Eli S. and {Sabater}, Jose and {Sakurikar}, Parikshit and {Salgado}, Jes{'u}s and {Sanghi}, Aniket and {Saunders}, Nicholas and {Savchenko}, Volodymyr and {Schwardt}, Ludwig and {Seifert-Eckert}, Michael and {Shih}, Albert Y. and {Jain}, Anany Shrey and {Shukla}, Gyanendra and {Sick}, Jonathan and {Simpson}, Chris and {Singanamalla}, Sudheesh and {Singer}, Leo P. and {Singhal}, Jaladh and {Sinha}, Manodeep and {Sip{H{o}}cz}, Brigitta M. and {Spitler}, Lee R. and {Stansby}, David and {Streicher}, Ole and {{{S}}umak}, Jani and {Swinbank}, John D. and {Taranu}, Dan S. and {Tewary}, Nikita and {Tremblay}, Grant R. and {Val-Borro}, Miguel de and {Van Kooten}, Samuel J. and {Vasovi{'c}}, Zlatan and {Verma}, Shresth and {de Miranda Cardoso}, Jos{'e} Vin{'i}cius and {Williams}, Peter K.~G. and {Wilson}, Tom J. and {Winkel}, Benjamin and {Wood-Vasey}, W.~M. and {Xue}, Rui and {Yoachim}, Peter and {Zhang}, Chen and {Zonca}, Andrea and {Astropy Project Contributors}},
	title = "{The Astropy Project: Sustaining and Growing a Community-oriented Open-source Project and the Latest Major Release (v5.0) of the Core Package}",
	journal = {apj},
	year = 2022,
	month = aug,
	volume = {935},
	number = {2},
	eid = {167},
	pages = {167},
	doi = {10.3847/1538-4357/ac7c74},
	archiveprefix = {arXiv},
	eprint = {2206.14220},
	primaryclass = {astro-ph.IM},
	adsurl = {https://ui.adsabs.harvard.edu/abs/2022ApJ...935..167A}
}

@article{baldwin1981,
	author = {{Baldwin}, J.~A. and {Phillips}, M.~M. and {Terlevich}, R.},
	title = "{Classification parameters for the emission-line spectra of extragalactic objects}",
	journal = {\pasp},
	year = 1981,
	month = feb,
	volume = 93,
	pages = {5--19},
	doi = {10.1086/130766},
	adsurl = {http://adsabs.harvard.edu/abs/1981PASP...93....5B}
}

@software{baryonsweep,
	author = {{Hutchison}, Taylor A. and {Welch}, Brian D. and {Rigby}, Jane R. and {Olivier}, Grace M. and {Birkin}, Jack E. and {Phadke}, Kedar A. and {Khullar}, Gourav and {Rauscher}, Bernard J. and {Sharon}, Keren and {Aravena}, Manuel and {Bayliss}, Matthew B. and {Elicker}, Lauren A. and {Kim}, Seonwoo and {Solimano}, Manuel and {Vieira}, Joaquin D. and {Vizgan}, David},
	title = "{baryon-sweep: Outlier rejection algorithm for JWST/NIRSpec IFS data}",
	howpublished = {Astrophysics Source Code Library, record ascl:2401.012},
	year = 2024,
	month = jan,
	eid = {ascl:2401.012},
	adsurl = {https://ui.adsabs.harvard.edu/abs/2024ascl.soft01012H}
}

@article{becker2021,
	author = {{Becker}, George D. and {D'Aloisio}, Anson and {Christenson}, Holly M. and {Zhu}, Yongda and {Worseck}, G{\'a}bor and {Bolton}, James S.},
	title = "{The mean free path of ionizing photons at 5 < z < 6: evidence for rapid evolution near reionization}",
	journal = {\mnras},
	year = 2021,
	month = dec,
	volume = {508},
	number = {2},
	pages = {1853--1869},
	doi = {10.1093/mnras/stab2696},
	archiveprefix = {arXiv},
	eprint = {2103.16610},
	primaryclass = {astro-ph.CO},
	adsurl = {https://ui.adsabs.harvard.edu/abs/2021MNRAS.508.1853B}
}

@article{behrens2014,
	author = {{Behrens}, C. and {Dijkstra}, M. and {Niemeyer}, J.~C.},
	title = "{Beamed Ly{$\alpha$} emission through outflow-driven cavities}",
	journal = {\aap},
	archiveprefix = "arXiv",
	eprint = {1401.4860},
	year = 2014,
	month = mar,
	volume = 563,
	eid = {A77},
	pages = {A77},
	doi = {10.1051/0004-6361/201322949},
	adsurl = {http://adsabs.harvard.edu/abs/2014A\%26A...563A..77B}
}

@article{bergvall2006,
	author = {{Bergvall}, N. and {Zackrisson}, E. and {Andersson}, B.-G. and {Arnberg}, D. and {Masegosa}, J. and {{\"O}stlin}, G.},
	title = "{First detection of Lyman continuum escape from a local starburst galaxy. I. Observations of the luminous blue compact galaxy Haro 11 with the Far Ultraviolet Spectroscopic Explorer (FUSE)}",
	journal = {\aap},
	year = 2006,
	month = mar,
	volume = 448,
	pages = {513--524},
	doi = {10.1051/0004-6361:20053788},
	adsurl = {http://adsabs.harvard.edu/abs/2006A\%26A...448..513B}
}

@article{bezanson2017julia,
	title = {Julia: A fresh approach to numerical computing},
	author = {Bezanson, Jeff and Edelman, Alan and Karpinski, Stefan and Shah, Viral B},
	journal = {SIAM review},
	volume = {59},
	number = {1},
	pages = {65--98},
	year = {2017},
	publisher = {SIAM},
	url = {https://doi.org/10.1137/141000671}
}

@article{bian2017,
	author = {{Bian}, F. and {Fan}, X. and {McGreer}, I. and {Cai}, Z. and {Jiang}, L.},
	title = "{High Lyman Continuum Escape Fraction in a Lensed Young Compact Dwarf Galaxy at z = 2.5}",
	journal = {\apjl},
	archiveprefix = "arXiv",
	eprint = {1702.06540},
	year = 2017,
	month = mar,
	volume = 837,
	eid = {L12},
	pages = {L12},
	doi = {10.3847/2041-8213/aa5ff7},
	adsurl = {http://adsabs.harvard.edu/abs/2017ApJ...837L..12B}
}

@ARTICLE{bik2022,
	author = {{Bik}, A. and {{\"O}stlin}, G. and {Hayes}, M. and {Melinder}, J. and {Menacho}, V.},
	title = "{Spatially resolved gas and stellar kinematics in compact starburst galaxies}",
	journal = {\aap},
	year = 2022,
	month = oct,
	volume = {666},
	eid = {A161},
	pages = {A161},
	doi = {10.1051/0004-6361/202243739},
	archivePrefix = {arXiv},
	eprint = {2208.10876},
	primaryClass = {astro-ph.GA},
	adsurl = {https://ui.adsabs.harvard.edu/abs/2022A&A...666A.161B}
}

@ARTICLE{boeker2023,
	author = {{B{\"o}ker}, T. and {Beck}, T.~L. and {Birkmann}, S.~M. and {Giardino}, G. and {Keyes}, C. and {Kumari}, N. and {Muzerolle}, J. and {Rawle}, T. and {Zeidler}, P. and {Abul-Huda}, Y. and {Alves de Oliveira}, C. and {Arribas}, S. and {Bechtold}, K. and {Bhatawdekar}, R. and {Bonaventura}, N. and {Bunker}, A.~J. and {Cameron}, A.~J. and {Carniani}, S. and {Charlot}, S. and {Curti}, M. and {Espinoza}, N. and {Ferruit}, P. and {Franx}, M. and {Jakobsen}, P. and {Karakla}, D. and {L{\'o}pez-Caniego}, M. and {L{\"u}tzgendorf}, N. and {Maiolino}, R. and {Manjavacas}, E. and {Marston}, A.~P. and {Moseley}, S.~H. and {Ogle}, P. and {Perna}, M. and {Pe{\~n}a-Guerrero}, M. and {Pirzkal}, N. and {Plesha}, R. and {Proffitt}, C.~R. and {Rauscher}, B.~J. and {Rix}, H. -W. and {Rodr{\'\i}guez del Pino}, B. and {Rustamkulov}, Z. and {Sabbi}, E. and {Sing}, D.~K. and {Sirianni}, M. and {te Plate}, M. and {{\'U}beda}, L. and {Wahlgren}, G.~M. and {Wislowski}, E. and {Wu}, R. and {Willott}, Chris J.},
	title = "{In-orbit Performance of the Near-infrared Spectrograph NIRSpec on the James Webb Space Telescope}",
	journal = {\pasp},
	year = 2023,
	month = mar,
	volume = {135},
	number = {1045},
	eid = {038001},
	pages = {038001},
	doi = {10.1088/1538-3873/acb846},
	archivePrefix = {arXiv},
	eprint = {2301.13766},
	primaryClass = {astro-ph.IM},
	adsurl = {https://ui.adsabs.harvard.edu/abs/2023PASP..135c8001B}
}

@article{borthakur2014,
	author = {{Borthakur}, S. and {Heckman}, T.~M. and {Leitherer}, C. and {Overzier}, R.~A.},
	title = "{A local clue to the reionization of the universe}",
	journal = {Science},
	archiveprefix = "arXiv",
	eprint = {1410.3511},
	year = 2014,
	month = oct,
	volume = 346,
	pages = {216--219},
	doi = {10.1126/science.1254214},
	adsurl = {http://adsabs.harvard.edu/abs/2014Sci...346..216B}
}

@MISC{bushouse23_pipeine1p11p4,
	author = {{Bushouse}, Howard and {Eisenhamer}, Jonathan and {Dencheva}, Nadia and {Davies}, James and {Greenfield}, Perry and {Morrison}, Jane and {Hodge}, Phil and {Simon}, Bernie and {Grumm}, David and {Droettboom}, Michael and {Slavich}, Edward and {Sosey}, Megan and {Pauly}, Tyler and {Miller}, Todd and {Jedrzejewski}, Robert and {Hack}, Warren and {Davis}, David and {Crawford}, Steven and {Law}, David and {Gordon}, Karl and {Regan}, Michael and {Cara}, Mihai and {MacDonald}, Ken and {Bradley}, Larry and {Shanahan}, Clare and {Jamieson}, William and {Teodoro}, Mairan and {Williams}, Thomas},
	title = "{JWST Calibration Pipeline}",
	year = 2023,
	month = aug,
	eid = {10.5281/zenodo.8247246},
	doi = {10.5281/zenodo.8247246},
	version = {1.11.4},
	publisher = {Zenodo},
	adsurl = {https://ui.adsabs.harvard.edu/abs/2023zndo...8247246B}
}

@article{calzetti2000,
	author = {{Calzetti}, Daniela and {Armus}, Lee and {Bohlin}, Ralph C. and {Kinney}, Anne L. and {Koornneef}, Jan and {Storchi-Bergmann}, Thaisa},
	title = "{The Dust Content and Opacity of Actively Star-forming Galaxies}",
	journal = {\apj},
	year = 2000,
	month = apr,
	volume = {533},
	number = {2},
	pages = {682--695},
	doi = {10.1086/308692},
	archiveprefix = {arXiv},
	eprint = {astro-ph/9911459},
	primaryclass = {astro-ph},
	adsurl = {https://ui.adsabs.harvard.edu/abs/2000ApJ...533..682C}
}

@ARTICLE{capellari2003,
	author = {{Cappellari}, Michele and {Copin}, Yannick},
	title = "{Adaptive spatial binning of integral-field spectroscopic data using Voronoi tessellations}",
	journal = {\mnras},
	year = 2003,
	month = jun,
	volume = {342},
	number = {2},
	pages = {345-354},
	doi = {10.1046/j.1365-8711.2003.06541.x},
	archivePrefix = {arXiv},
	eprint = {astro-ph/0302262},
	primaryClass = {astro-ph},
	adsurl = {https://ui.adsabs.harvard.edu/abs/2003MNRAS.342..345C}
}

@article{cardamone2009,
	author = {{Cardamone}, C. and {Schawinski}, K. and {Sarzi}, M. and {Bamford}, S.~P. and {Bennert}, N. and {Urry}, C.~M. and {Lintott}, C. and {Keel}, W.~C. and {Parejko}, J. and {Nichol}, R.~C. and {Thomas}, D. and {Andreescu}, D. and {Murray}, P. and {Raddick}, M.~J. and {Slosar}, A. and {Szalay}, A. and {Vandenberg}, J.},
	title = "{Galaxy Zoo Green Peas: discovery of a class of compact extremely star-forming galaxies}",
	journal = {\mnras},
	archiveprefix = "arXiv",
	eprint = {0907.4155},
	year = 2009,
	month = nov,
	volume = 399,
	pages = {1191--1205},
	doi = {10.1111/j.1365-2966.2009.15383.x},
	adsurl = {http://adsabs.harvard.edu/abs/2009MNRAS.399.1191C}
}

@article{chisholm2017,
	author = {{Chisholm}, J. and {Orlitov{\'a}}, I. and {Schaerer}, D. and {Verhamme}, A. and {Worseck}, G. and {Izotov}, Y.~I. and {Thuan}, T.~X. and {Guseva}, N.~G.},
	title = "{Do galaxies that leak ionizing photons have extreme outflows?}",
	journal = {\aap},
	archiveprefix = "arXiv",
	eprint = {1707.01913},
	year = 2017,
	month = sep,
	volume = 605,
	eid = {A67},
	pages = {A67},
	doi = {10.1051/0004-6361/201730610},
	adsurl = {http://adsabs.harvard.edu/abs/2017A\%26A...605A..67C}
}

@article{chisholm2018a,
	author = {{Chisholm}, J. and {Gazagnes}, S. and {Schaerer}, D. and {Verhamme}, A. and {Rigby}, J.~R. and {Bayliss}, M. and {Sharon}, K. and {Gladders}, M. and {Dahle}, H.},
	title = "{Accurately predicting the escape fraction of ionizing photons using rest-frame ultraviolet absorption lines}",
	journal = {\aap},
	year = 2018,
	month = aug,
	volume = {616},
	eid = {A30},
	pages = {A30},
	doi = {10.1051/0004-6361/201832758},
	primaryclass = {astro-ph.GA},
	adsurl = {https://ui.adsabs.harvard.edu/\#abs/2018A\&A...616A..30C}
}

@article{chisholm2019,
	author = {{Chisholm}, J. and {Rigby}, J.~R. and {Bayliss}, M. and {Berg}, D.~A. and {Dahle}, H. and {Gladders}, M. and {Sharon}, K.},
	title = "{Constraining the Metallicities, Ages, Star Formation Histories, and Ionizing Continua of Extragalactic Massive Star Populations}",
	journal = {\apj},
	year = "2019",
	month = "Sep",
	volume = {882},
	number = {2},
	eid = {182},
	pages = {182},
	doi = {10.3847/1538-4357/ab3104},
	archiveprefix = {arXiv},
	eprint = {1905.04314},
	primaryclass = {astro-ph.GA},
	adsurl = {https://ui.adsabs.harvard.edu/abs/2019ApJ...882..182C}
}

@ARTICLE{choe2025,
	author = {{Choe}, S. and {Emil Rivera-Thorsen}, T. and {Dahle}, H. and {Sharon}, K. and {Owens}, M. Riley and {Rigby}, J.~R. and {Bayliss}, M.~B. and {Hayes}, M.~J. and {Hutchison}, T. and {Welch}, B. and {Chisholm}, J. and {Gladders}, M.~D. and {Khullar}, G. and {Kim}, K.},
	title = "{The Sunburst Arc with JWST: II. Observations of an Eta Carinae analog at z = 2.37}",
	journal = {\aap},
	year = 2025,
	month = jun,
	volume = {698},
	eid = {A16},
	pages = {A16},
	doi = {10.1051/0004-6361/202450685},
	archivePrefix = {arXiv},
	eprint = {2405.06953},
	primaryClass = {astro-ph.GA},
	adsurl = {https://ui.adsabs.harvard.edu/abs/2025A&A...698A..16C}
}

@article{cooke2014,
	author = {{Cooke}, J. and {Ryan-Weber}, E.~V. and {Garel}, T. and {D{\'\i}az}, C.~G.},
	title = "{Lyman-continuum galaxies and the escape fraction of Lyman-break galaxies}",
	journal = {\mnras},
	year = 2014,
	month = jun,
	volume = {441},
	number = {1},
	pages = {837--851},
	doi = {10.1093/mnras/stu635},
	archiveprefix = {arXiv},
	eprint = {1404.0125},
	primaryclass = {astro-ph.GA},
	adsurl = {https://ui.adsabs.harvard.edu/abs/2014MNRAS.441..837C}
}

@article{dahle2016,
	author = {{Dahle}, H. and {Aghanim}, N. and {Guennou}, L. and {Hudelot}, P. and {Kneissl}, R. and {Pointecouteau}, E. and {Beelen}, A. and {Bayliss}, M. and {Douspis}, M. and {Nesvadba}, N. and {Hempel}, A. and {Gronke}, M. and {Burenin}, R. and {Dole}, H. and {Harrison}, D. and {Mazzotta}, P. and {Sunyaev}, R.},
	title = "{Discovery of an exceptionally bright giant arc at z = 2.369, gravitationally lensed by the Planck cluster PSZ1 G311.65-18.48}",
	journal = {\aap},
	year = 2016,
	month = may,
	volume = 590,
	eid = {L4},
	pages = {L4},
	doi = {10.1051/0004-6361/201628297},
	adsurl = {http://adsabs.harvard.edu/abs/2016A\%26A...590L...4D}
}

@misc{dai2024arxiv,
	title = {Partially Ionized Gas at Equilibrium Powered by Ultraviolet Irradiation},
	author = {Liang Dai},
	year = {2024},
	eprint = {2410.09753},
	archivePrefix = {arXiv},
	primaryClass = {astro-ph.GA},
	url = {https://arxiv.org/abs/2410.09753}
}

@article{davies2021,
	author = {{Davies}, Frederick B. and {Bosman}, Sarah E.~I. and {Furlanetto}, Steven R. and {Becker}, George D. and {D'Aloisio}, Anson},
	title = "{The Predicament of Absorption-dominated Reionization: Increased Demands on Ionizing Sources}",
	journal = {\apjl},
	year = 2021,
	month = sep,
	volume = {918},
	number = {2},
	eid = {L35},
	pages = {L35},
	doi = {10.3847/2041-8213/ac1ffb},
	archiveprefix = {arXiv},
	eprint = {2105.10518},
	primaryclass = {astro-ph.CO},
	adsurl = {https://ui.adsabs.harvard.edu/abs/2021ApJ...918L..35D}
}

@article{debarros2016,
	author = {{de Barros}, S. and {Vanzella}, E. and {Amor{\'{\i}}n}, R. and {Castellano}, M. and {Siana}, B. and {Grazian}, A. and {Suh}, H. and {Balestra}, I. and {Vignali}, C. and {Verhamme}, A. and {Zamorani}, G. and {Mignoli}, M. and {Hasinger}, G. and {Comastri}, A. and {Pentericci}, L. and {P{\'e}rez-Montero}, E. and {Fontana}, A. and {Giavalisco}, M. and {Gilli}, R.},
	title = "{An extreme [O III] emitter at z = 3.2: a low metallicity Lyman continuum source}",
	journal = {\aap},
	archiveprefix = "arXiv",
	eprint = {1507.06648},
	year = 2016,
	month = jan,
	volume = 585,
	eid = {A51},
	pages = {A51},
	doi = {10.1051/0004-6361/201527046},
	adsurl = {http://adsabs.harvard.edu/abs/2016A\%26A...585A..51D}
}

@ARTICLE{diego2022,
	author = {{Diego}, J.~M. and {Pascale}, M. and {Kavanagh}, B.~J. and {Kelly}, P. and {Dai}, L. and {Frye}, B. and {Broadhurst}, T.},
	title = "{Godzilla, a monster lurks in the Sunburst galaxy}",
	journal = {\aap},
	year = 2022,
	month = sep,
	volume = {665},
	eid = {A134},
	pages = {A134},
	doi = {10.1051/0004-6361/202243605},
	archivePrefix = {arXiv},
	eprint = {2203.08158},
	primaryClass = {astro-ph.GA},
	adsurl = {https://ui.adsabs.harvard.edu/abs/2022A&A...665A.134D}
}

@article{dijkstra2016,
	author = {{Dijkstra}, M. and {Gronke}, M. and {Venkatesan}, A.},
	title = "{The Ly{$\alpha$}-LyC Connection: Evidence for an Enhanced Contribution of UV-faint Galaxies to Cosmic Reionization}",
	journal = {\apj},
	archiveprefix = "arXiv",
	eprint = {1604.08208},
	year = 2016,
	month = sep,
	volume = 828,
	eid = {71},
	pages = {71},
	doi = {10.3847/0004-637X/828/2/71},
	adsurl = {http://adsabs.harvard.edu/abs/2016ApJ...828...71D}
}

@article{dijkstrarev,
	author = {{Dijkstra}, M.},
	title = "{Ly{$\alpha$} Emitting Galaxies as a Probe of Reionisation}",
	journal = {\pasa},
	archiveprefix = "arXiv",
	eprint = {1406.7292},
	year = 2014,
	month = oct,
	volume = 31,
	eid = {e040},
	pages = {40},
	doi = {10.1017/pasa.2014.33},
	adsurl = {http://adsabs.harvard.edu/abs/2014PASA...31...40D}
}

@ARTICLE{dominguez2013,
	author = {{Dom{\'\i}nguez}, A. and {Siana}, B. and {Henry}, A.~L. and {Scarlata}, C. and {Bedregal}, A.~G. and {Malkan}, M. and {Atek}, H. and {Ross}, N.~R. and {Colbert}, J.~W. and {Teplitz}, H.~I. and {Rafelski}, M. and {McCarthy}, P. and {Bunker}, A. and {Hathi}, N.~P. and {Dressler}, A. and {Martin}, C.~L. and {Masters}, D.},
	title = "{Dust Extinction from Balmer Decrements of Star-forming Galaxies at 0.75 <= z <= 1.5 with Hubble Space Telescope/Wide-Field-Camera 3 Spectroscopy from the WFC3 Infrared Spectroscopic Parallel Survey}",
	journal = {\apj},
	year = 2013,
	month = feb,
	volume = {763},
	number = {2},
	eid = {145},
	pages = {145},
	doi = {10.1088/0004-637X/763/2/145},
	archivePrefix = {arXiv},
	eprint = {1206.1867},
	primaryClass = {astro-ph.CO},
	adsurl = {https://ui.adsabs.harvard.edu/abs/2013ApJ...763..145D}
}

@ARTICLE{duan2025,
	author = {{Duan}, Qiao and {Conselice}, Christopher J. and {Li}, Qiong and {Austin}, Duncan and {Harvey}, Thomas and {Adams}, Nathan J. and {Duncan}, Kenneth J. and {Trussler}, James and {Ferreira}, Leonardo and {Westcott}, Lewi and et al.},
	title = "{Galaxy mergers in the epoch of reionization {\textendash} I. A JWST study of pair fractions, merger rates, and stellar mass accretion rates at z = 4.5{\textendash}11.5}",
	journal = {\mnras},
	year = 2025,
	month = jun,
	volume = {540},
	number = {1},
	pages = {774-805},
	doi = {10.1093/mnras/staf638},
	archivePrefix = {arXiv},
	eprint = {2407.09472},
	primaryClass = {astro-ph.GA},
	adsurl = {https://ui.adsabs.harvard.edu/abs/2025MNRAS.540..774D}
}

@article{erb2016,
	author = {{Erb}, Dawn K. and {Pettini}, Max and {Steidel}, Charles C. and {Strom}, Allison L. and {Rudie}, Gwen C. and {Trainor}, Ryan F. and {Shapley}, Alice E. and {Reddy}, Naveen A.},
	title = "{A High Fraction of Ly{\ensuremath{\alpha}} Emitters among Galaxies with Extreme Emission Line Ratios at z \raisebox{-0.5ex}\textasciitilde2}",
	journal = {\apj},
	year = 2016,
	month = oct,
	volume = {830},
	number = {1},
	eid = {52},
	pages = {52},
	doi = {10.3847/0004-637X/830/1/52},
	archiveprefix = {arXiv},
	eprint = {1605.04919},
	primaryclass = {astro-ph.GA},
	adsurl = {https://ui.adsabs.harvard.edu/abs/2016ApJ...830...52E}
}

@article{fauchergiguere2020,
	author = {{Faucher-Gigu{\`e}re}, Claude-Andr{\'e}},
	title = "{A cosmic UV/X-ray background model update}",
	journal = {\mnras},
	year = 2020,
	month = apr,
	volume = {493},
	number = {2},
	pages = {1614--1632},
	doi = {10.1093/mnras/staa302},
	archiveprefix = {arXiv},
	eprint = {1903.08657},
	primaryclass = {astro-ph.CO},
	adsurl = {https://ui.adsabs.harvard.edu/abs/2020MNRAS.493.1614F}
}

@article{fletcher2018,
	author = {{Fletcher}, Thomas J. and {Tang}, Mengtao and {Robertson}, Brant E. and {Nakajima}, Kimihiko and {Ellis}, Richard S. and {Stark}, Daniel P. and {Inoue}, Akio},
	title = "{The Lyman Continuum Escape Survey: Ionizing Radiation from [O III]-strong Sources at a Redshift of 3.1}",
	journal = {\apj},
	year = "2019",
	month = "Jun",
	volume = {878},
	number = {2},
	eid = {87},
	pages = {87},
	doi = {10.3847/1538-4357/ab2045},
	archiveprefix = {arXiv},
	eprint = {1806.01741},
	primaryclass = {astro-ph.GA},
	adsurl = {https://ui.adsabs.harvard.edu/abs/2019ApJ...878...87F}
}

@article{flury2022a,
	author = {{Flury}, Sophia R. and {Jaskot}, Anne E. and {Ferguson}, Harry C. and {Worseck}, G{\'a}bor and {Makan}, Kirill and {Chisholm}, John and {Saldana-Lopez}, Alberto and {Schaerer}, Daniel and {McCandliss}, Stephan and {Wang}, Bingjie and {Ford}, N.~M. and {Heckman}, Timothy and {Ji}, Zhiyuan and {Giavalisco}, Mauro and {Amorin}, Ricardo and {Atek}, Hakim and {Blaizot}, Jeremy and {Borthakur}, Sanchayeeta and {Carr}, Cody and {Castellano}, Marco and {Cristiani}, Stefano and {De Barros}, Stephane and {Dickinson}, Mark and {Finkelstein}, Steven L. and {Fleming}, Brian and {Fontanot}, Fabio and {Garel}, Thibault and {Grazian}, Andrea and {Hayes}, Matthew and {Henry}, Alaina and {Mauerhofer}, Valentin and {Micheva}, Genoveva and {Oey}, M.~S. and {Ostlin}, Goran and {Papovich}, Casey and {Pentericci}, Laura and {Ravindranath}, Swara and {Rosdahl}, Joakim and {Rutkowski}, Michael and {Santini}, Paola and {Scarlata}, Claudia and {Teplitz}, Harry and {Thuan}, Trinh and {Trebitsch}, Maxime and {Vanzella}, Eros and {Verhamme}, Anne and {Xu}, Xinfeng},
	title = "{The Low-redshift Lyman Continuum Survey. I. New, Diverse Local Lyman Continuum Emitters}",
	journal = {\apjs},
	year = 2022,
	month = may,
	volume = {260},
	number = {1},
	eid = {1},
	pages = {1},
	doi = {10.3847/1538-4365/ac5331},
	archiveprefix = {arXiv},
	eprint = {2201.11716},
	primaryclass = {astro-ph.GA},
	adsurl = {https://ui.adsabs.harvard.edu/abs/2022ApJS..260....1F}
}

@article{flury2022b,
	author = {{Flury}, Sophia R. and {Jaskot}, Anne E. and {Ferguson}, Harry C. and {Worseck}, G{\'a}bor and {Makan}, Kirill and {Chisholm}, John and {Saldana-Lopez}, Alberto and {Schaerer}, Daniel and {McCandliss}, Stephan R. and {Xu}, Xinfeng and {Wang}, Bingjie and {Oey}, M.~S. and {Ford}, N.~M. and {Heckman}, Timothy and {Ji}, Zhiyuan and {Giavalisco}, Mauro and {Amor{\'\i}n}, Ricardo and {Atek}, Hakim and {Blaizot}, Jeremy and {Borthakur}, Sanchayeeta and {Carr}, Cody and {Castellano}, Marco and {Barros}, Stephane De and {Dickinson}, Mark and {Finkelstein}, Steven L. and {Fleming}, Brian and {Fontanot}, Fabio and {Garel}, Thibault and {Grazian}, Andrea and {Hayes}, Matthew and {Henry}, Alaina and {Mauerhofer}, Valentin and {Micheva}, Genoveva and {Ostlin}, Goran and {Papovich}, Casey and {Pentericci}, Laura and {Ravindranath}, Swara and {Rosdahl}, Joakim and {Rutkowski}, Michael and {Santini}, Paola and {Scarlata}, Claudia and {Teplitz}, Harry and {Thuan}, Trinh and {Trebitsch}, Maxime and {Vanzella}, Eros and {Verhamme}, Anne},
	title = "{The Low-redshift Lyman Continuum Survey. II. New Insights into LyC Diagnostics}",
	journal = {\apj},
	year = 2022,
	month = may,
	volume = {930},
	number = {2},
	eid = {126},
	pages = {126},
	doi = {10.3847/1538-4357/ac61e4},
	archiveprefix = {arXiv},
	eprint = {2203.15649},
	primaryclass = {astro-ph.GA},
	adsurl = {https://ui.adsabs.harvard.edu/abs/2022ApJ...930..126F}
}

@ARTICLE{gardner2023,
	author = {{Gardner}, Jonathan P. and {Mather}, John C. and {Abbott}, Randy and {Abell}, James S. and {Abernathy}, Mark and {Abney}, Faith E. and {Abraham}, John G. and {Abraham}, Roberto and {Abul-Huda}, Yasin M. and {Acton}, Scott and {Adams}, Cynthia K. and {Adams}, Evan and {Adler}, David S. and {Adriaensen}, Maarten and {Aguilar}, Jonathan Albert and {Ahmed}, Mansoor and {Ahmed}, Nasif S. and {Ahmed}, Tanjira and {Albat}, R{\"u}deger and {Albert}, Lo{\"\i}c and {Alberts}, Stacey and {Aldridge}, David and {Allen}, Mary Marsha and {Allen}, Shaune S. and {Altenburg}, Martin and {Altunc}, Serhat and {Alvarez}, Jose Lorenzo and {{\'A}lvarez-M{\'a}rquez}, Javier and {Alves de Oliveira}, Catarina and {Ambrose}, Leslie L. and {Anandakrishnan}, Satya M. and {Andersen}, Gregory C. and {Anderson}, Harry James and {Anderson}, Jay and {Anderson}, Kristen and {Anderson}, Sara M. and {Aprea}, Julio and {Archer}, Benita J. and {Arenberg}, Jonathan W. and {Argyriou}, Ioannis and {Arribas}, Santiago and {Artigau}, {\'E}tienne and {Arvai}, Amanda Rose and {Atcheson}, Paul and {Atkinson}, Charles B. and {Averbukh}, Jesse and {Aymergen}, Cagatay and {Bacinski}, John J. and {Baggett}, Wayne E. and {Bagnasco}, Giorgio and {Baker}, Lynn L. and {Balzano}, Vicki Ann and {Banks}, Kimberly A. and {Baran}, David A. and {Barker}, Elizabeth A. and {Barrett}, Larry K. and {Barringer}, Bruce O. and {Barto}, Allison and {Bast}, William and {Baudoz}, Pierre and {Baum}, Stefi and {Beatty}, Thomas G. and {Beaulieu}, Mathilde and {Bechtold}, Kathryn and {Beck}, Tracy and {Beddard}, Megan M. and {Beichman}, Charles and {Bellagama}, Larry and {Bely}, Pierre and {Berger}, Timothy W. and {Bergeron}, Louis E. and {Bernier}, Antoine-Darveau and {Bertch}, Maria D. and {Beskow}, Charlotte and {Betz}, Laura E. and {Biagetti}, Carl P. and {Birkmann}, Stephan and {Bjorklund}, Kurt F. and {Blackwood}, James D. and {Blazek}, Ronald Paul and {Blossfeld}, Stephen and {Bluth}, Marcel and {Boccaletti}, Anthony and {Boegner}, Jr., Martin E. and {Bohlin}, Ralph C. and {Boia}, John Joseph and {B{\"o}ker}, Torsten and {Bonaventura}, N. and {Bond}, Nicholas A. and {Bosley}, Kari Ann and {Boucarut}, Rene A. and {Bouchet}, Patrice and {Bouwman}, Jeroen and {Bower}, Gary and {Bowers}, Ariel S. and {Bowers}, Charles W. and {Boyce}, Leslye A. and {Boyer}, Christine T. and {Boyer}, Martha L. and {Boyer}, Michael and {Boyer}, Robert and {Bradley}, Larry D. and {Brady}, Gregory R. and {Brandl}, Bernhard R. and {Brannen}, Judith L. and {Breda}, David and {Bremmer}, Harold G. and {Brennan}, David and {Bresnahan}, Pamela A. and {Bright}, Stacey N. and {Broiles}, Brian J. and {Bromenschenkel}, Asa and {Brooks}, Brian H. and {Brooks}, Keira J. and {Brown}, Bob and {Brown}, Bruce and {Brown}, Thomas M. and {Bruce}, Barry W. and {Bryson}, Jonathan G. and {Bujanda}, Edwin D. and {Bullock}, Blake M. and {Bunker}, A.~J. and {Bureo}, Rafael and {Burt}, Irving J. and {Bush}, James Aaron and {Bushouse}, Howard A. and {Bussman}, Marie C. and {Cabaud}, Olivier and {Cale}, Steven and {Calhoon}, Charles D. and {Calvani}, Humberto and {Canipe}, Alicia M. and {Caputo}, Francis M. and {Cara}, Mihai and {Carey}, Larkin and {Case}, Michael Eli and {Cesari}, Thaddeus and {Cetorelli}, Lee D. and {Chance}, Don R. and {Chandler}, Lynn and {Chaney}, Dave and {Chapman}, George N. and {Charlot}, S. and {Chayer}, Pierre and {Cheezum}, Jeffrey I. and {Chen}, Bin and {Chen}, Christine H. and {Cherinka}, Brian and {Chichester}, Sarah C. and {Chilton}, Zachary S. and {Chittiraibalan}, Dharini and {Clampin}, Mark and {Clark}, Charles R. and {Clark}, Kerry W. and {Clark}, Stephanie M. and {Claybrooks}, Edward E. and {Cleveland}, Keith A. and {Cohen}, Andrew L. and {Cohen}, Lester M. and {Col{\'o}n}, Knicole D. and {Coleman}, Benee L. and {Colina}, Luis and {Comber}, Brian J. and {Comeau}, Thomas M. and {Comer}, Thomas and {Conde Reis}, Alain and {Connolly}, Dennis C. and {Conroy}, Kyle E. and {Contos}, Adam R. and {Contreras}, James and {Cook}, Neil J. and {Cooper}, James L. and {Cooper}, Rachel Aviva and {Correia}, Michael F. and {Correnti}, Matteo and {Cossou}, Christophe and {Costanza}, Brian F. and {Coulais}, Alain and {Cox}, Colin R. and {Coyle}, Ray T. and {Cracraft}, Misty M. and {Crew}, Keith A. and {Curtis}, Gary J. and {Cusveller}, Bianca and {Da Costa Maciel}, Cleyciane and {Dailey}, Christopher T. and {Daugeron}, Fr{\'e}d{\'e}ric and {Davidson}, Greg S. and {Davies}, James E. and {Davis}, Katherine Anne and {Davis}, Michael S. and {Day}, Ratna and {de Chambure}, Daniel and {de Jong}, Pauline and {De Marchi}, Guido and {Dean}, Bruce H. and {Decker}, John E. and {Delisa}, Amy S. and {Dell}, Lawrence C. and {Dellagatta}, Gail},
	title = "{The James Webb Space Telescope Mission}",
	journal = {\pasp},
	year = 2023,
	month = jun,
	volume = {135},
	number = {1048},
	eid = {068001},
	pages = {068001},
	doi = {10.1088/1538-3873/acd1b5},
	archivePrefix = {arXiv},
	eprint = {2304.04869},
	primaryClass = {astro-ph.IM},
	adsurl = {https://ui.adsabs.harvard.edu/abs/2023PASP..135f8001G}
}

@article{gazagnes2018,
	author = {{Gazagnes}, S. and {Chisholm}, J. and {Schaerer}, D. and {Verhamme}, A. and {Rigby}, J.~R. and {Bayliss}, M.},
	title = "{Neutral gas properties of Lyman continuum emitting galaxies: Column densities and covering fractions from UV absorption lines}",
	journal = {\aap},
	year = 2018,
	month = aug,
	volume = {616},
	eid = {A29},
	pages = {A29},
	doi = {10.1051/0004-6361/201832759},
	archiveprefix = {arXiv},
	eprint = {1802.06378},
	primaryclass = {astro-ph.GA},
	adsurl = {https://ui.adsabs.harvard.edu/abs/2018A\&A...616A..29G}
}

@ARTICLE{glazebrook2013,
	author = {{Glazebrook}, Karl},
	title = "{The Dawes Review 1: Kinematic Studies of Star-Forming Galaxies Across Cosmic Time}",
	journal = {\pasa},
	year = 2013,
	month = nov,
	volume = {30},
	eid = {e056},
	pages = {e056},
	doi = {10.1017/pasa.2013.34},
	archivePrefix = {arXiv},
	eprint = {1305.2469},
	primaryClass = {astro-ph.CO},
	adsurl = {https://ui.adsabs.harvard.edu/abs/2013PASA...30...56G}
}

@article{gronke2016letter,
	author = {{Gronke}, M. and {Dijkstra}, M. and {McCourt}, M. and {Oh}, S.~P.},
	title = "{From Mirrors to Windows: Lyman-alpha Radiative Transfer in a Very Clumpy Medium}",
	journal = {\apjl},
	archiveprefix = "arXiv",
	eprint = {1611.01161},
	year = 2016,
	month = dec,
	volume = 833,
	eid = {L26},
	pages = {L26},
	doi = {10.3847/2041-8213/833/2/L26},
	adsurl = {http://adsabs.harvard.edu/abs/2016ApJ...833L..26G}
}

@article{haardt2012,
	author = {{Haardt}, Francesco and {Madau}, Piero},
	title = "{Radiative Transfer in a Clumpy Universe. IV. New Synthesis Models of the Cosmic UV/X-Ray Background}",
	journal = {\apj},
	year = 2012,
	month = feb,
	volume = {746},
	eid = {125},
	pages = {125},
	doi = {10.1088/0004-637X/746/2/125},
	archiveprefix = {arXiv},
	eprint = {1105.2039},
	primaryclass = {astro-ph.CO},
	adsurl = {https://ui.adsabs.harvard.edu/\#abs/2012ApJ...746..125H}
}

@article{hashimoto2018nature,
	author = {{Hashimoto}, Takuya and {Laporte}, Nicolas and {Mawatari}, Ken and {Ellis}, Richard S. and {Inoue}, Akio K. and {Zackrisson}, Erik and {Roberts-Borsani}, Guido and {Zheng}, Wei and {Tamura}, Yoichi and {Bauer}, Franz E. and {Fletcher}, Thomas and {Harikane}, Yuichi and {Hatsukade}, Bunyo and {Hayatsu}, Natsuki H. and {Matsuda}, Yuichi and {Matsuo}, Hiroshi and {Okamoto}, Takashi and {Ouchi}, Masami and {Pell{\'o}}, Roser and {Rydberg}, Claes-Erik and {Shimizu}, Ikkoh and {Taniguchi}, Yoshiaki and {Umehata}, Hideki and {Yoshida}, Naoki},
	title = "{The onset of star formation 250 million years after the Big Bang}",
	journal = {\nat},
	year = 2018,
	month = may,
	volume = {557},
	number = {7705},
	pages = {392--395},
	doi = {10.1038/s41586-018-0117-z},
	archiveprefix = {arXiv},
	eprint = {1805.05966},
	primaryclass = {astro-ph.GA},
	adsurl = {https://ui.adsabs.harvard.edu/abs/2018Natur.557..392H}
}

@ARTICLE{hayes2025,
	author = {{Hayes}, Matthew J. and {Saldana-Lopez}, Alberto and {Citro}, Annalisa and {James}, Bethan L. and {Mingozzi}, Matilde and {Scarlata}, Claudia and {Martinez}, Zorayda and {Berg}, Danielle A.},
	title = "{On the Average Ultraviolet Emission-line Spectra of High-redshift Galaxies: Hot and Cold, Carbon-poor, Nitrogen Modest, and Oozing Ionizing Photons}",
	journal = {\apj},
	year = 2025,
	month = mar,
	volume = {982},
	number = {1},
	eid = {14},
	pages = {14},
	doi = {10.3847/1538-4357/adaea1},
	archivePrefix = {arXiv},
	eprint = {2411.09262},
	primaryClass = {astro-ph.GA},
	adsurl = {https://ui.adsabs.harvard.edu/abs/2025ApJ...982...14H}
}

@article{heckman2011,
	author = {{Heckman}, T.~M. and {Borthakur}, S. and {Overzier}, R. and {Kauffmann}, G. and {Basu-Zych}, A. and {Leitherer}, C. and {Sembach}, K. and {Martin}, D.~C. and {Rich}, R.~M. and {Schiminovich}, D. and {Seibert}, M.},
	title = "{Extreme Feedback and the Epoch of Reionization: Clues in the Local Universe}",
	journal = {\apj},
	archiveprefix = "arXiv",
	eprint = {1101.4219},
	primaryclass = "astro-ph.CO",
	year = 2011,
	month = mar,
	volume = 730,
	eid = {5},
	pages = {5},
	doi = {10.1088/0004-637X/730/1/5},
	adsurl = {http://adsabs.harvard.edu/abs/2011ApJ...730....5H}
}

@article{herenz2016,
	author = {{Herenz}, Edmund Christian and {Gruyters}, Pieter and {Orlitova}, Ivana and {Hayes}, Matthew and {{\"O}stlin}, G{\"o}ran and {Cannon}, John M. and {Roth}, Martin M. and {Bik}, Arjan and {Pardy}, Stephen and {Ot{\'\i}-Floranes}, H{\'e}ctor and {Mas-Hesse}, J. Miguel and {Adamo}, Angela and {Atek}, Hakim and {Duval}, Florent and {Guaita}, Lucia and {Kunth}, Daniel and {Laursen}, Peter and {Melinder}, Jens and {Puschnig}, Johannes and {Rivera-Thorsen}, Th{\o}ger E. and {Schaerer}, Daniel and {Verhamme}, Anne},
	title = "{The Lyman alpha reference sample. VII. Spatially resolved H{\ensuremath{\alpha}} kinematics}",
	journal = {\aap},
	year = 2016,
	month = mar,
	volume = {587},
	eid = {A78},
	pages = {A78},
	doi = {10.1051/0004-6361/201527373},
	archiveprefix = {arXiv},
	eprint = {1511.05406},
	primaryclass = {astro-ph.GA},
	adsurl = {https://ui.adsabs.harvard.edu/abs/2016A\&A...587A..78H}
}

@ARTICLE{herenz2025,
	author = {{Herenz}, E.~C. and {Schaible}, A. and {Laursen}, P. and {Runnholm}, A. and {Melinder}, J. and {Le Reste}, A. and {Hayes}, M.~J. and {{\"O}stlin}, G. and {Cannon}, J. and {Micheva}, G. and {Roth}, M. and {Saha}, K.},
	title = "{The Lyman alpha reference sample: XV. Relating ionised gas kinematics with Lyman-{\ensuremath{\alpha}} observables}",
	journal = {\aap},
	year = 2025,
	month = jan,
	volume = {693},
	eid = {A252},
	pages = {A252},
	doi = {10.1051/0004-6361/202451012},
	archivePrefix = {arXiv},
	eprint = {2406.03956},
	primaryClass = {astro-ph.GA},
	adsurl = {https://ui.adsabs.harvard.edu/abs/2025A&A...693A.252H}
}

@ARTICLE{hutchison23_sigmaclip,
	author = {{Hutchison}, Taylor A. and {Welch}, Brian D. and {Rigby}, Jane R. and {Olivier}, Grace M. and {Birkin}, Jack E. and {Phadke}, Kedar A. and {Khullar}, Gourav and {Rauscher}, Bernard J. and {Sharon}, Keren and {Aravena}, Manuel and {Bayliss}, Matthew B. and {Elicker}, Lauren A. and {Kim}, Seonwoo and {Solimano}, Manuel and {Vieira}, Joaquin D. and {Vizgan}, David},
	title = "{TEMPLATES: A Robust Outlier Rejection Method for JWST/NIRSpec Integral Field Spectroscopy}",
	journal = {arXiv e-prints},
	year = 2023,
	month = dec,
	eid = {arXiv:2312.12518},
	pages = {arXiv:2312.12518},
	doi = {10.48550/arXiv.2312.12518},
	archivePrefix = {arXiv},
	eprint = {2312.12518},
	primaryClass = {astro-ph.IM},
	adsurl = {https://ui.adsabs.harvard.edu/abs/2023arXiv231212518H}
}

@article{inoue2014,
	author = {{Inoue}, Akio K. and {Shimizu}, Ikkoh and {Iwata}, Ikuru and {Tanaka}, Masayuki},
	title = "{An updated analytic model for attenuation by the intergalactic medium}",
	journal = {\mnras},
	year = 2014,
	month = aug,
	volume = {442},
	number = {2},
	pages = {1805--1820},
	doi = {10.1093/mnras/stu936},
	archiveprefix = {arXiv},
	eprint = {1402.0677},
	primaryclass = {astro-ph.CO},
	adsurl = {https://ui.adsabs.harvard.edu/abs/2014MNRAS.442.1805I}
}

@article{ipython2007,
	author = {P\'erez, Fernando and Granger, Brian E.},
	title = {{IP}ython: a System for Interactive Scientific Computing},
	journal = {Computing in Science and Engineering},
	volume = {9},
	number = {3},
	pages = {21--29},
	month = may,
	year = 2007,
	url = "http://ipython.org",
	issn = "1521-9615",
	doi = {10.1109/MCSE.2007.53},
	publisher = {IEEE Computer Society}
}

@article{james2014,
	author = {{James}, B.~L. and {Aloisi}, A. and {Heckman}, T. and {Sohn}, S.~T. and {Wolfe}, M.~A.},
	title = "{Investigating Nearby Star-forming Galaxies in the Ultraviolet with HST/COS Spectroscopy. I. Spectral Analysis and Interstellar Abundance Determinations}",
	journal = {\apj},
	archiveprefix = "arXiv",
	eprint = {1408.4420},
	year = 2014,
	month = nov,
	volume = 795,
	eid = {109},
	pages = {109},
	doi = {10.1088/0004-637X/795/2/109},
	adsurl = {http://adsabs.harvard.edu/abs/2014ApJ...795..109J}
}

@article{jaskot2014,
	author = {{Jaskot}, A.~E. and {Oey}, M.~S.},
	title = "{Linking Ly{$\alpha$} and Low-ionization Transitions at Low Optical Depth}",
	journal = {\apjl},
	archiveprefix = "arXiv",
	eprint = {1406.4413},
	year = 2014,
	month = aug,
	volume = 791,
	eid = {L19},
	pages = {L19},
	doi = {10.1088/2041-8205/791/2/L19},
	adsurl = {http://adsabs.harvard.edu/abs/2014ApJ...791L..19J}
}

@article{jaskot2019,
	author = {{Jaskot}, Anne E. and {Dowd}, Tara and {Oey}, M.~S. and {Scarlata}, Claudia and {McKinney}, Jed},
	title = "{New Insights on Ly{\ensuremath{\alpha}} and Lyman Continuum Radiative Transfer in the Greenest Peas}",
	journal = {\apj},
	year = 2019,
	month = nov,
	volume = {885},
	number = {1},
	eid = {96},
	pages = {96},
	doi = {10.3847/1538-4357/ab3d3b},
	archiveprefix = {arXiv},
	eprint = {1908.09763},
	primaryclass = {astro-ph.GA},
	adsurl = {https://ui.adsabs.harvard.edu/abs/2019ApJ...885...96J}
}

@ARTICLE{jaskot2024a,
	author = {{Jaskot}, Anne E. and {Silveyra}, Anneliese C. and {Plantinga}, Anna and {Flury}, Sophia R. and {Hayes}, Matthew and {Chisholm}, John and {Heckman}, Timothy and {Pentericci}, Laura and {Schaerer}, Daniel and {Trebitsch}, Maxime and {Verhamme}, Anne and {Carr}, Cody and {Ferguson}, Henry C. and {Ji}, Zhiyuan and {Giavalisco}, Mauro and {Henry}, Alaina and {Marques-Chaves}, Rui and {{\"O}stlin}, G{\"o}ran and {Saldana-Lopez}, Alberto and {Scarlata}, Claudia and {Worseck}, G{\'a}bor and {Xu}, Xinfeng},
	title = "{Multivariate Predictors of Lyman Continuum Escape. II. Predicting Lyman Continuum Escape Fractions for High-redshift Galaxies}",
	journal = {\apj},
	year = 2024,
	month = oct,
	volume = {973},
	number = {2},
	eid = {111},
	pages = {111},
	doi = {10.3847/1538-4357/ad5557},
	archivePrefix = {arXiv},
	eprint = {2406.10179},
	primaryClass = {astro-ph.GA},
	adsurl = {https://ui.adsabs.harvard.edu/abs/2024ApJ...973..111J}
}

@ARTICLE{jaskot2024b,
	author = {{Jaskot}, Anne E. and {Silveyra}, Anneliese C. and {Plantinga}, Anna and {Flury}, Sophia R. and {Hayes}, Matthew and {Chisholm}, John and {Heckman}, Timothy and {Pentericci}, Laura and {Schaerer}, Daniel and {Trebitsch}, Maxime and {Verhamme}, Anne and {Carr}, Cody and {Ferguson}, Henry C. and {Ji}, Zhiyuan and {Giavalisco}, Mauro and {Henry}, Alaina and {Marques-Chaves}, Rui and {{\"O}stlin}, G{\"o}ran and {Saldana-Lopez}, Alberto and {Scarlata}, Claudia and {Worseck}, G{\'a}bor and {Xu}, Xinfeng},
	title = "{Multivariate Predictors of Lyman Continuum Escape. II. Predicting Lyman Continuum Escape Fractions for High-redshift Galaxies}",
	journal = {\apj},
	year = 2024,
	month = oct,
	volume = {973},
	number = {2},
	eid = {111},
	pages = {111},
	doi = {10.3847/1538-4357/ad5557},
	archivePrefix = {arXiv},
	eprint = {2406.10179},
	primaryClass = {astro-ph.GA},
	adsurl = {https://ui.adsabs.harvard.edu/abs/2024ApJ...973..111J}
}

@article{ji2020,
	author = {{Ji}, Zhiyuan and {Giavalisco}, Mauro and {Vanzella}, Eros and {Siana}, Brian and {Pentericci}, Laura and {Jaskot}, Anne and {Liu}, Teng and {Nonino}, Mario and {Ferguson}, Henry C. and {Castellano}, Marco and {Mannucci}, Filippo and {Schaerer}, Daniel and {Fynbo}, Johan Peter Uldall and {Papovich}, Casey and {Carnall}, Adam C. and {Amorin}, Ricardo and {Simons}, Raymond C. and {Hathi}, Nimish and {Cullen}, Fergus and {McLeod}, Derek},
	title = "{HST Imaging of the Ionizing Radiation from a Star-forming Galaxy at z = 3.794}",
	journal = {\apj},
	year = 2020,
	month = jan,
	volume = {888},
	number = {2},
	eid = {109},
	pages = {109},
	doi = {10.3847/1538-4357/ab5fdc},
	archiveprefix = {arXiv},
	eprint = {1908.00556},
	primaryclass = {astro-ph.GA},
	adsurl = {https://ui.adsabs.harvard.edu/abs/2020ApJ...888..109J}
}

@ARTICLE{ji2025arxiv,
	author = {{Ji}, Xihan and {Belokurov}, Vasily and {Maiolino}, Roberto and {Monty}, Stephanie and {Isobe}, Yuki and {Kravtsov}, Andrey and {McClymont}, William and {{\"U}bler}, Hannah},
	title = "{Connecting JWST discovered N/O-enhanced galaxies to globular clusters: Evidence from chemical imprints}",
	journal = {arXiv e-prints},
	year = 2025,
	month = may,
	eid = {arXiv:2505.12505},
	pages = {arXiv:2505.12505},
	doi = {10.48550/arXiv.2505.12505},
	archivePrefix = {arXiv},
	eprint = {2505.12505},
	primaryClass = {astro-ph.GA},
	adsurl = {https://ui.adsabs.harvard.edu/abs/2025arXiv250512505J}
}

@article{jones2013,
	author = {{Jones}, T.~A. and {Ellis}, R.~S. and {Schenker}, M.~A. and {Stark}, D.~P.},
	title = "{Keck Spectroscopy of Gravitationally Lensed z \~{}= 4 Galaxies: Improved Constraints on the Escape Fraction of Ionizing Photons}",
	journal = {\apj},
	archiveprefix = "arXiv",
	eprint = {1304.7015},
	primaryclass = "astro-ph.CO",
	year = 2013,
	month = dec,
	volume = 779,
	eid = {52},
	pages = {52},
	doi = {10.1088/0004-637X/779/1/52},
	adsurl = {http://adsabs.harvard.edu/abs/2013ApJ...779...52J}
}

@ARTICLE{jupyter2021,
	author = {Granger, Brian E. and Pérez, Fernando},
	journal = {Computing in Science \& Engineering},
	title = {Jupyter: Thinking and Storytelling With Code and Data},
	year = {2021},
	volume = {23},
	number = {2},
	pages = {7-14},
	doi = {10.1109/MCSE.2021.3059263}
}

@article{kakiichi2019,
	author = {{Kakiichi}, Koki and {Gronke}, Max},
	title = "{Lyman Radiation Hydrodynamics of Turbulent H II Regions in Molecular Clouds: A Physical Origin of LyC Leakage and the Associated Ly$\alpha$ Spectra}",
	journal = {arXiv e-prints},
	year = 2019,
	month = may,
	eid = {arXiv:1905.02480},
	pages = {arXiv:1905.02480},
	archiveprefix = {arXiv},
	eprint = {1905.02480},
	primaryclass = {astro-ph.GA},
	adsurl = {https://ui.adsabs.harvard.edu/abs/2019arXiv190502480K}
}

@article{kakiichi2021,
	author = {{Kakiichi}, Koki and {Gronke}, Max},
	title = "{Radiation Hydrodynamics of Turbulent H II Regions in Molecular Clouds: A Physical Origin of LyC Leakage and the Associated Ly{\ensuremath{\alpha}} Spectra}",
	journal = {\apj},
	year = 2021,
	month = feb,
	volume = {908},
	number = {1},
	eid = {30},
	pages = {30},
	doi = {10.3847/1538-4357/abc2d9},
	adsurl = {https://ui.adsabs.harvard.edu/abs/2021ApJ...908...30K}
}

@article{kauffmann2003,
	author = {{Kauffmann}, G. and {Heckman}, T.~M. and {Tremonti}, C. and {Brinchmann}, J. and {Charlot}, S. and {White}, S.~D.~M. and {Ridgway}, S.~E. and {Brinkmann}, J. and {Fukugita}, M. and {Hall}, P.~B. and {Ivezi{\'c}}, {\v Z}. and {Richards}, G.~T. and {Schneider}, D.~P.},
	title = "{The host galaxies of active galactic nuclei}",
	journal = {\mnras},
	eprint = {astro-ph/0304239},
	year = 2003,
	month = dec,
	volume = 346,
	pages = {1055--1077},
	doi = {10.1111/j.1365-2966.2003.07154.x},
	adsurl = {http://adsabs.harvard.edu/abs/2003MNRAS.346.1055K}
}

@ARTICLE{kerutt2024,
	author = {{Kerutt}, J. and {Oesch}, P.~A. and {Wisotzki}, L. and {Verhamme}, A. and {Atek}, H. and {Herenz}, E.~C. and {Illingworth}, G.~D. and {Kusakabe}, H. and {Matthee}, J. and {Mauerhofer}, V. and {Montes}, M. and {Naidu}, R.~P. and {Nelson}, E. and {Reddy}, N. and {Schaye}, J. and {Simmonds}, C. and {Urrutia}, T. and {Vitte}, E.},
	title = "{Lyman continuum leaker candidates at z {\ensuremath{\sim}} 3-4 in the HDUV based on a spectroscopic sample of MUSE LAEs}",
	journal = {\aap},
	year = 2024,
	month = apr,
	volume = {684},
	eid = {A42},
	pages = {A42},
	doi = {10.1051/0004-6361/202346656},
	archivePrefix = {arXiv},
	eprint = {2312.08791},
	primaryClass = {astro-ph.GA},
	adsurl = {https://ui.adsabs.harvard.edu/abs/2024A&A...684A..42K}
}

@article{kewley2001,
	author = {{Kewley}, L.~J. and {Dopita}, M.~A. and {Sutherland}, R.~S. and {Heisler}, C.~A. and {Trevena}, J.},
	title = "{Theoretical Modeling of Starburst Galaxies}",
	journal = {\apj},
	eprint = {astro-ph/0106324},
	year = 2001,
	month = jul,
	volume = 556,
	pages = {121--140},
	doi = {10.1086/321545},
	adsurl = {http://adsabs.harvard.edu/abs/2001ApJ...556..121K}
}

@article{kewley2006,
	author = {{Kewley}, L.~J. and {Groves}, B. and {Kauffmann}, G. and {Heckman}, T.},
	title = "{The host galaxies and classification of active galactic nuclei}",
	journal = {\mnras},
	eprint = {astro-ph/0605681},
	year = 2006,
	month = nov,
	volume = 372,
	pages = {961--976},
	doi = {10.1111/j.1365-2966.2006.10859.x},
	adsurl = {http://adsabs.harvard.edu/abs/2006MNRAS.372..961K}
}

@ARTICLE{kim2020,
	author = {{Kim}, Keunho and {Malhotra}, Sangeeta and {Rhoads}, James E. and {Butler}, Nathaniel R. and {Yang}, Huan},
	title = "{The Importance of Star Formation Intensity in Ly{\ensuremath{\alpha}} Escape from Green Pea Galaxies and Lyman Break Galaxy Analogs}",
	journal = {\apj},
	year = 2020,
	month = apr,
	volume = {893},
	number = {2},
	eid = {134},
	pages = {134},
	doi = {10.3847/1538-4357/ab7895},
	archivePrefix = {arXiv},
	eprint = {2002.08961},
	primaryClass = {astro-ph.GA},
	adsurl = {https://ui.adsabs.harvard.edu/abs/2020ApJ...893..134K}
}

@ARTICLE{kim2023,
	author = {{Kim}, Keunho J. and {Bayliss}, Matthew B. and {Rigby}, Jane R. and {Gladders}, Michael D. and {Chisholm}, John and {Sharon}, Keren and {Dahle}, H{\r{a}}kon and {Rivera-Thorsen}, T. Emil and {Florian}, Michael K. and {Khullar}, Gourav and {Mahler}, Guillaume and {Mainali}, Ramesh and {Napier}, Kate A. and {Navarre}, Alexander and {Owens}, M. Riley and {Roberson}, Joshua},
	title = "{Small Region, Big Impact: Highly Anisotropic Lyman-continuum Escape from a Compact Starburst Region with Extreme Physical Properties}",
	journal = {\apjl},
	year = 2023,
	month = sep,
	volume = {955},
	number = {1},
	eid = {L17},
	pages = {L17},
	doi = {10.3847/2041-8213/acf0c5},
	archivePrefix = {arXiv},
	eprint = {2305.13405},
	primaryClass = {astro-ph.GA},
	adsurl = {https://ui.adsabs.harvard.edu/abs/2023ApJ...955L..17K}
}

@ARTICLE{komarova2024,
	author = {{Komarova}, Lena and {Oey}, M.~S. and {Hernandez}, Svea and {Adamo}, Angela and {Sirressi}, Mattia and {Leitherer}, Claus and {Mas-Hesse}, J.~M. and {{\"O}stlin}, G{\"o}ran and {Hodges-Kluck}, Edmund and {Bik}, Arjan and {Hayes}, Matthew J. and {Jaskot}, Anne E. and {Kunth}, Daniel and {Laursen}, Peter and {Melinder}, Jens and {Rivera-Thorsen}, T. Emil},
	title = "{Haro 11: The Spatially Resolved Lyman Continuum Sources}",
	journal = {\apj},
	year = 2024,
	month = jun,
	volume = {967},
	number = {2},
	eid = {117},
	pages = {117},
	doi = {10.3847/1538-4357/ad3962},
	archivePrefix = {arXiv},
	eprint = {2404.01435},
	primaryClass = {astro-ph.GA},
	adsurl = {https://ui.adsabs.harvard.edu/abs/2024ApJ...967..117K}
}

@ARTICLE{kostyuk2024,
	author = {{Kostyuk}, Ivan and {Ciardi}, Benedetta},
	title = "{Influence of mergers on LyC escape of high redshift galaxies}",
	journal = {arXiv e-prints},
	year = 2024,
	month = dec,
	eid = {arXiv:2412.04348},
	pages = {arXiv:2412.04348},
	doi = {10.48550/arXiv.2412.04348},
	archivePrefix = {arXiv},
	eprint = {2412.04348},
	primaryClass = {astro-ph.GA},
	adsurl = {https://ui.adsabs.harvard.edu/abs/2024arXiv241204348K}
}

@ARTICLE{law23_3ddrizzle,
	author = {{Law}, David R. and {E. Morrison}, Jane and {Argyriou}, Ioannis and {Patapis}, Polychronis and {{\'A}lvarez-M{\'a}rquez}, J. and {Labiano}, Alvaro and {Vandenbussche}, Bart},
	title = "{A 3D Drizzle Algorithm for JWST and Practical Application to the MIRI Medium Resolution Spectrometer}",
	journal = {\aj},
	year = 2023,
	month = aug,
	volume = {166},
	number = {2},
	eid = {45},
	pages = {45},
	doi = {10.3847/1538-3881/acdddc},
	archivePrefix = {arXiv},
	eprint = {2306.05520},
	primaryClass = {astro-ph.IM},
	adsurl = {https://ui.adsabs.harvard.edu/abs/2023AJ....166...45L}
}

@article{leitet2011,
	author = {{Leitet}, E. and {Bergvall}, N. and {Piskunov}, N. and {Andersson}, B.-G.},
	title = "{Analyzing low signal-to-noise FUSE spectra. Confirmation of Lyman continuum escape from Haro 11}",
	journal = {\aap},
	archiveprefix = "arXiv",
	eprint = {1106.1178},
	year = 2011,
	month = aug,
	volume = 532,
	eid = {A107},
	pages = {A107},
	doi = {10.1051/0004-6361/201015654},
	adsurl = {http://adsabs.harvard.edu/abs/2011A\%26A...532A.107L}
}

@article{leitherer2016,
	author = {{Leitherer}, C. and {Hernandez}, S. and {Lee}, J.~C. and {Oey}, M.~S.},
	title = "{Direct Detection of Lyman Continuum Escape from Local Starburst Galaxies with the Cosmic Origins Spectrograph}",
	journal = {\apj},
	archiveprefix = "arXiv",
	eprint = {1603.06779},
	year = 2016,
	month = may,
	volume = 823,
	eid = {64},
	pages = {64},
	doi = {10.3847/0004-637X/823/1/64},
	adsurl = {http://adsabs.harvard.edu/abs/2016ApJ...823...64L}
}

@ARTICLE{lereste2024,
	author = {{Le Reste}, Alexandra and {Cannon}, John M. and {Hayes}, Matthew J. and {Inoue}, John L. and {Kepley}, Amanda A. and {Melinder}, Jens and {Menacho}, Veronica and {Adamo}, Angela and {Bik}, Arjan and {Ejdetj{\"a}rn}, Timmy and {J{\'o}zsa}, Gyula I.~G. and {{\"O}stlin}, G{\"o}ran and {Taft}, Sarah H.},
	title = "{Tidally offset neutral gas in Lyman continuum emitting galaxy Haro 11}",
	journal = {\mnras},
	year = 2024,
	month = feb,
	volume = {528},
	number = {1},
	pages = {757-770},
	doi = {10.1093/mnras/stad3910},
	archivePrefix = {arXiv},
	eprint = {2301.02676},
	primaryClass = {astro-ph.GA},
	adsurl = {https://ui.adsabs.harvard.edu/abs/2024MNRAS.528..757L}
}

@ARTICLE{liu2023,
	author = {{Liu}, Yuchen and {Jiang}, Linhua and {Windhorst}, Rogier A. and {Guo}, Yucheng and {Zheng}, Zhen-Ya},
	title = "{Lyman Continuum Emission from Spectroscopically Confirmed Ly{\ensuremath{\alpha}} Emitters at z   3.1}",
	journal = {\apj},
	year = 2023,
	month = nov,
	volume = {958},
	number = {1},
	eid = {22},
	pages = {22},
	doi = {10.3847/1538-4357/acf9fa},
	archivePrefix = {arXiv},
	eprint = {2310.07283},
	primaryClass = {astro-ph.GA},
	adsurl = {https://ui.adsabs.harvard.edu/abs/2023ApJ...958...22L}
}

@misc{lmfit2014,
	author = {Newville, Matthew and Stensitzki, Till and Allen, Daniel B. and Ingargiola, Antonino},
	title = {{LMFIT: Non-Linear Least-Square Minimization and Curve-Fitting for Python\textparagraph{}}},
	month = sep,
	year = 2014,
	doi = {10.5281/zenodo.11813},
	url = {http://dx.doi.org/10.5281/zenodo.11813}
}

@software{lmfit_zenodo,
	author = {Newville, Matthew and
                  Otten, Renee and
                  Nelson, Andrew and
                  Stensitzki, Till and
                  Ingargiola, Antonino and
                  Allan, Daniel and
                  Fox, Austin and
                  Carter, Faustin and
                  Rawlik, Michal},
	title = {LMFIT: Non-Linear Least-Squares Minimization and
                   Curve-Fitting for Python
                  },
	month = jul,
	year = 2025,
	publisher = {Zenodo},
	version = {1.3.4},
	doi = {10.5281/zenodo.16175987},
	url = {https://doi.org/10.5281/zenodo.16175987},
	swhid = {swh:1:dir:76742b0e41b1d2bff5a3716dd2376531f3a21db8
                   ;origin=https://doi.org/10.5281/zenodo.598352;visi
                   t=swh:1:snp:89f98ce93be53a573d85de0145f9174c827b7e
                   92;anchor=swh:1:rel:9528e5133d2d78c4036335f023b212
                   487cf1b632;path=lmfit-lmfit-py-0566445
                  }
}

@ARTICLE{mainali2022,
	author = {{Mainali}, Ramesh and {Rigby}, Jane R. and {Chisholm}, John and {Bayliss}, Matthew and {Bordoloi}, Rongmon and {Gladders}, Michael D. and {Rivera-Thorsen}, T. Emil and {Dahle}, H{\r{a}}kon and {Sharon}, Keren and {Florian}, Michael and {Berg}, Danielle A. and {Sharma}, Soniya and {Owens}, M. Riley and {Kjellgren}, Karin and {Kim}, Keunho J. and {Wayne}, Julia},
	title = "{The Connection Between Galactic Outflows and the Escape of Ionizing Photons}",
	journal = {\apj},
	year = 2022,
	month = dec,
	volume = {940},
	number = {2},
	eid = {160},
	pages = {160},
	doi = {10.3847/1538-4357/ac9cd6},
	archivePrefix = {arXiv},
	eprint = {2210.11575},
	primaryClass = {astro-ph.GA},
	adsurl = {https://ui.adsabs.harvard.edu/abs/2022ApJ...940..160M}
}

@article{malkan2021b,
	author = {{Malkan}, Matthew A. and {Malkan}, Brian K.},
	title = "{Lyman Continuum Emission Escaping from Luminous Green Pea Galaxies at z = 0.5}",
	journal = {\apj},
	year = 2021,
	month = mar,
	volume = {909},
	number = {1},
	eid = {92},
	pages = {92},
	doi = {10.3847/1538-4357/abd84e},
	adsurl = {https://ui.adsabs.harvard.edu/abs/2021ApJ...909...92M}
}

@article{marqueschaves2021,
	author = {{Marques-Chaves}, R. and {Schaerer}, D. and {{\'A}lvarez-M{\'a}rquez}, J. and {Colina}, L. and {Dessauges-Zavadsky}, M. and {P{\'e}rez-Fournon}, I. and {Saldana-Lopez}, A. and {Verhamme}, A.},
	title = "{The UV-brightest Lyman continuum emitting star-forming galaxy}",
	journal = {\mnras},
	year = 2021,
	month = oct,
	volume = {507},
	number = {1},
	pages = {524--538},
	doi = {10.1093/mnras/stab2187},
	archiveprefix = {arXiv},
	eprint = {2107.12313},
	primaryclass = {astro-ph.GA},
	adsurl = {https://ui.adsabs.harvard.edu/abs/2021MNRAS.507..524M}
}

@article{marqueschaves2022,
	author = {{Marques-Chaves}, R. and {Schaerer}, D. and {Alvarez-Marquez}, J. and {Verhamme}, A. and {Ceverino}, D. and {Chisholm}, J. and {Colina}, L. and {Dessauges-Zavadsky}, M. and {Perez-Fournon}, I. and {Saldana-Lopez}, A. and {Upadhyaya}, A. and {Vanzella}, E.},
	title = "{An extreme blue nugget, UV-bright starburst at z=3.613 with ninety per cent of Lyman continuum photon escape}",
	journal = {arXiv e-prints},
	year = 2022,
	month = oct,
	eid = {arXiv:2210.02392},
	pages = {arXiv:2210.02392},
	archiveprefix = {arXiv},
	eprint = {2210.02392},
	primaryclass = {astro-ph.GA},
	adsurl = {https://ui.adsabs.harvard.edu/abs/2022arXiv221002392M}
}

@article{matplotlib,
	author = {Hunter, J.D.},
	journal = {Computing in Science Engineering},
	title = {Matplotlib: A 2D Graphics Environment},
	year = {2007},
	month = {May},
	volume = {9},
	number = {3},
	pages = {90--95},
	doi = {10.1109/MCSE.2007.55},
	issn = {1521-9615}
}

@ARTICLE{mestric2023,
	author = {{Me{\v{s}}tri{\'c}}, U. and {Vanzella}, E. and {Upadhyaya}, A. and {Martins}, F. and {Marques-Chaves}, R. and {Schaerer}, D. and {Guibert}, J. and {Zanella}, A. and {Grillo}, C. and {Rosati}, P. and {Calura}, F. and {Caminha}, G.~B. and {Bolamperti}, A. and {Meneghetti}, M. and {Bergamini}, P. and {Mercurio}, A. and {Nonino}, M. and {Pascale}, R.},
	title = "{Clues on the presence and segregation of very massive stars in the Sunburst Lyman-continuum cluster at z = 2.37}",
	journal = {\aap},
	year = 2023,
	month = may,
	volume = {673},
	eid = {A50},
	pages = {A50},
	doi = {10.1051/0004-6361/202345895},
	archivePrefix = {arXiv},
	eprint = {2301.04672},
	primaryClass = {astro-ph.GA},
	adsurl = {https://ui.adsabs.harvard.edu/abs/2023A&A...673A..50M}
}

@article{mostardi2015,
	author = {{Mostardi}, R.~E. and {Shapley}, A.~E. and {Steidel}, C.~C. and {Trainor}, R.~F. and {Reddy}, N.~A. and {Siana}, B.},
	title = "{A High-Resolution Hubble Space Telescope Study of Apparent Lyman Continuum Leakers at z\raisebox{-0.5ex}\textasciitilde3}",
	journal = {\apj},
	year = 2015,
	month = sep,
	volume = {810},
	number = {2},
	eid = {107},
	pages = {107},
	doi = {10.1088/0004-637X/810/2/107},
	archiveprefix = {arXiv},
	eprint = {1506.08201},
	primaryclass = {astro-ph.GA},
	adsurl = {https://ui.adsabs.harvard.edu/abs/2015ApJ...810..107M}
}

@article{naidu2020a,
	author = {{Naidu}, Rohan P. and {Tacchella}, Sandro and {Mason}, Charlotte A. and {Bose}, Sownak and {Oesch}, Pascal A. and {Conroy}, Charlie},
	title = "{Rapid Reionization by the Oligarchs: The Case for Massive, UV-bright, Star-forming Galaxies with High Escape Fractions}",
	journal = {\apj},
	year = 2020,
	month = apr,
	volume = {892},
	number = {2},
	eid = {109},
	pages = {109},
	doi = {10.3847/1538-4357/ab7cc9},
	archiveprefix = {arXiv},
	eprint = {1907.13130},
	primaryclass = {astro-ph.GA},
	adsurl = {https://ui.adsabs.harvard.edu/abs/2020ApJ...892..109N}
}

@article{nakajima2020,
	author = {{Nakajima}, Kimihiko and {Ellis}, Richard S. and {Robertson}, Brant E. and {Tang}, Mengtao and {Stark}, Daniel P.},
	title = "{The Lyman Continuum Escape Survey. II. Ionizing Radiation as a Function of the [O III]/[O II] Line Ratio}",
	journal = {\apj},
	year = 2020,
	month = feb,
	volume = {889},
	number = {2},
	eid = {161},
	pages = {161},
	doi = {10.3847/1538-4357/ab6604},
	archiveprefix = {arXiv},
	eprint = {1909.07396},
	primaryclass = {astro-ph.GA},
	adsurl = {https://ui.adsabs.harvard.edu/abs/2020ApJ...889..161N}
}

@misc{nist_asd,
	author = {A.~Kramida and {Yu.~Ralchenko} and J.~Reader and {and NIST ASD Team}},
	howpublished = {{NIST Atomic Spectra Database (ver. 5.1), [Online]. Available: {\tt{http://physics.nist.gov/asd}} [2014, March 24]. National Institute of Standards and Technology, Gaithersburg, MD.}},
	year = {2013}
}

@book{osterbrock,
	author = {{Osterbrock}, D.~E. and {Ferland}, G.~J.},
	title = "{Astrophysics of gaseous nebulae and active galactic nuclei}",
	booktitle = {Astrophysics of gaseous nebulae and active galactic nuclei, 2nd.~ed.~by D.E.~Osterbrock and G.J.~Ferland.~Sausalito, CA: University Science Books, 2006},
	year = 2006,
	adsurl = {http://adsabs.harvard.edu/abs/2006agna.book.....O}
}

@article{ostlin2015,
	author = {{{\"O}stlin}, G. and {Marquart}, T. and {Cumming}, R.~J. and {Fathi}, K. and {Bergvall}, N. and {Adamo}, A. and {Amram}, P. and {Hayes}, M.},
	title = "{Kinematics of Haro 11: The miniature Antennae}",
	journal = {\aap},
	archiveprefix = "arXiv",
	eprint = {1508.00541},
	year = 2015,
	month = nov,
	volume = 583,
	eid = {A55},
	pages = {A55},
	doi = {10.1051/0004-6361/201323233},
	adsurl = {http://adsabs.harvard.edu/abs/2015A\%26A...583A..55O}
}

@article{ostlin2021,
	author = {{{\"O}stlin}, G{\"o}ran and {Rivera-Thorsen}, T. Emil and {Menacho}, Veronica and {Hayes}, Matthew and {Runnholm}, Axel and {Micheva}, Genoveva and {Oey}, M.~S. and {Adamo}, Angela and {Bik}, Arjan and {Cannon}, John M. and {Gronke}, Max and {Kunth}, Daniel and {Laursen}, Peter and {Mas-Hesse}, J. Miguel and {Melinder}, Jens and {Messa}, Matteo and {Sirressi}, Mattia and {Smith}, Linda},
	title = "{The Source of Leaking Ionizing Photons from Haro11: Clues from HST/COS Spectroscopy of Knots A, B, and C}",
	journal = {\apj},
	year = 2021,
	month = may,
	volume = {912},
	number = {2},
	eid = {155},
	pages = {155},
	doi = {10.3847/1538-4357/abf1e8},
	archiveprefix = {arXiv},
	eprint = {2103.15854},
	primaryclass = {astro-ph.GA},
	adsurl = {https://ui.adsabs.harvard.edu/abs/2021ApJ...912..155O}
}

@article{pahl2021,
	author = {{Pahl}, Anthony J. and {Shapley}, Alice and {Steidel}, Charles C. and {Chen}, Yuguang and {Reddy}, Naveen A.},
	title = "{An uncontaminated measurement of the escaping Lyman continuum at z   3}",
	journal = {\mnras},
	year = 2021,
	month = aug,
	volume = {505},
	number = {2},
	pages = {2447--2467},
	doi = {10.1093/mnras/stab1374},
	archiveprefix = {arXiv},
	eprint = {2104.02081},
	primaryclass = {astro-ph.GA},
	adsurl = {https://ui.adsabs.harvard.edu/abs/2021MNRAS.505.2447P}
}

@inproceedings{pandas,
	author = {Wes McKinney},
	title = {Data Structures for Statistical Computing in Python},
	booktitle = {Proceedings of the 9th Python in Science Conference},
	pages = {51--56},
	year = {2010},
	editor = {St\'efan van der Walt and Jarrod Millman}
}

@ARTICLE{pascale2023,
	author = {{Pascale}, Massimo and {Dai}, Liang and {McKee}, Christopher F. and {Tsang}, Benny T. -H.},
	title = "{Nitrogen-enriched, Highly Pressurized Nebular Clouds Surrounding a Super Star Cluster at Cosmic Noon}",
	journal = {\apj},
	year = 2023,
	month = nov,
	volume = {957},
	number = {2},
	eid = {77},
	pages = {77},
	doi = {10.3847/1538-4357/acf75c},
	archivePrefix = {arXiv},
	eprint = {2301.10790},
	primaryClass = {astro-ph.GA},
	adsurl = {https://ui.adsabs.harvard.edu/abs/2023ApJ...957...77P}
}

@article{pignataro2021,
	author = {{Pignataro}, G.~V. and {Bergamini}, P. and {Meneghetti}, M. and {Vanzella}, E. and {Calura}, F. and {Grillo}, C. and {Rosati}, P. and {Angora}, G. and {Brammer}, G. and {Caminha}, G.~B. and {Mercurio}, A. and {Nonino}, M. and {Tozzi}, P.},
	title = "{A strong lensing model of the galaxy cluster PSZ1 G311.65-18.48}",
	journal = {\aap},
	year = 2021,
	month = nov,
	volume = {655},
	eid = {A81},
	pages = {A81},
	doi = {10.1051/0004-6361/202141586},
	archiveprefix = {arXiv},
	eprint = {2106.10286},
	primaryclass = {astro-ph.GA},
	adsurl = {https://ui.adsabs.harvard.edu/abs/2021A\&A...655A..81P}
}

@article{plancksz2014,
	author = {{Planck Collaboration} and {Ade}, P.~A.~R. and {Aghanim}, N. and {Armitage-Caplan}, C. and {Arnaud}, M. and {Ashdown}, M. and {Atrio-Barandela}, F. and {Aumont}, J. and {Aussel}, H. and {Baccigalupi}, C. and et al.},
	title = "{Planck 2013 results. XXIX. The Planck catalogue of Sunyaev-Zeldovich sources}",
	journal = {\aap},
	archiveprefix = "arXiv",
	eprint = {1303.5089},
	year = 2014,
	month = nov,
	volume = 571,
	eid = {A29},
	pages = {A29},
	doi = {10.1051/0004-6361/201321523},
	adsurl = {http://adsabs.harvard.edu/abs/2014A\%26A...571A..29P}
}

@article{puschnig2017,
	author = {{Puschnig}, J. and {Hayes}, M. and {{\"O}stlin}, G. and {Rivera-Thorsen}, T.~E. and {Melinder}, J. and {Cannon}, J.~M. and {Menacho}, V. and {Zackrisson}, E. and {Bergvall}, N. and {Leitet}, E.},
	title = "{The Lyman continuum escape and ISM properties in Tololo 1247-232 - new insights from HST and VLA$^{★}$}",
	journal = {\mnras},
	archiveprefix = "arXiv",
	eprint = {1704.05943},
	year = 2017,
	month = aug,
	volume = 469,
	pages = {3252--3269},
	doi = {10.1093/mnras/stx951},
	adsurl = {http://adsabs.harvard.edu/abs/2017MNRAS.469.3252P}
}

@ARTICLE{puskas2025,
	author = {{Pusk{\'a}s}, D{\'a}vid and {Tacchella}, Sandro and {Simmonds}, Charlotte and {Hainline}, Kevin and {D'Eugenio}, Francesco and {Alberts}, Stacey and {Arribas}, Santiago and {Baker}, William M. and {Bunker}, Andrew J. and {Carniani}, Stefano and {Charlot}, St{\'e}phane and {Duan}, Qiao and {Eisenstein}, Daniel J. and {Ji}, Zhiyuan and {Johnson}, Benjamin D. and {Jones}, Gareth C. and {Maiolino}, Roberto and {McClymont}, William and {Rieke}, Marcia and {Rinaldi}, Pierluigi and {Robertson}, Brant and {{\"U}bler}, Hannah and {Williams}, Christina C. and {Willmer}, Christopher N.~A. and {Willott}, Chris and {Witstok}, Joris},
	title = "{Constraining the major merger history of z \raisebox{-0.5ex}\textasciitilde 3-9 galaxies using JADES: dominant in situ star formation}",
	journal = {\mnras},
	year = 2025,
	month = jul,
	volume = {540},
	number = {3},
	pages = {2146-2175},
	doi = {10.1093/mnras/staf813},
	archivePrefix = {arXiv},
	eprint = {2502.01721},
	primaryClass = {astro-ph.GA},
	adsurl = {https://ui.adsabs.harvard.edu/abs/2025MNRAS.540.2146P}
}

@ARTICLE{rauscher23_nsclean,
	author = {{Rauscher}, Bernard J.},
	title = "{NSClean: An Algorithm for Removing Correlated Noise from JWST NIRSpec Images}",
	journal = {arXiv e-prints},
	year = 2023,
	month = jun,
	eid = {arXiv:2306.03250},
	pages = {arXiv:2306.03250},
	doi = {10.48550/arXiv.2306.03250},
	archivePrefix = {arXiv},
	eprint = {2306.03250},
	primaryClass = {astro-ph.IM},
	adsurl = {https://ui.adsabs.harvard.edu/abs/2023arXiv230603250R}
}

@ARTICLE{rieke2023,
	author = {{Rieke}, Marcia J. and {Kelly}, Douglas M. and {Misselt}, Karl and {Stansberry}, John and {Boyer}, Martha and {Beatty}, Thomas and {Egami}, Eiichi and {Florian}, Michael and {Greene}, Thomas P. and {Hainline}, Kevin and {Leisenring}, Jarron and {Roellig}, Thomas and {Schlawin}, Everett and {Sun}, Fengwu and {Tinnin}, Lee and {Williams}, Christina C. and {Willmer}, Christopher N.~A. and {Wilson}, Debra and {Clark}, Charles R. and {Rohrbach}, Scott and {Brooks}, Brian and {Canipe}, Alicia and {Correnti}, Matteo and {DiFelice}, Audrey and {Gennaro}, Mario and {Girard}, Julien H. and {Hartig}, George and {Hilbert}, Bryan and {Koekemoer}, Anton M. and {Nikolov}, Nikolay K. and {Pirzkal}, Norbert and {Rest}, Armin and {Robberto}, Massimo and {Sunnquist}, Ben and {Telfer}, Randal and {Wu}, Chi Rai and {Ferry}, Malcolm and {Lewis}, Dan and {Baum}, Stefi and {Beichman}, Charles and {Doyon}, Ren{\'e} and {Dressler}, Alan and {Eisenstein}, Daniel J. and {Ferrarese}, Laura and {Hodapp}, Klaus and {Horner}, Scott and {Jaffe}, Daniel T. and {Johnstone}, Doug and {Krist}, John and {Martin}, Peter and {McCarthy}, Donald W. and {Meyer}, Michael and {Rieke}, George H. and {Trauger}, John and {Young}, Erick T.},
	title = "{Performance of NIRCam on JWST in Flight}",
	journal = {\pasp},
	year = 2023,
	month = feb,
	volume = {135},
	number = {1044},
	eid = {028001},
	pages = {028001},
	doi = {10.1088/1538-3873/acac53},
	archivePrefix = {arXiv},
	eprint = {2212.12069},
	primaryClass = {astro-ph.IM},
	adsurl = {https://ui.adsabs.harvard.edu/abs/2023PASP..135b8001R}
}

@ARTICLE{rigby23_overview,
	author = {{Rigby}, Jane R. and {Vieira}, Joaquin D. and {Phadke}, Kedar A. and {Hutchison}, Taylor A. and {Welch}, Brian and {Cathey}, Jared and {Spilker}, Justin S. and {Gonzalez}, Anthony H. and {Adhikari}, Prasanna and {Aravena}, M. and {Bayliss}, Matthew B. and {Birkin}, Jack E. and {Bursk}, Emmy and {Chapman}, Scott C. and {Dahle}, H{\r{a}}kon and {Elicker}, Lauren A. and {Fischer}, Travis C. and {Florian}, Michael K. and {Gladders}, Michael D. and {Hayward}, Christopher C. and {Hewald}, Rose and {Kettler}, Lily A. and {Khullar}, Gourav and {Kim}, Seonwoo and {Law}, David R. and {Mahler}, Guillaume and {Malhotra}, Sangeeta and {Murphy}, Eric J. and {Narayanan}, Desika and {Olivier}, Grace M. and {Rhoads}, James E. and {Sharon}, Keren and {Solimano}, Manuel and {Thiruvengadam}, Athish and {Vizgan}, David and {Younker}, Nikolas},
	title = "{JWST Early Release Science Program TEMPLATES: Targeting Extremely Magnified Panchromatic Lensed Arcs and their Extended Star formation}",
	journal = {arXiv e-prints},
	year = 2023,
	month = dec,
	eid = {arXiv:2312.10465},
	pages = {arXiv:2312.10465},
	doi = {10.48550/arXiv.2312.10465},
	archivePrefix = {arXiv},
	eprint = {2312.10465},
	primaryClass = {astro-ph.GA},
	adsurl = {https://ui.adsabs.harvard.edu/abs/2023arXiv231210465R}
}

@article{riverathorsen2017,
	author = {{Rivera-Thorsen}, T.~E. and {{\"O}stlin}, G. and {Hayes}, M. and {Puschnig}, J.},
	title = "{Neutral ISM, Ly{$\alpha$}, and Lyman-continuum in the Nearby Starburst Haro11}",
	journal = {\apj},
	archiveprefix = "arXiv",
	eprint = {1701.08024},
	year = 2017,
	month = mar,
	volume = 837,
	eid = {29},
	pages = {29},
	doi = {10.3847/1538-4357/aa5d0a},
	adsurl = {http://adsabs.harvard.edu/abs/2017ApJ...837...29R}
}

@article{riverathorsen2019,
	author = {{Rivera-Thorsen}, T. Emil and {Dahle}, H{\r{a}}kon and {Chisholm}, John and {Florian}, Michael K. and {Gronke}, Max and {Rigby}, Jane R. and {Gladders}, Michael D. and {Mahler}, Guillaume and {Sharon}, Keren and {Bayliss}, Matthew},
	title = "{Gravitational lensing reveals ionizing ultraviolet photons escaping from a distant galaxy}",
	journal = {Science},
	year = "2019",
	month = "Nov",
	volume = {366},
	number = {6466},
	pages = {738--741},
	doi = {10.1126/science.aaw0978},
	archiveprefix = {arXiv},
	eprint = {1904.08186},
	primaryclass = {astro-ph.GA},
	adsurl = {https://ui.adsabs.harvard.edu/abs/2019Sci...366..738R}
}

@article{riverathorsen2022,
	author = {Rivera-Thorsen, T. E. and Hayes, M. and Melinder, J.},
	title = {A bottom-up search for Lyman-continuum leakage in the Hubble Ultra Deep Field},
	journal = {\aap},
	year = 2022,
	volume = 666,
	month = {Oct},
	pages = {A145},
	issn = {1432-0746},
	doi = {10.1051/0004-6361/202243678},
	url = {http://dx.doi.org/10.1051/0004-6361/202243678},
	publisher = {EDP Sciences}
}

@ARTICLE{riverathorsen2024,
	author = {{Rivera-Thorsen}, T. Emil and {Chisholm}, J. and {Welch}, B. and {Rigby}, J.~R. and {Hutchison}, T. and {Florian}, M. and {Sharon}, K. and {Choe}, S. and {Dahle}, H. and {Bayliss}, M.~B. and {Khullar}, G. and {Gladders}, M. and {Hayes}, M. and {Adamo}, A. and {Owens}, M.~R. and {Kim}, K.},
	title = "{The Sunburst Arc with JWST: I. Detection of Wolf-Rayet stars injecting nitrogen into a low-metallicity, z = 2.37 proto-globular cluster leaking ionizing photons}",
	journal = {\aap},
	year = 2024,
	month = oct,
	volume = {690},
	eid = {A269},
	pages = {A269},
	doi = {10.1051/0004-6361/202450359},
	archivePrefix = {arXiv},
	eprint = {2404.08884},
	primaryClass = {astro-ph.GA},
	adsurl = {https://ui.adsabs.harvard.edu/abs/2024A&A...690A.269R}
}

@software{riverathorsen2025cubefitter,
	author = {T. Emil Rivera-Thorsen},
	title = {thriveth/CubeFitter.jl: Initialize DOI and add
                   logo
                  },
	month = jul,
	year = 2025,
	publisher = {Zenodo},
	version = {v0.3.1},
	doi = {10.5281/zenodo.15919371},
	url = {https://doi.org/10.5281/zenodo.15919371},
	swhid = {swh:1:dir:1261b4fcb32ad7b35cd450f9357abe3bc74cbbfb
                   ;origin=https://doi.org/10.5281/zenodo.15919370;vi
                   sit=swh:1:snp:d22fdd4df0d0fb9bf563bad2aadae4fcfd45
                   c501;anchor=swh:1:rel:a370ff3461889939d81e9c49117e
                   c7d82592996b;path=thriveth-CubeFitter.jl-eddea1c
                  }
}

@article{robertson2015,
	author = {{Robertson}, Brant E. and {Ellis}, Richard S. and {Furlanetto}, Steven R. and {Dunlop}, James S.},
	title = "{Cosmic Reionization and Early Star-forming Galaxies: A Joint Analysis of New Constraints from Planck and the Hubble Space Telescope}",
	journal = {\apjl},
	year = 2015,
	month = apr,
	volume = {802},
	number = {2},
	eid = {L19},
	pages = {L19},
	doi = {10.1088/2041-8205/802/2/L19},
	archiveprefix = {arXiv},
	eprint = {1502.02024},
	primaryclass = {astro-ph.CO},
	adsurl = {https://ui.adsabs.harvard.edu/abs/2015ApJ...802L..19R}
}

@ARTICLE{rosdahl2018,
	author = {{Rosdahl}, Joakim and {Katz}, Harley and {Blaizot}, J{\'e}r{\'e}my and {Kimm}, Taysun and {Michel-Dansac}, L{\'e}o and {Garel}, Thibault and {Haehnelt}, Martin and {Ocvirk}, Pierre and {Teyssier}, Romain},
	title = "{The SPHINX cosmological simulations of the first billion years: the impact of binary stars on reionization}",
	journal = {\mnras},
	year = 2018,
	month = sep,
	volume = {479},
	number = {1},
	pages = {994-1016},
	doi = {10.1093/mnras/sty1655},
	archivePrefix = {arXiv},
	eprint = {1801.07259},
	primaryClass = {astro-ph.GA},
	adsurl = {https://ui.adsabs.harvard.edu/abs/2018MNRAS.479..994R}
}

@article{saha2020,
	author = {{Saha}, Kanak and {Tandon}, Shyam N. and {Simmonds}, Charlotte and {Verhamme}, Anne and {Paswan}, Abhishek and {Schaerer}, Daniel and {Rutkowski}, Michael and {Borgohain}, Anshuman and {Elmegreen}, Bruce and {Inoue}, Akio K. and {Combes}, Francoise and {Elmegreen}, Debra and {Paalvast}, Mieke},
	title = "{AstroSat detection of Lyman continuum emission from a z = 1.42 galaxy}",
	journal = {Nature Astronomy},
	year = 2020,
	month = aug,
	doi = {10.1038/s41550-020-1173-5},
	archiveprefix = {arXiv},
	eprint = {2008.11394},
	primaryclass = {astro-ph.GA},
	adsurl = {https://ui.adsabs.harvard.edu/abs/2020NatAs.tmp..193S}
}

@article{saxena2022,
	author = {{Saxena}, A. and {Pentericci}, L. and {Ellis}, R.~S. and {Guaita}, L. and {Calabr{\`o}}, A. and {Schaerer}, D. and {Vanzella}, E. and {Amor{\'\i}n}, R. and {Bolzonella}, M. and {Castellano}, M. and {Fontanot}, F. and {Hathi}, N.~P. and {Hibon}, P. and {Llerena}, M. and {Mannucci}, F. and {Saldana-Lopez}, A. and {Talia}, M. and {Zamorani}, G.},
	title = "{No strong dependence of Lyman continuum leakage on physical properties of star-forming galaxies at {\ensuremath{\lesssim}} z {\ensuremath{\lesssim}} 3.5}",
	journal = {\mnras},
	year = 2022,
	month = mar,
	volume = {511},
	number = {1},
	pages = {120--138},
	doi = {10.1093/mnras/stab3728},
	archiveprefix = {arXiv},
	eprint = {2109.03662},
	primaryclass = {astro-ph.GA},
	adsurl = {https://ui.adsabs.harvard.edu/abs/2022MNRAS.511..120S}
}

@ARTICLE{schreiber2020,
	author = {{F{\"o}rster Schreiber}, Natascha M. and {Wuyts}, Stijn},
	title = "{Star-Forming Galaxies at Cosmic Noon}",
	journal = {\araa},
	year = 2020,
	month = aug,
	volume = {58},
	pages = {661-725},
	doi = {10.1146/annurev-astro-032620-021910},
	archivePrefix = {arXiv},
	eprint = {2010.10171},
	primaryClass = {astro-ph.GA},
	adsurl = {https://ui.adsabs.harvard.edu/abs/2020ARA&A..58..661F}
}

@ARTICLE{shajib2025arxiv,
	author = {{Shajib}, Anowar J. and {Treu}, Tommaso and {Melo}, Alejandra and {Roberts-Borsani}, Guido and {Knabel}, Shawn and {Cappellari}, Michele and {Frieman}, Joshua A.},
	title = "{An accurate measurement of the spectral resolution of the JWST Near Infrared Spectrograph}",
	journal = {arXiv e-prints},
	year = 2025,
	month = jul,
	eid = {arXiv:2507.03746},
	pages = {arXiv:2507.03746},
	doi = {10.48550/arXiv.2507.03746},
	archivePrefix = {arXiv},
	eprint = {2507.03746},
	primaryClass = {astro-ph.IM},
	adsurl = {https://ui.adsabs.harvard.edu/abs/2025arXiv250703746S}
}

@article{shapley2016,
	author = {{Shapley}, A.~E. and {Steidel}, C.~C. and {Strom}, A.~L. and {Bogosavljevi{\'c}}, M. and {Reddy}, N.~A. and {Siana}, B. and {Mostardi}, R.~E. and {Rudie}, G.~C.},
	title = "{Q1549-C25: A Clean Source of Lyman-Continuum Emission at z = 3.15}",
	journal = {\apjl},
	archiveprefix = "arXiv",
	eprint = {1606.00443},
	year = 2016,
	month = aug,
	volume = 826,
	eid = {L24},
	pages = {L24},
	doi = {10.3847/2041-8205/826/2/L24},
	adsurl = {http://adsabs.harvard.edu/abs/2016ApJ...826L..24S}
}

@ARTICLE{sharon2022,
	author = {{Sharon}, Keren and {Mahler}, Guillaume and {Rivera-Thorsen}, T. Emil and {Dahle}, H{\r{a}}kon and {Gladders}, Michael D. and {Bayliss}, Matthew B. and {Florian}, Michael K. and {Kim}, Keunho J. and {Khullar}, Gourav and {Mainali}, Ramesh and {Napier}, Kate A. and {Navarre}, Alexander and {Rigby}, Jane R. and {Remolina Gonz{\'a}lez}, Juan David and {Sharma}, Soniya},
	title = "{The Cosmic Telescope That Lenses the Sunburst Arc, PSZ1 G311.65-18.48: Strong Gravitational Lensing Model and Source Plane Analysis}",
	journal = {\apj},
	year = 2022,
	month = dec,
	volume = {941},
	number = {2},
	eid = {203},
	pages = {203},
	doi = {10.3847/1538-4357/ac927a},
	archivePrefix = {arXiv},
	eprint = {2209.03417},
	primaryClass = {astro-ph.GA},
	adsurl = {https://ui.adsabs.harvard.edu/abs/2022ApJ...941..203S}
}

@ARTICLE{solhaug2025,
	author = {{Solhaug}, Erik and {Chen}, Hsiao-Wen and {Chen}, Mandy C. and {Zahedy}, Fakhri and {Gronke}, Max and {Hamel-Bravo}, Magdalena J. and {Bayliss}, Matthew B. and {Gladders}, Michael D. and {L{\'o}pez}, Sebasti{\'a}n and {Tejos}, Nicol{\'a}s},
	title = "{Deciphering Spatially Resolved Lyman-Alpha Profiles in Reionization Analogs: The Sunburst Arc at Cosmic Noon}",
	journal = {The Open Journal of Astrophysics},
	year = 2025,
	month = apr,
	volume = {8},
	eid = {35},
	pages = {35},
	doi = {10.33232/001c.134065},
	archivePrefix = {arXiv},
	eprint = {2409.10604},
	primaryClass = {astro-ph.GA},
	adsurl = {https://ui.adsabs.harvard.edu/abs/2025OJAp....8E..35S}
}

@article{steidel2018,
	author = {{Steidel}, Charles C. and {Bogosavljevi{\'c}}, Milan and {Shapley}, Alice E. and {Reddy}, Naveen A. and {Rudie}, Gwen C. and {Pettini}, Max and {Trainor}, Ryan F. and {Strom}, Allison L.},
	title = "{The Keck Lyman Continuum Spectroscopic Survey (KLCS): The Emergent Ionizing Spectrum of Galaxies at z {\ensuremath{\sim}} 3}",
	journal = {\apj},
	year = 2018,
	month = dec,
	volume = {869},
	number = {2},
	eid = {123},
	pages = {123},
	doi = {10.3847/1538-4357/aaed28},
	archiveprefix = {arXiv},
	eprint = {1805.06071},
	primaryclass = {astro-ph.GA},
	adsurl = {https://ui.adsabs.harvard.edu/abs/2018ApJ...869..123S}
}

@article{sunburst2017,
	author = {{Rivera-Thorsen}, T.~E. and {Dahle}, H. and {Gronke}, M. and {Bayliss}, M. and {Rigby}, J.~R. and {Simcoe}, R. and {Bordoloi}, R. and {Turner}, M. and {Furesz}, G.},
	title = "{The Sunburst Arc: Direct Lyman {$\alpha$} escape observed in the brightest known lensed galaxy}",
	journal = {\aap},
	archiveprefix = "arXiv",
	eprint = {1710.09482},
	year = 2017,
	month = nov,
	volume = 608,
	eid = {L4},
	pages = {L4},
	doi = {10.1051/0004-6361/201732173},
	adsurl = {http://adsabs.harvard.edu/abs/2017A\%26A...608L...4R}
}

@article{trebitsch2017,
	author = {{Trebitsch}, Maxime and {Blaizot}, J{\'e}r{\'e}my and {Rosdahl}, Joakim and {Devriendt}, Julien and {Slyz}, Adrianne},
	title = "{Fluctuating feedback-regulated escape fraction of ionizing radiation in low-mass, high-redshift galaxies}",
	journal = {\mnras},
	year = 2017,
	month = sep,
	volume = {470},
	number = {1},
	pages = {224--239},
	doi = {10.1093/mnras/stx1060},
	archiveprefix = {arXiv},
	eprint = {1705.00941},
	primaryclass = {astro-ph.GA},
	adsurl = {https://ui.adsabs.harvard.edu/abs/2017MNRAS.470..224T}
}

@ARTICLE{vanhoof2018,
	author = {{van Hoof}, Peter A.~M.},
	title = "{Recent Development of the Atomic Line List}",
	journal = {Galaxies},
	year = 2018,
	month = jun,
	volume = {6},
	number = {2},
	eid = {63},
	pages = {63},
	doi = {10.3390/galaxies6020063},
	adsurl = {https://ui.adsabs.harvard.edu/abs/2018Galax...6...63V}
}

@article{vanzella2012,
	author = {{Vanzella}, Eros and {Guo}, Yicheng and {Giavalisco}, Mauro and {Grazian}, Andrea and {Castellano}, Marco and {Cristiani}, Stefano and {Dickinson}, Mark and {Fontana}, Adriano and {Nonino}, Mario and {Giallongo}, Emanuele and {Pentericci}, Laura and {Galametz}, Audrey and {Faber}, S.~M. and {Ferguson}, Henry C. and {Grogin}, Norman A. and {Koekemoer}, Anton M. and {Newman}, Jeffrey and {Siana}, Brian D.},
	title = "{On the Detection of Ionizing Radiation Arising from Star-forming Galaxies at Redshift z \raisebox{-0.5ex}\textasciitilde 3-4: Looking for Analogs of ``Stellar Re-ionizers''}",
	journal = {\apj},
	year = 2012,
	month = may,
	volume = {751},
	number = {1},
	eid = {70},
	pages = {70},
	doi = {10.1088/0004-637X/751/1/70},
	archiveprefix = {arXiv},
	eprint = {1201.5642},
	primaryclass = {astro-ph.CO},
	adsurl = {https://ui.adsabs.harvard.edu/abs/2012ApJ...751...70V}
}

@article{vanzella2016,
	author = {{Vanzella}, E. and {de Barros}, S. and {Vasei}, K. and {Alavi}, A. and {Giavalisco}, M. and {Siana}, B. and {Grazian}, A. and {Hasinger}, G. and {Suh}, H. and {Cappelluti}, N. and {Vito}, F. and {Amorin}, R. and {Balestra}, I. and {Brusa}, M. and {Calura}, F. and {Castellano}, M. and {Comastri}, A. and {Fontana}, A. and {Gilli}, R. and {Mignoli}, M. and {Pentericci}, L. and {Vignali}, C. and {Zamorani}, G.},
	title = "{Hubble Imaging of the Ionizing Radiation from a Star-forming Galaxy at Z=3.2 with fescundefined\%}",
	journal = {\apj},
	archiveprefix = "arXiv",
	eprint = {1602.00688},
	year = 2016,
	month = jul,
	volume = 825,
	eid = {41},
	pages = {41},
	doi = {10.3847/0004-637X/825/1/41},
	adsurl = {http://adsabs.harvard.edu/abs/2016ApJ...825...41V}
}

@article{vanzella2018,
	author = {{Vanzella}, E. and {Nonino}, M. and {Cupani}, G. and {Castellano}, M. and {Sani}, E. and {Mignoli}, M. and {Calura}, F. and {Meneghetti}, M. and {Gilli}, R. and {Comastri}, A. and {Mercurio}, A. and {Caminha}, G.~B. and {Caputi}, K. and {Rosati}, P. and {Grillo}, C. and {Cristiani}, S. and {Balestra}, I. and {Fontana}, A. and {Giavalisco}, M.},
	title = "{Direct Lyman continuum and Ly {$\alpha$} escape observed at redshift 4}",
	journal = {\mnras},
	archiveprefix = "arXiv",
	eprint = {1712.07661},
	year = 2018,
	month = may,
	volume = 476,
	pages = {L15-L19},
	doi = {10.1093/mnrasl/sly023},
	adsurl = {http://adsabs.harvard.edu/abs/2018MNRAS.476L..15V}
}

@article{vanzella2020,
	author = {{Vanzella}, E. and {Meneghetti}, M. and {Pastorello}, A. and {Calura}, F. and {Sani}, E. and {Cupani}, G. and {Caminha}, G.~B. and {Castellano}, M. and {Rosati}, P. and {D'Odorico}, V. and {Cristiani}, S. and {Grillo}, C. and {Mercurio}, A. and {Nonino}, M. and {Brammer}, G.~B. and {Hartman}, H.},
	title = "{Probing the circumstellar medium 2.8 Gyr after the big bang: detection of Bowen fluorescence in the Sunburst arc}",
	journal = {\mnras},
	year = 2020,
	month = sep,
	volume = {499},
	number = {1},
	pages = {L67-L71},
	doi = {10.1093/mnrasl/slaa163},
	archiveprefix = {arXiv},
	eprint = {2004.08400},
	primaryclass = {astro-ph.GA},
	adsurl = {https://ui.adsabs.harvard.edu/abs/2020MNRAS.499L..67V}
}

@article{verhamme2006,
	author = {{Verhamme}, A. and {Schaerer}, D. and {Maselli}, A.},
	title = "{3D Ly{$\alpha$} radiation transfer. I. Understanding Ly{\$\alpha\$} line profile morphologies}",
	journal = {\aap},
	eprint = {astro-ph/0608075},
	year = 2006,
	month = dec,
	volume = 460,
	pages = {397--413},
	doi = {10.1051/0004-6361:20065554},
	adsurl = {http://adsabs.harvard.edu/abs/2006A\%26A...460..397V}
}

@article{verhamme2008,
	author = {{Verhamme}, A. and {Schaerer}, D. and {Atek}, H. and {Tapken}, C.},
	title = "{3D Ly{$\alpha$} radiation transfer. III. Constraints on gas and stellar properties of z \~{} 3 Lyman break galaxies (LBG) and implications for high-z LBGs and Ly{\$\alpha\$} emitters}",
	journal = {\aap},
	archiveprefix = "arXiv",
	eprint = {0805.3601},
	year = 2008,
	month = nov,
	volume = 491,
	pages = {89--111},
	doi = {10.1051/0004-6361:200809648},
	adsurl = {http://adsabs.harvard.edu/abs/2008A\%26A...491...89V}
}

@article{verhamme2015,
	author = {{Verhamme}, A. and {Orlitov{\'a}}, I. and {Schaerer}, D. and {Hayes}, M.},
	title = "{Using Lyman-{$\alpha$} to detect galaxies that leak Lyman continuum}",
	journal = {\aap},
	archiveprefix = "arXiv",
	eprint = {1404.2958},
	year = 2015,
	month = jun,
	volume = 578,
	eid = {A7},
	pages = {A7},
	doi = {10.1051/0004-6361/201423978},
	adsurl = {http://adsabs.harvard.edu/abs/2015A\%26A...578A...7V}
}

@article{wang2019,
	author = {{Wang}, Bingjie and {Heckman}, Timothy M. and {Leitherer}, Claus and {Alexandroff}, Rachel and {Borthakur}, Sanchayeeta and {Overzier}, Roderik A.},
	title = "{A New Technique for Finding Galaxies Leaking Lyman-continuum Radiation: [S II]-deficiency}",
	journal = {\apj},
	year = 2019,
	month = nov,
	volume = {885},
	number = {1},
	eid = {57},
	pages = {57},
	doi = {10.3847/1538-4357/ab418f},
	archiveprefix = {arXiv},
	eprint = {1909.01368},
	primaryclass = {astro-ph.GA},
	adsurl = {https://ui.adsabs.harvard.edu/abs/2019ApJ...885...57W}
}

@article{wang2021,
	author = {{Wang}, Bingjie and {Heckman}, Timothy M. and {Amor{\'\i}n}, Ricardo and {Borthakur}, Sanchayeeta and {Chisholm}, John and {Ferguson}, Harry and {Flury}, Sophia and {Giavalisco}, Mauro and {Grazian}, Andrea and {Hayes}, Matthew and {Henry}, Alaina and {Jaskot}, Anne and {Ji}, Zhiyuan and {Makan}, Kirill and {McCandliss}, Stephan and {Oey}, M.~S. and {{\"O}stlin}, G{\"o}ran and {Saldana-Lopez}, Alberto and {Schaerer}, Daniel and {Thuan}, Trinh and {Worseck}, G{\'a}bor and {Xu}, Xinfeng},
	title = "{The Low-redshift Lyman-continuum Survey: [S II] Deficiency and the Leakage of Ionizing Radiation}",
	journal = {\apj},
	year = 2021,
	month = jul,
	volume = {916},
	number = {1},
	eid = {3},
	pages = {3},
	doi = {10.3847/1538-4357/ac0434},
	archiveprefix = {arXiv},
	eprint = {2104.03432},
	primaryclass = {astro-ph.GA},
	adsurl = {https://ui.adsabs.harvard.edu/abs/2021ApJ...916....3W}
}

@ARTICLE{welch2025,
	author = {{Welch}, Brian and {Rivera-Thorsen}, T. Emil and {Rigby}, Jane R. and {Hutchison}, Taylor A. and {Olivier}, Grace M. and {Berg}, Danielle A. and {Sharon}, Keren and {Dahle}, H{\r{a}}kon and {Owens}, M. Riley and {Bayliss}, Matthew B. and {Khullar}, Gourav and {Chisholm}, John and {Hayes}, Matthew and {Kim}, Keunho J.},
	title = "{The Sunburst Arc with JWST. III. An Abundance of Direct Chemical Abundances}",
	journal = {\apj},
	year = 2025,
	month = feb,
	volume = {980},
	number = {1},
	eid = {33},
	pages = {33},
	doi = {10.3847/1538-4357/ada76c},
	adsurl = {https://ui.adsabs.harvard.edu/abs/2025ApJ...980...33W}
}

@ARTICLE{xu2023,
	author = {{Xu}, Xinfeng and {Henry}, Alaina and {Heckman}, Timothy and {Chisholm}, John and {Marques-Chaves}, Rui and {Leclercq}, Floriane and {Berg}, Danielle A. and {Jaskot}, Anne and {Schaerer}, Daniel and {Worseck}, G{\'a}bor and {Amor{\'\i}n}, Ricardo O. and {Atek}, Hakim and {Hayes}, Matthew and {Ji}, Zhiyuan and {{\"O}stlin}, G{\"o}ran and {Saldana-Lopez}, Alberto and {Thuan}, Trinh},
	title = "{The Low-Redshift Lyman Continuum Survey: Optically Thin and Thick Mg II Lines as Probes of Lyman Continuum Escape}",
	journal = {arXiv e-prints},
	year = 2023,
	month = jan,
	eid = {arXiv:2301.04087},
	pages = {arXiv:2301.04087},
	archivePrefix = {arXiv},
	eprint = {2301.04087},
	primaryClass = {astro-ph.GA},
	adsurl = {https://ui.adsabs.harvard.edu/abs/2023arXiv230104087X}
}

@article{zackrisson2013,
	author = {{Zackrisson}, E. and {Inoue}, A.~K. and {Jensen}, H.},
	title = "{The Spectral Evolution of the First Galaxies. II. Spectral Signatures of Lyman Continuum Leakage from Galaxies in the Reionization Epoch}",
	journal = {\apj},
	archiveprefix = "arXiv",
	eprint = {1304.6404},
	year = 2013,
	month = nov,
	volume = 777,
	eid = {39},
	pages = {39},
	doi = {10.1088/0004-637X/777/1/39},
	adsurl = {http://adsabs.harvard.edu/abs/2013ApJ...777...39Z}
}

@ARTICLE{zhu2025,
	author = {{Zhu}, Shuairu and {Zheng}, Zhen-Ya and {Yuan}, Fang-Ting and {Jiang}, Chunyan and {Lin}, Ruqiu},
	title = "{Lyman Continuum Leakers at z > 3 in the GOODS-S Field: Mergers Dominated}",
	journal = {\apjl},
	year = 2025,
	month = apr,
	volume = {982},
	number = {2},
	eid = {L58},
	pages = {L58},
	doi = {10.3847/2041-8213/adc125},
	archivePrefix = {arXiv},
	eprint = {2412.08395},
	primaryClass = {astro-ph.GA},
	adsurl = {https://ui.adsabs.harvard.edu/abs/2025ApJ...982L..58Z}
}

\begin{appendix}

\section{Masks for error estimation and source identification}
\label{sec:org7811acb}
\begin{figure}[htbp]
\centering
\includegraphics[width=.9\linewidth]{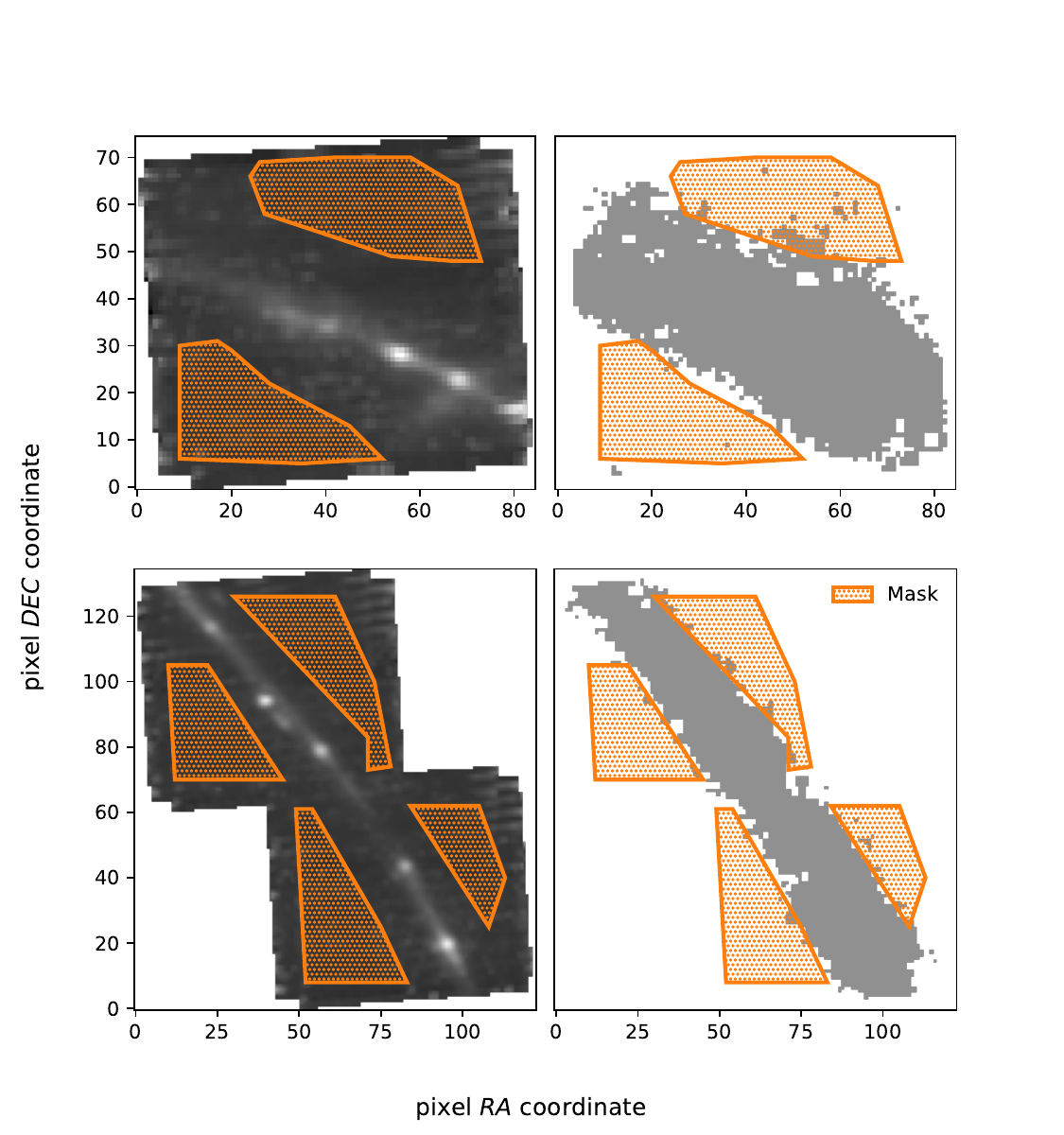}
\caption{\label{fig:noisemasks}Empty sky masks used for noise estimation. \textbf{Upper panels:} Pointing 1. \textbf{Lower Panels:} Pointing 2+3. \textbf{Left panels:} Median image of the F100LP/G140H cubes, showing stellar continuum emission, scaled as in \autoref{fig:nircamrgb}. \textbf{Right panels:} S/N < 3 masks for stacked [\ion{O}{3}] 4960+5008 \AA{} (see Sect. \ref{sec:org1427a4a}).}
\end{figure}

\begin{figure}[htbp]
\centering
\includegraphics[width=.9\linewidth]{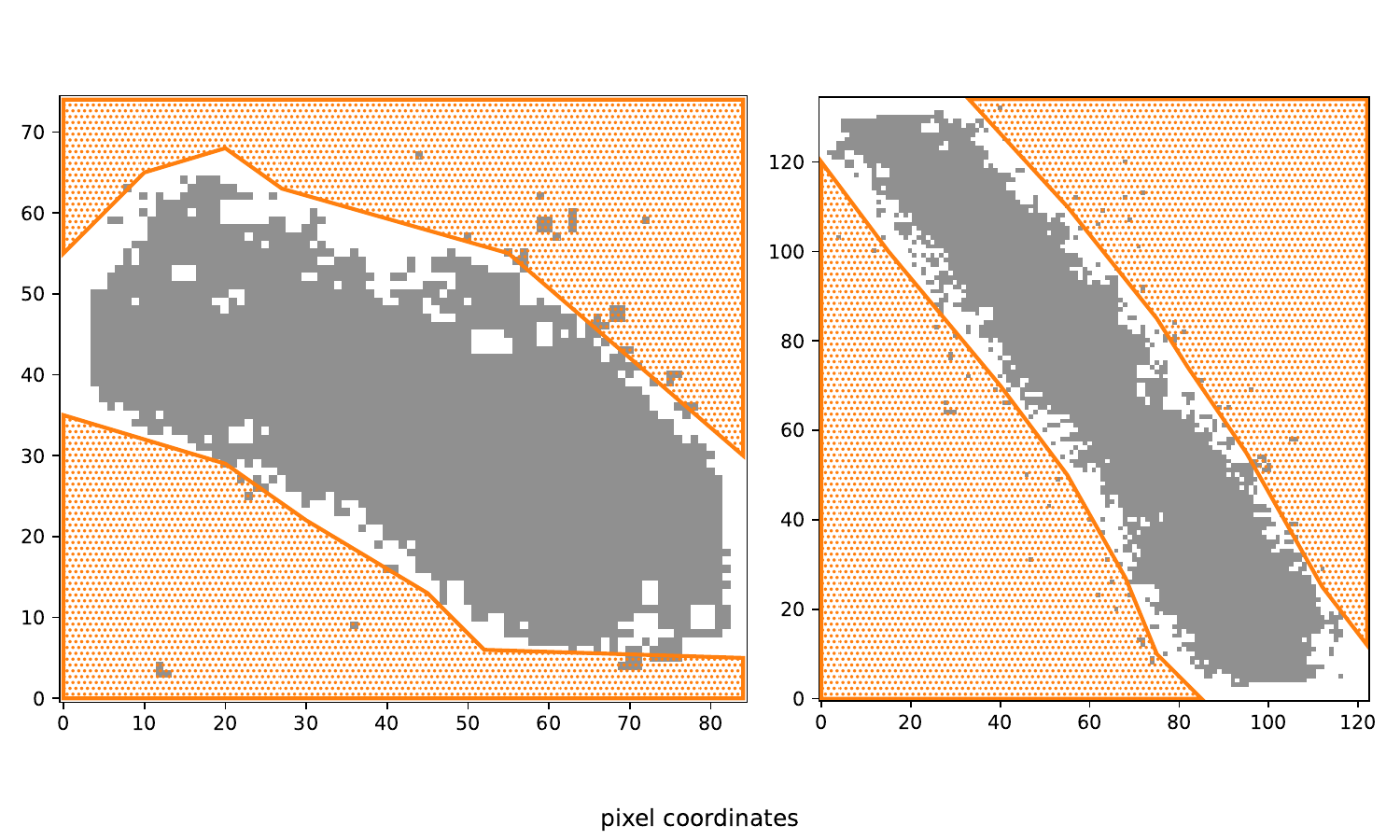}
\caption{\label{fig:manualmask}Manual and S/N based masks for P1 (\textbf{left}) and P2+3 (\textbf{right}). Gray pixels meet the requirement of S/N ([\ion{O}{3}]) > 3. Orange hatched areas are the hand-drawn masks. Only gray pixels outside these masks are included in subsequent computations.}
\end{figure}

Here, we include the masks used for noise estimation (\autoref{fig:noisemasks}) and the manual and S/N based mask identifying the observed boundaries of the arc (\autoref{fig:manualmask}), created as described in Sects. \ref{sec:errcorr} and \ref{sec:masking}.

\section{Stacking of the LCE spectrum and emission line identification in the stack \label{app:lines}}
\label{sec:orgf1e8034}
\begin{figure*}
    \centering
    \includegraphics[width=.90\linewidth]{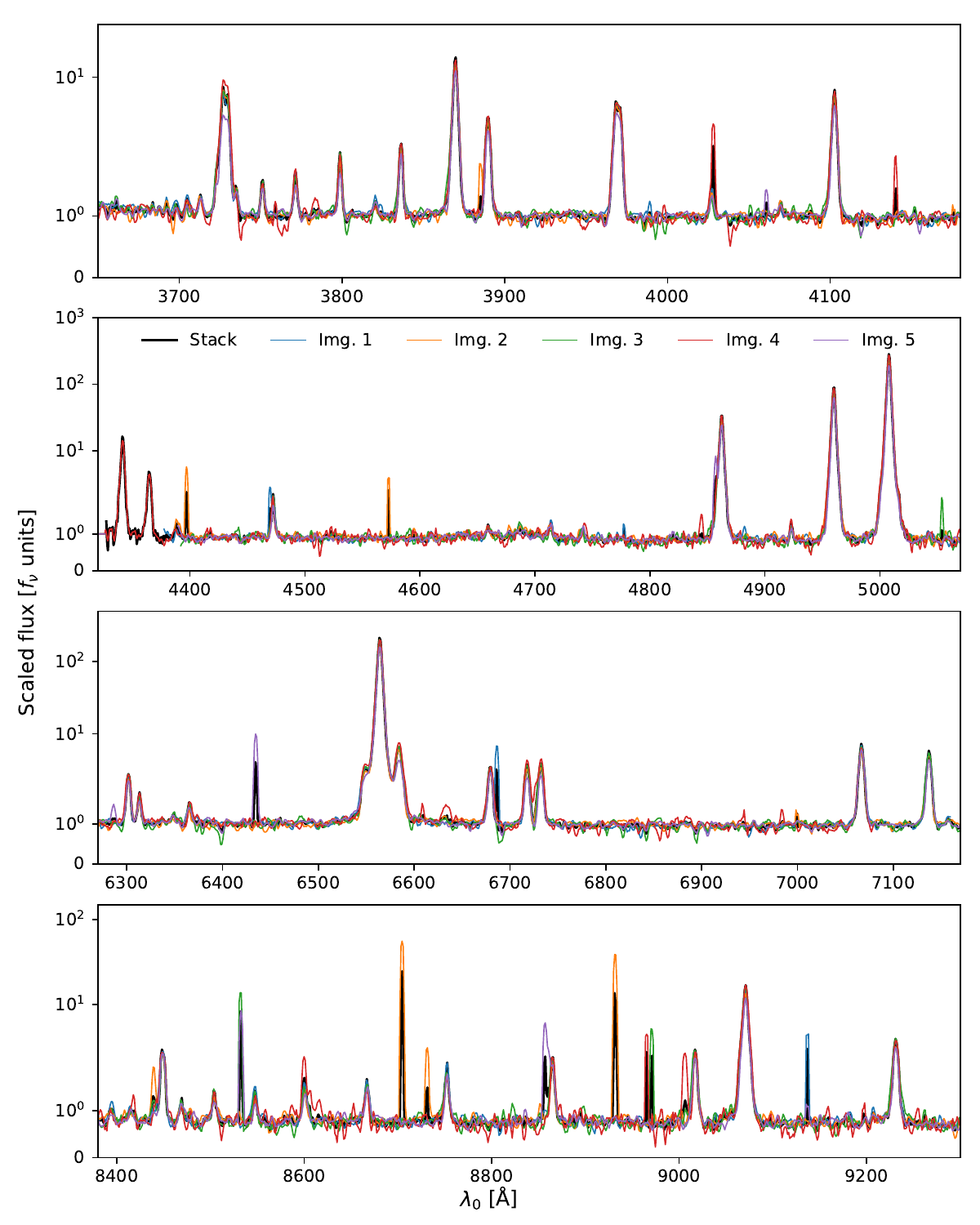}
    \caption{\label{fig:unstacked}Segments of the median-normalized, extracted spectra of the five lensed images of the LCE cluster. The stacked spectrum is shown in black for comparison. The spectra have been normalized by a single multiplicative factor each. The spectra have been smoothed by a 3 pixel boxcar kernel to dampen noise, and are shown on a ``symlog'' scale, which is linear between 0 and 1, and logarithmic above 1.}
\end{figure*}

Here, we first show in \autoref{fig:unstacked} the individual, median-normalized spectra included in the stack, along with the stacked spectrum for comparison. We have opted to show selected wavelength ranges rather than the whole covered range, to allow for better comparison between e.g. line shapes and other narrow features. We do however note that the extracted spectra all are normalized by a single multiplicative factor; the remarkable agreement between both emission lines and average continuum shape across the entire wavelength range is strong evidence that these images are indeed of the same object, and the stacking does not erase information. Features due to hot pixels and other outliers are easily identified with a single colored peak above the black peak, while the other spectra are featureless in the same wavelength range. We have not attempted to remove these outliers, as they do not interfere with any of our science goals.

\begin{figure*}
\centering
\includegraphics[width=0.9\textwidth]{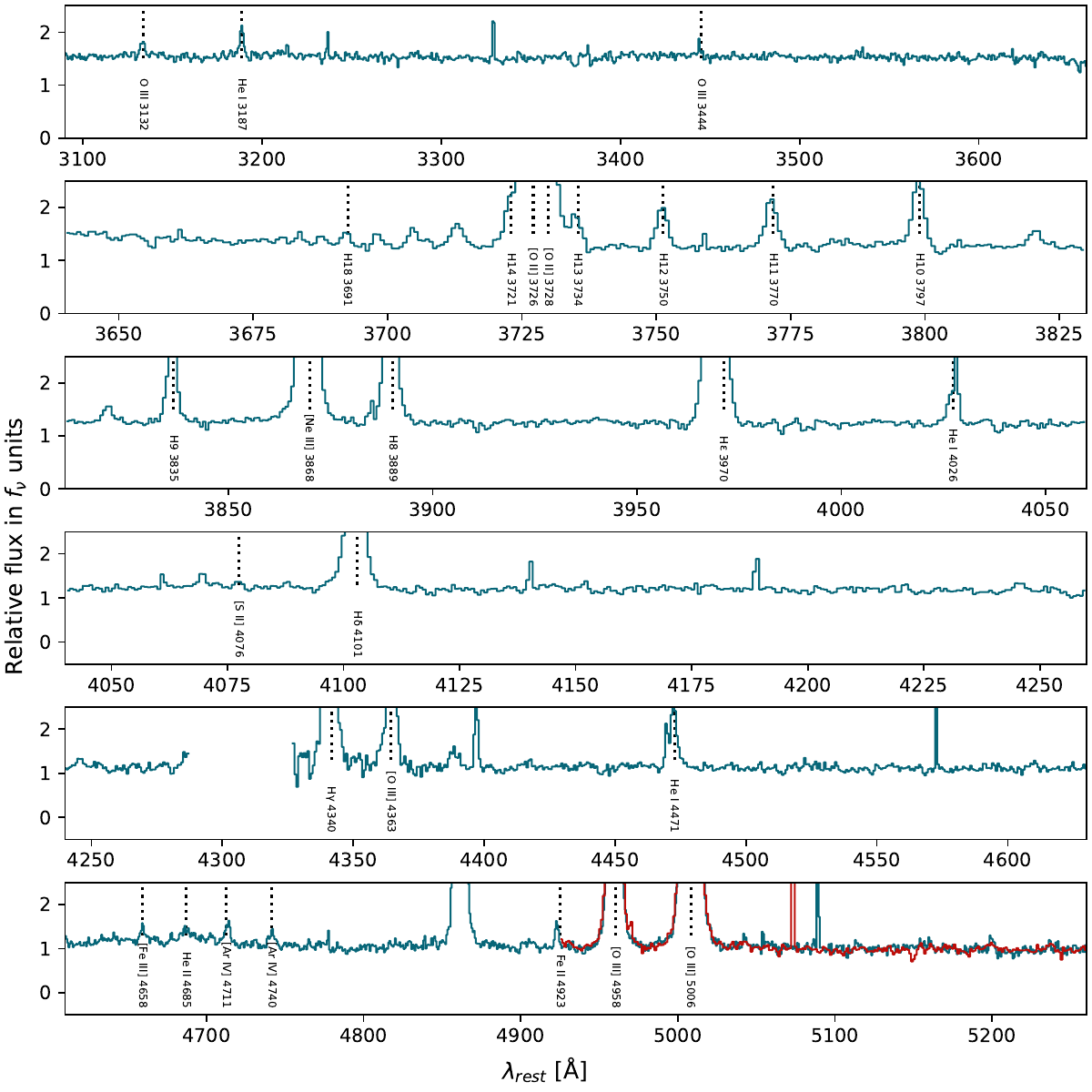}
\caption{\label{fig:lineid1}Overview of identified emission lines in F100L/G140H. Data from F170L/G235H is overlaid in the overlap region.}
\end{figure*}

\begin{figure*}
\centering
\includegraphics[width=0.9\textwidth]{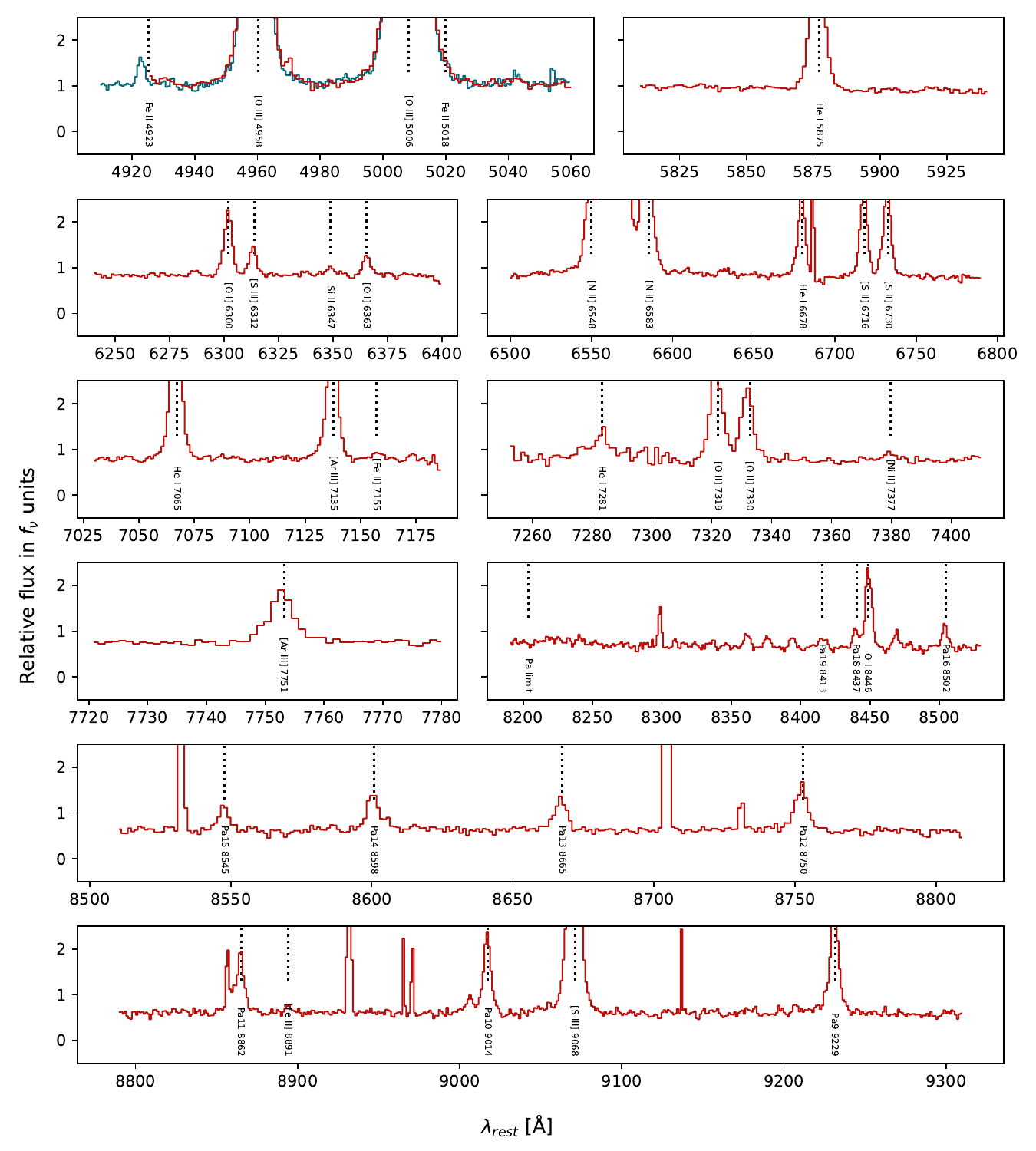}
\caption{\label{fig:lineid2}Overview of identified emission lines in F170L/G235H. Data from F100L/G140H is overlaid in the overlap region.}
\end{figure*}

Next, we show detailed line identification plots of the stacked spectrum of the Sunburst LCE. We count 60 visually identified lines, plus a number of higher-order \((n \sim 20)\) Balmer and Paschen lines which have not been labeled here. \autoref{fig:lineid1} shows the line ID plot for F100LP/G140H, while \autoref{fig:lineid2} shows F170LP/G235H.

In addition, we include in \autoref{tab:linefluxes} a full list of line fluxes measured in the stacked LCE spectrum as described in sect. \ref{sec:lcestack}. The spectra have been normalized to arbitrary units during stacking, and all fluxes are relative to H\(\beta\), for which we report the measured in Img. 4 in the table notes.  

\begin{deluxetable*}{lcc|lcc}
\tablecaption{Measured emission line fluxes in the stacked LCE spectrum.}
\tablehead{
\colhead{Line} & \colhead{\(\lambda_{\text{rest}}\)} & \colhead{F/F(H\(\beta\))} & \colhead{Line} & \colhead{\(\lambda_{\text{rest}}\)} & \colhead{F/F(H\(\beta\))}
}
\colnumbers
\startdata
{}He~I~3187     & $3188.67$ & $0.021  \pm 0.003$  & [O~I]~6302    & $6302.05$ & $0.026 \pm 0.004$   \\
{}[Ne~III]~3343 & $3343.50$ & $0.0033 \pm 0.0008$ & [S~III]~631   & $6313.80$ & $0.010 \pm 0.001$   \\
{}O~III~3444    & $3445.04$ & $0.005 \pm 0.001$   & Si~II~6347    & $6348.86$ & $0.0031 \pm 0.0009$ \\
{}H18           & $3692.61$ & $0.004  \pm 0.002$  & [O~I]~6365    & $6365.54$ & $0.0031 \pm 0.0009$ \\
{}H17           & $3698.20$ & $0.003  \pm 0.002$  & [N~II]~6550   & $6549.85$ & $0.040 \pm 0.005$   \\
{}H16           & $3704.91$ & $0.007  \pm 0.002$  & H$\alpha$     & $6564.61$ & $3.036 \pm 0.293$   \\
{}H15           & $3713.03$ & $0.014  \pm 0.003$  & [N~II]~6585   & $6585.28$ & $0.113 \pm 0.005$   \\
{}[O~II]~3727   & $3727.09$ & $0.28 \pm 0.03$     & He~I~6679     & $6680.00$ & $0.04 \pm 0.02$     \\
{}[O~II]~3730   & $3729.88$ & $0.21 \pm 0.03$     & [S~II]~6718   & $6718.29$ & $0.040 \pm 0.003$   \\
{}H12           & $3751.22$ & $0.020 \pm 0.002$   & [S~II]~6733   & $6732.67$ & $0.044 \pm 0.003$   \\
{}H11           & $3771.70$ & $0.026 \pm 0.002$   & He~I          & $7064.21$ & $0.08 \pm 0.01$     \\
{}H10           & $3798.98$ & $0.036 \pm 0.003$   & [Ar~III]~7138 & $7137.80$ & $0.062 \pm 0.007$   \\
{}H9            & $3836.47$ & $0.052 \pm 0.003$   & He~I~7281     & $7283.36$ & $0.0135 \pm 0.003$  \\
{}[Ne~III]      & $3869.86$ & $0.56 \pm 0.03$     & [O~II]~7322   & $7322.01$ & $0.031 \pm 0.002$   \\
{}H8            & $3890.15$ & $0.134 \pm 0.007$   & [O~II]~7332   & $7331.68$ & $0.023 \pm 0.001$   \\
{}[Ne~III]~3969 & $3968.59$ & $0.17 \pm 0.02$     & [Ni~II]~7377  & $7379.86$ & $0.0025 \pm 0.0008$ \\
{}H$\epsilon$   & $3971.20$ & $0.14 \pm 0.03$     & [Ar~III]~7753 & $7753.20$ & $0.0148 \pm 0.0009$ \\
{}He~I~4025     & $4025.12$ & $0.022 \pm 0.003$   & Pa~22         & $8361.30$ & $0.0029 \pm 0.0005$ \\
{}[S~II]~4069   & $4069.75$ & $0.0057 \pm 0.0009$ & Pa~21         & $8376.78$ & $0.0023 \pm 0.0005$ \\
{}S~II]~4076    & $4075.79$ & $0.0025 \pm 0.0008$ & Pa~20         & $8394.70$ & $0.0021 \pm 0.0005$ \\
{}H$\delta$     & $4102.89$ & $0.24 \pm 0.02$     & Pa~19         & $8415.63$ & $0.0028 \pm 0.0006$ \\
{}H$\gamma$     & $4341.68$ & $0.52 \pm 0.06$     & Pa~18         & $8440.27$ & $0.0045 \pm 0.0008$ \\
{}[Fe~II]~4360  & $4359.59$ & $0.010 \pm 0.002$   & O~I~8446      & $8448.68$ & $0.021 \pm 0.001$   \\
{}[O~III]~4363  & $4364.44$ & $0.135 \pm 0.006$   & Pa~17         & $8469.58$ & $0.0032 \pm 0.0007$ \\
{}He~I~4473     & $4472.70$ & $0.040 \pm 0.010$   & Pa~16         & $8504.82$ & $0.0052 \pm 0.0008$ \\
{}[Fe~III]~4658 & $4659.35$ & $0.008 \pm 0.002$   & Pa~15         & $8547.73$ & $0.0052 \pm 0.0007$ \\
{}[Ar~IV]~4713  & $4712.69$ & $0.010 \pm 0.003$   & Pa~14         & $8600.75$ & $0.0090 \pm 0.0009$ \\
{}[Ar~IV]~4741  & $4741.45$ & $0.010 \pm 0.002$   & Pa~13         & $8667.40$ & $0.0076 \pm 0.0008$ \\
{}H$\beta$\tablenotemark{$\star$}      & $4862.68$ & $1.000 \pm 0.113$   & Pa~12         & $8752.88$ & $0.0110 \pm 0.0009$ \\
{}Fe~II~4923    & $4923.57$ & $0.010 \pm 0.002$   & Pa~11         & $8865.22$ & $0.014 \pm 0.001$   \\
{}[O~III]~4960  & $4960.30$ & $2.22 \pm 0.32$     & Pa~10         & $9017.39$ & $0.018 \pm 0.001$   \\
{}[O~III]~5008  & $5008.24$ & $6.84 \pm 0.43$     & [S~III]~9071  & $9071.10$ & $0.151 \pm 0.007$   \\
He~I~5878       & $5877.59$ & $0.143 \pm 0.007$   & Pa~9          & $9231.55$ & $0.028 \pm 0.002$   \\
\enddata
\tablenotetext{\star}{$F(\text{H}\beta) \times \mu \text{(Img. 4)} = 1.01 \pm 0.03 \times 10^{-17} \text{erg s}^{-1} \text{cm}^{-1}$.
% \tablenotetext{}{
The demagnfied flux is \(6.6 \pm 0.2 \times 10^{-19} \text{erg s}^{-1} \text{cm}^{-1}\), 
assuming the calculated magnfication of \(\mu= 15.3\) from K. Sharon et al. (2022).}
\end{deluxetable*}
\label{tab:linefluxes}

\end{appendix}
\end{document}